\documentclass[aps,physrev,onecolumn,superscriptaddress,nofootinbib, amsmath,amssymb]{revtex4-2}

\usepackage{braket}
\usepackage{amsmath} 
\usepackage{graphicx} 
\usepackage{float} 

\usepackage[ruled,linesnumbered,longend]{algorithm2e}

\usepackage{siunitx} 

\usepackage[colorlinks,
           linkcolor=blue,
            anchorcolor=blue,
            citecolor=blue
            ]{hyperref}

\newcommand{\OurMethod}{ACQST} 
\newcommand{\SpecialProperty}{Self-Retaining Cancellation} 
\newcommand{\SelectiveSummation}{selective summation} 

\newcommand{\phasecontrol}{Phase Control} 
\newcommand{\singlecontrol}{Single Control} 
\newcommand{\doublecontrol}{Double Control} 

\newcommand{\signdecomposition}{negative-signs decomposition} 

\newcommand{\heterobit}{hetero-bit} 
\newcommand{\homobit}{homo-bit} 

\newcommand{\GateMatrix}{Gate Matrix} 
\newcommand{\GateMatrixproblem}{Gate Matrix Problem} 

\begin{document}


\title{Efficient quantum state tomography with auxiliary systems}


\author{Wenlong Zhao}
\author{Da Zhang}
    \affiliation{Center for Quantum Technology Research and Key Laboratory of Advanced Optoelectronic Quantum Architecture and Measurements (MOE), \\ School of Physics, Beijing Institute of Technology, Beijing 100081, China}
\author{Huili Zhang}
 \affiliation{Beijing Key Laboratory of Fault-Tolerant Quantum Computing, Beijing Academy of Quantum Information Sciences, Beijing
100193, China}
\author{Haifeng Yu}
     \affiliation{Beijing Key Laboratory of Fault-Tolerant Quantum Computing, Beijing Academy of Quantum Information Sciences, Beijing 100193, China}
     \affiliation{Hefei National Laboratory, Hefei 230088, China}

\author{Zhang-qi Yin}
    \email{zqyin@bit.edu.cn}
    \affiliation{Center for Quantum Technology Research and Key Laboratory of Advanced Optoelectronic Quantum Architecture and Measurements (MOE), \\ School of Physics, Beijing Institute of Technology, Beijing 100081, China}


\date{\today}

\begin{abstract}
Quantum state tomography is a technique in quantum information science used to reconstruct the density matrix of an unknown quantum state, providing complete information about the quantum state. It is of significant importance in fields such as quantum computation, quantum communication, and quantum simulation. However, as the size of the quantum system increases, the number of measurement settings and sampling requirements for quantum state tomography grow exponentially with the number of qubits. This not only makes experimental design and implementation more complex, but also exacerbates the consumption of experimental resources. These limitations severely hinder the application of state tomography in large-scale quantum systems.
To reduce measurement settings and improve sampling efficiency, this study proposes a state tomography method based on auxiliary systems. This method can be implemented through either entanglement between the quantum system to be measured and a quantum auxiliary system or through correlation between the quantum system and a probabilistic classical auxiliary system. Measurements on the entire joint system enable more efficient extraction of information about the quantum state to be measured. This method relies on standard quantum gate operations and requires only two measurement settings, with a total sampling complexity of \( O(d^2) \), significantly simplifying experimental operations and measurement processes. Additionally, this study provides two schemes for measuring purity based on the proposed circuit, one of which achieves measurement precision at the Heisenberg limit. This study validates the effectiveness of the proposed method through a detailed theoretical analysis, a series of numerical simulations, and experiments.
\end{abstract}


\maketitle


\section{Introduction} \label{sec:1}
With the development of quantum computing technology, the scale of quantum systems continues to expand, and the number of controllable qubits is steadily increasing. This advancement accelerates the practical applications of quantum computing but also imposes higher demands on the measurement and control of quantum systems. In quantum information processing, the measurement of qubits is fundamental and crucial.

Quantum state tomography (QST) is a technique used to reconstruct quantum states \cite{paris2004quantum,Toninelli:19}. When a quantum state is not pure, it can only be described using a density matrix. The density matrix provides a complete description of the quantum state, including its probability distribution, quantum entanglement, coherence, and other related information. Its form is shown in Eq.~\eqref{eq:rho}. Quantum state tomography extracts sufficient information by measuring a large number of identical copies of the quantum state, thereby reconstructing the density matrix.
\begin{equation}
\rho =
\begin{pmatrix}
P_0 & \alpha_{10} - i \beta_{10} & \alpha_{20} - i \beta_{20} & \cdots & \alpha_{(d-1)0} - i \beta_{(d-1)0} \\
\alpha_{10} + i \beta_{10} & P_1 & \alpha_{21} - i \beta_{21} & \cdots & \alpha_{(d-1)1} - i \beta_{(d-1)1} \\
\alpha_{20} + i \beta_{20} & \alpha_{21} + i \beta_{21} & P_2 & \cdots & \alpha_{(d-1)2} - i \beta_{(d-1)2} \\
\vdots & \vdots & \vdots & \ddots & \vdots \\
\alpha_{(d-1)0} + i \beta_{(d-1)0} & \alpha_{(d-1)1} + i \beta_{(d-1)1} & \alpha_{(d-1)2} + i \beta_{(d-1)2} & \cdots & P_{d-1}
\end{pmatrix}.
\label{eq:rho}
\end{equation}

\begin{figure}
    \centering
    \includegraphics[width=0.9\textwidth]{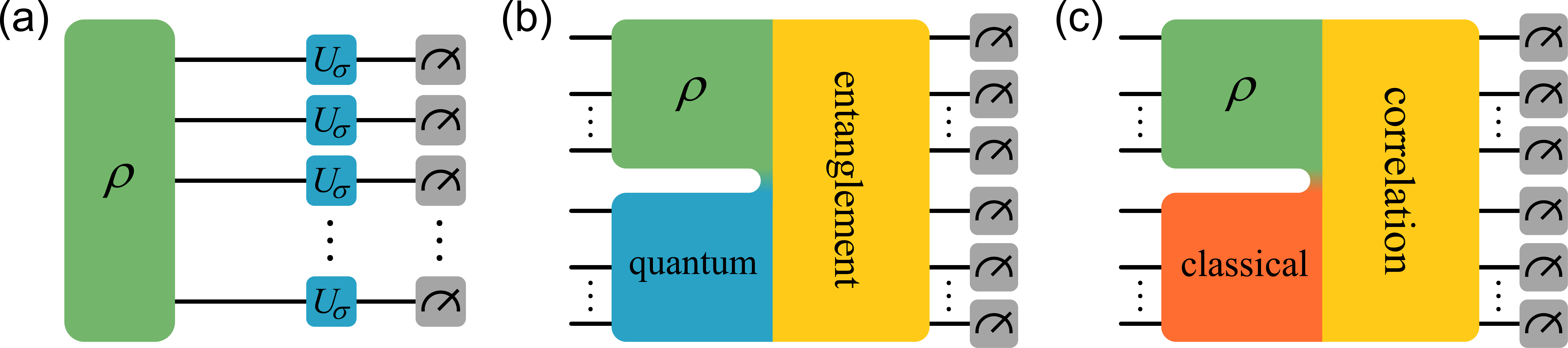}
    \caption{
        The three diagrams above illustrate the quantum circuit representations for state tomography. Panel (a) depicts the circuit for SQST, where the measurement bases for each qubit need to be adjusted to enable measurements in the X, Y, and Z directions. There are total of $3^n$ different combinations. Panel (b) shows the quantum scheme circuit for \OurMethod, which requires entangling the quantum system to be measured with a quantum auxiliary system, followed by measurements on the entire joint system. Panel (c) illustrates the classical scheme circuit for \OurMethod, where the quantum system to be measured is correlated with a probabilistic classical auxiliary system, and the entire joint system is subsequently measured.
    }
    \label{fig:circuit-diagram}
\end{figure}

However, quantum state tomography is challenging in practice. As the number of qubits increases, the complexity of the system grows exponentially. As shown in Eq.~\eqref{eq:rho}, for a system with \( n \) qubits, the dimension of its density matrix is \( d = 2^n \), which requires the measurement of \( 4^n \) unknown parameters \( P_i \), \( \alpha_{ij} \), and \( \beta_{ij} \). To fully determine these parameters, standard quantum state tomography (SQST) requires \( 3^n \) groups of experiments under different Pauli measurement bases to obtain sufficient information to reconstruct the density matrix, as shown in Fig.~\ref{fig:circuit-diagram}(a).

The exponentially growing number of measurement settings and samples required for QST drastically escalates operational costs and time overhead. To address these challenges, researchers have proposed various improved QST schemes. QST based on Mutually Unbiased Bases (MUBs) can reduce the number of experimental setups to \(2^n + 1\) and improve measurement accuracy \cite{WOOTTERS1989363,PhysRevLett.105.030406,PhysRevLett.110.143601,Lima:11}. Some methods apply compressed sensing techniques to QST to further reduce the number of experimental setups. However, these methods are more suitable for low-rank density matrix \cite{PhysRevLett.105.150401,Flammia_2012,KUENG201788,riofrio2017experimental}. Using Symmetric Informationally Complete Positive Operator-Valued Measures (SIC-POVMs) can greatly reduce the number of samples and be realized with only a single experimental setup \cite{10.1063/1.1737053,PhysRevX.5.041006,PRXQuantum.3.040310}. However, the existence of SIC-POVMs has not yet been rigorously proven in density matrix with arbitrary  dimensions. Besides, in Ref. \cite{PhysRevLett.92.120402},   auxiliary quantum systems are introduced to perform QST by solving the density matrix from the joint measurement probability equations, although no general solution has been provided for all cases. These methods can obtain sufficient information to reconstruct the density matrix with a single measurement setup, greatly simplifying the experimental process. When the quantum state has specific assumptions or symmetries, some methods can significantly improve the efficiency of QST  \cite{PhysRevLett.118.020401,wu2024investigatingpurestateuniqueness,cramer2010efficient,lanyon2017efficient,PhysRevLett.105.250403,Moroder_2012,PhysRevA.87.012109}. There are also methods that perform joint measurements on multiple copies of a quantum state to significantly reduce the number of samples \cite{10.1145/2897518.2897544,10.1145/2897518.2897585}. However, these methods have high experimental requirements, and the resources needed are difficult to meet at the current stage. Recent research has also introduced machine learning algorithms to improve the efficiency of QST \cite{quek2021adaptive,PhysRevLett.127.140502,schmale2022efficient,2019Local,torlai2018neural,PhysRevA.106.012409}. Although the transfer of these methods between different dimensions of density matrix usually faces certain difficulties.

This study is dedicated to measuring the complete density matrix in a single process and ultimately proposes a method to reconstruct the density matrix using two measurement settings. The research has produced an algorithm that simultaneously measures all off-diagonal parameters of the density matrix with equal precision, followed by an additional trivial measurement to determine all diagonal parameters of $\rho $, thereby completing the reconstruction of the entire density matrix. By introducing a quantum auxiliary system and entangling the quantum system to be measured with the auxiliary quantum system, the measurement dimension can be expanded to extract more information. Solving the system of probability equations derived from the measurement outcomes yields the density matrix of the quantum state, an idea that has also been explored in other study \cite{PhysRevLett.92.120402}.

However, this study finds that dimension expansion can be achieved not only through entanglement with a quantum auxiliary system but also through correlations, similar to entanglement, with a probabilistic classical system, avoiding the consumption of valuable qubit resources, as illustrated in Fig.~\ref{fig:circuit-diagram}(b) and Fig.~\ref{fig:circuit-diagram}(c). Furthermore, the probability equations of the algorithm exhibit the property of \SpecialProperty, where all parameters of $\rho $ can be obtained by performing simple summation operations on the measurement results, eliminating the need to solve complex equations. This makes the execution of the algorithm highly efficient. Moreover, the method of separately measuring the diagonal and off-diagonal parameters can achieve uniform error distribution. Meanwhile, the measurement performance does not degrade with the decrease of the purity of the auxiliary system, thus exhibiting greater robustness.

In the context of reconstructing density matrices, this study proposes a method called Auxiliary-Correlated Quantum State Tomography (\OurMethod{}), with a sampling complexity of $O(d^2)$ for the Frobenius norm. For the trace distance, the sampling complexity is \(O(d^3)\), matching the known lower bound for single-copy measurements \cite{gebhart_learning_2023,KUENG201788,10353129}. In quantum state tomography, MUBs and SIC-POVMs are considered optimal measurement approaches \cite{WOOTTERS1989363,Scott_2006}. The performance of the \OurMethod{} is comparable to these optimal methods. For classical schemes, if the probabilistic classical system is regarded as sampling from a series of different unitary evolutions, the process of the circuit implementation of this study is equivalent to classical shadow \cite{huang2020predicting}. Furthermore, in the special case where the number of ancillary qubits equals the number of qubits to be measured, the algorithm is equivalent to performing measurements within a set of MUBs \cite{PhysRevApplied.21.064001}.

\OurMethod{} requires only two measurement settings, significantly reducing experimental complexity. Moreover, \OurMethod{} is universally applicable, as it imposes no prior assumptions on the quantum states to be measured, thereby covering all possible quantum states. The circuit of \OurMethod{} is implemented with standard quantum gates and is directly scalable to systems with arbitrary numbers of qubits, offering significant experimental feasibility advantages.

Classical shadows enable rapid estimation of various quantum state parameters, and \OurMethod{} exhibits analogous potential in this regard. The circuit counts of \OurMethod{} can reconstruct the full density matrix, indicating that the counts contain complete information about the quantum state. Simultaneously, the method features a deterministic system of probability equations, which makes it possible to directly estimate various parameters of the quantum state through counts. This study demonstrates this with purity as an example. Based on the \OurMethod{} circuit, we propose two purity measurement methods, both with $O(d)$ complexity, where one method even achieves Heisenberg-limited precision.

This study comprehensively explores the proposed method through theoretical analysis, numerical simulations, and experimental validations. The structure of this paper is organized as follows: Section~\ref{sec:algorithm} provides a detailed introduction to the circuit design and theoretical foundations of the algorithm, including the algorithm principles in Sec.~\ref{subsec:principle} and circuit structure in Sec.~\ref{subsec:circuit}. Section~\ref{sec:algorithm-process} describes the specific steps of the algorithm, covering the density matrix reconstruction algorithm in Sec.~\ref{subsec:density-reconstruction} and the circuit-based purity computation algorithm in Sec.~\ref{subsec:purity}. Section~\ref{sec:simulation-verification} verifies the performance and validity of the algorithm through numerical simulations. Finally, Section~\ref{sec:experimental-results} presents experimental results to further evaluate the effectiveness and feasibility of the proposed method.

\section{Theoretical Foundation and Circuits for Algorithm} \label{sec:algorithm}

\subsection{Algorithm Principle} \label{subsec:principle}

In the process of quantum measurement, the data directly obtained from experiments are the counts of various measurement outcomes, namely the frequency distribution of the observed results. Based on these experimentally measured frequency distributions, combined with specific measurement settings and the measurement theory of quantum mechanics, we can infer information about the quantum state. For the tomography of an $n$-qubit multipartite state, it is impossible to reconstruct its density matrix using a single measurement setting. This is because an $n$-qubit system has only $2^n$ possible measurement outcomes in total, while its density matrix contains $4^n - 1$ independent parameters. This implies that in the process of inferring quantum state information based on probability distributions, the number of equations is fewer than the number of unknowns, which fails to meet the necessary conditions for a well-posed system of equations, thereby rendering it unsolvable for the density matrix of the quantum state. 

Therefore, to reconstruct the density matrix of a quantum state using a single measurement setting, the quantum system needs to be entangled with other systems to expand the dimensionality of the entire system, thereby increasing the number of equations. Only when the number of equations exceeds the number of unknown parameters can the density matrix of the quantum state be uniquely determined. This also requires that the number of qubits in the auxiliary system must be greater than or equal to \( n \).

In the process of \OurMethod{}, the off-diagonal elements of the density matrix can be determined in a single measurement setting. This is achieved through entanglement with the auxiliary system and the ingenious design of the method for setting quantum gates, ultimately constructing a special system of probability equations. The structure of this equation system is symmetric and evenly distributed, possessing the \SpecialProperty{} property. Such a property completely avoids the complexity of solving the equation system; it only requires selectively summing the equations to directly obtain the off-diagonal parameters of the density matrix. To further illustrate this property and process, a two-qubit example is presented below.

\begin{figure}[H]
    \centering
    \includegraphics[width=\textwidth]{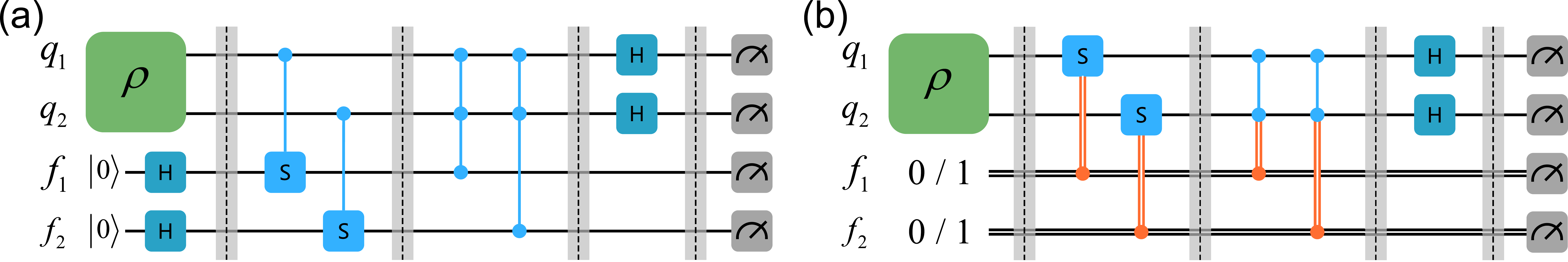}
    \caption{
        (a) shows the circuit diagram of the \OurMethod{} where all auxiliary bits use qubits, and (b) shows the circuit diagram of the \OurMethod{} where all auxiliary bits use classical bits. $q_1 q_2$ represent the qubits to be measured, and $f_1 f_2$ represent the auxiliary (qu)bits.
    }
    \label{fig:two-qubit-example}
\end{figure}

The circuit shown in Fig.~\ref{fig:two-qubit-example}(a) achieves entanglement between the quantum system to be measured and the quantum auxiliary system through a series of controlled quantum gates. In contrast, the circuit in Fig.~\ref{fig:two-qubit-example}(b) achieves correlation between the quantum system and the probabilistic classical auxiliary system via controlled quantum gates governed by a series of classical bits. Each of the two classical auxiliary bits has a probability of $1/2$ to be 0 and $1/2$ to be 1. The controlled quantum gates activated by each classical bit will only function when that specific bit is 1. The measurement results of the entire system for both circuits are identical and the probabilities of each measurement outcome \( P(q_1q_2f_1f_2) \) can be listed to derive the equation set in Eq.~\eqref{two-qubit-probability-equations}.

\begin{equation}
\label{two-qubit-probability-equations}
\begin{aligned}
P(0000) &= \frac{1}{16} + \frac{1}{8}{{\alpha }_{10}} + \frac{1}{8}{{\alpha }_{20}} + \frac{1}{8}{{\alpha }_{21}} + \frac{1}{8}{{\alpha }_{30}} + \frac{1}{8}{{\alpha }_{31}} + \frac{1}{8}{{\alpha }_{32}} \\
P(0001) &= \frac{1}{16} - \frac{1}{8}{{\beta }_{10}} + \frac{1}{8}{{\alpha }_{20}} + \frac{1}{8}{{\beta }_{21}} + \frac{1}{8}{{\beta }_{30}} - \frac{1}{8}{{\alpha }_{31}} + \frac{1}{8}{{\beta }_{32}} \\
P(0010) &= \frac{1}{16} + \frac{1}{8}{{\alpha }_{10}} - \frac{1}{8}{{\beta }_{20}} - \frac{1}{8}{{\beta }_{21}} + \frac{1}{8}{{\beta }_{30}} + \frac{1}{8}{{\beta }_{31}} - \frac{1}{8}{{\alpha }_{32}} \\
P(0011) &= \frac{1}{16} - \frac{1}{8}{{\beta }_{10}} - \frac{1}{8}{{\beta }_{20}} + \frac{1}{8}{{\alpha }_{21}} - \frac{1}{8}{{\alpha }_{30}} - \frac{1}{8}{{\beta }_{31}} - \frac{1}{8}{{\beta }_{32}} \\
P(0100) &= \frac{1}{16} - \frac{1}{8}{{\alpha }_{10}} + \frac{1}{8}{{\alpha }_{20}} - \frac{1}{8}{{\alpha }_{21}} - \frac{1}{8}{{\alpha }_{30}} + \frac{1}{8}{{\alpha }_{31}} - \frac{1}{8}{{\alpha }_{32}} \\
P(0101) &= \frac{1}{16} + \frac{1}{8}{{\beta }_{10}} + \frac{1}{8}{{\alpha }_{20}} - \frac{1}{8}{{\beta }_{21}} - \frac{1}{8}{{\beta }_{30}} - \frac{1}{8}{{\alpha }_{31}} - \frac{1}{8}{{\beta }_{32}} \\
P(0110) &= \frac{1}{16} - \frac{1}{8}{{\alpha }_{10}} - \frac{1}{8}{{\beta }_{20}} + \frac{1}{8}{{\beta }_{21}} - \frac{1}{8}{{\beta }_{30}} + \frac{1}{8}{{\beta }_{31}} + \frac{1}{8}{{\alpha }_{32}} \\
P(0111) &= \frac{1}{16} + \frac{1}{8}{{\beta }_{10}} - \frac{1}{8}{{\beta }_{20}} - \frac{1}{8}{{\alpha }_{21}} + \frac{1}{8}{{\alpha }_{30}} - \frac{1}{8}{{\beta }_{31}} + \frac{1}{8}{{\beta }_{32}} \\
P(1000) &= \frac{1}{16} + \frac{1}{8}{{\alpha }_{10}} - \frac{1}{8}{{\alpha }_{20}} - \frac{1}{8}{{\alpha }_{21}} - \frac{1}{8}{{\alpha }_{30}} - \frac{1}{8}{{\alpha }_{31}} + \frac{1}{8}{{\alpha }_{32}} \\
P(1001) &= \frac{1}{16} - \frac{1}{8}{{\beta }_{10}} - \frac{1}{8}{{\alpha }_{20}} - \frac{1}{8}{{\beta }_{21}} - \frac{1}{8}{{\beta }_{30}} + \frac{1}{8}{{\alpha }_{31}} + \frac{1}{8}{{\beta }_{32}} \\
P(1010) &= \frac{1}{16} + \frac{1}{8}{{\alpha }_{10}} + \frac{1}{8}{{\beta }_{20}} + \frac{1}{8}{{\beta }_{21}} - \frac{1}{8}{{\beta }_{30}} - \frac{1}{8}{{\beta }_{31}} - \frac{1}{8}{{\alpha }_{32}} \\
P(1011) &= \frac{1}{16} - \frac{1}{8}{{\beta }_{10}} + \frac{1}{8}{{\beta }_{20}} - \frac{1}{8}{{\alpha }_{21}} + \frac{1}{8}{{\alpha }_{30}} + \frac{1}{8}{{\beta }_{31}} - \frac{1}{8}{{\beta }_{32}} \\
P(1100) &= \frac{1}{16} - \frac{1}{8}{{\alpha }_{10}} - \frac{1}{8}{{\alpha }_{20}} + \frac{1}{8}{{\alpha }_{21}} + \frac{1}{8}{{\alpha }_{30}} - \frac{1}{8}{{\alpha }_{31}} - \frac{1}{8}{{\alpha }_{32}} \\
P(1101) &= \frac{1}{16} + \frac{1}{8}{{\beta }_{10}} - \frac{1}{8}{{\alpha }_{20}} + \frac{1}{8}{{\beta }_{21}} + \frac{1}{8}{{\beta }_{30}} + \frac{1}{8}{{\alpha }_{31}} - \frac{1}{8}{{\beta }_{32}} \\
P(1110) &= \frac{1}{16} - \frac{1}{8}{{\alpha }_{10}} + \frac{1}{8}{{\beta }_{20}} - \frac{1}{8}{{\beta }_{21}} + \frac{1}{8}{{\beta }_{30}} - \frac{1}{8}{{\beta }_{31}} + \frac{1}{8}{{\alpha }_{32}} \\
P(1111) &= \frac{1}{16} + \frac{1}{8}{{\beta }_{10}} + \frac{1}{8}{{\beta }_{20}} + \frac{1}{8}{{\alpha }_{21}} - \frac{1}{8}{{\alpha }_{30}} + \frac{1}{8}{{\beta }_{31}} + \frac{1}{8}{{\beta }_{32}}
\end{aligned}
\end{equation}

The probabilistic classical bits can match the qubits because the algorithm relies only on the entanglement property of the auxiliary system rather than its coherence. Therefore, completely mixed states or probabilistic classical bits can perfectly replace the auxiliary qubits in the algorithm.

Next, we explain the concept of the so-called \SpecialProperty{}. First, perform \SelectiveSummation{} over ${{\alpha }_{10}}$ and ${{\beta }_{10}}$, that is, summing the probability equations in Eq.~\eqref{two-qubit-probability-equations} containing $+{{\alpha }_{10}}$, $-{{\alpha }_{10}}$, $+{{\beta }_{10}}$, and $-{{\beta }_{10}}$, respectively, which yields:
\begin{equation}
\label{four-results}
\begin{aligned}
  & P\{+{{\alpha }_{10}}\} = P(0000) + P(0010) + P(1000) + P(1010) = \frac{1}{4} + \frac{1}{2}{{\alpha }_{10}} ,\\ 
  & P\{-{{\alpha }_{10}}\} = P(0100) + P(0110) + P(1100) + P(1110) = \frac{1}{4} - \frac{1}{2}{{\alpha }_{10}} ,\\ 
  & P\{+{{\beta }_{10}}\} = P(0101) + P(0111) + P(1101) + P(1111) = \frac{1}{4} + \frac{1}{2}{{\beta }_{10}} ,\\ 
  & P\{-{{\beta }_{10}}\} = P(0001) + P(0011) + P(1001) + P(1011) = \frac{1}{4} - \frac{1}{2}{{\beta }_{10}} .\\ 
\end{aligned}
\end{equation}
It can be observed that in the summation results, all parameters except for ${{\alpha }_{10}}$ and ${{\beta }_{10}}$ cancel each other out. In fact, any pair of ${{\alpha }_{ij}}$ and ${{\beta }_{ij}}$ in these equations exhibits this property. The probability equations are \SpecialProperty{}, meaning that for any pair of ${{\alpha }_{ij}}$ and ${{\beta }_{ij}}$, the \SelectiveSummation{} results retain only ${{\alpha }_{ij}}$ and ${{\beta }_{ij}}$ themselves, while all other parameters cancel out to zero.

For a general case, the algorithm is applied to an $n$-qubit system using $n_f$ auxiliary (qu)bits (which can be either quantum or classical), and its probability equations can be written as:
\begin{widetext}
\begin{equation}
\label{general-equation}
\resizebox{\textwidth}{!}{$
\renewcommand{\arraystretch}{0.7} 
\setlength{\arraycolsep}{1pt} 
\begin{array}{cccc}
   P({{q}_{1}}{{q}_{2}}\ldots {{q}_{n}}{{f}_{1}}{{f}_{2}}\ldots {{f}_{{{n}_{f}}}})=\frac{1}{{{2}^{n+{{n}_{f}}}}}+\frac{1}{{{2}^{n+{{n}_{f}}}}}2\operatorname{Re}\sum\limits_{i>j}{({{\alpha }_{ij}}+i{{\beta }_{ij}})} & {{(-1)}^{{{i}_{1}}{{q}_{1}}}}{{(-1)}^{{{j}_{1}}{{q}_{1}}}} & {{i}^{{{i}_{1}}{{f}_{1}}}}{{(-i)}^{{{j}_{1}}{{f}_{1}}}} & {{(-1)}^{{{i}_{1}}{{i}_{2}}{{x}_{1}}}}{{(-1)}^{{{j}_{1}}{{j}_{2}}{{x}_{1}}}}  \\
   {} & {{(-1)}^{{{i}_{2}}{{q}_{2}}}}{{(-1)}^{{{j}_{2}}{{q}_{2}}}} & {{i}^{{{i}_{2}}{{f}_{2}}}}{{(-i)}^{{{j}_{2}}{{f}_{2}}}} & {{(-1)}^{{{i}_{1}}{{i}_{3}}{{x}_{2}}}}{{(-1)}^{{{j}_{1}}{{j}_{3}}{{x}_{2}}}}  \\
   {} & \vdots  & \vdots  & \vdots   \\
   {} & {{(-1)}^{{{i}_{n}}{{q}_{n}}}}{{(-1)}^{{{j}_{n}}{{q}_{n}}}} & {{i}^{{{i}_{n}}{{f}_{n}}}}{{(-i)}^{{{j}_{n}}{{f}_{n}}}} & {}  \\
\end{array}.
$}
\end{equation}
\end{widetext}
The principle of \OurMethod{} lies in the density matrix utilizing auxiliary (qu)bits to modulate its own phase, thereby achieving the \SpecialProperty{}. Classical auxiliary bits share the same phase modulation mechanism as auxiliary qubits. By properly configuring the connections of various quantum gates, the phases in Eq.~\eqref{general-equation} can be adjusted, ensuring that the overall probability equations exhibit the property of \SpecialProperty{}. The specific configurations are detailed in the next section, \ref{subsec:circuit}. 

The data collected by the algorithm consist of counts, i.e., the frequencies of occurrences of readings for the qubits and auxiliary (qu)bits in repeated experiments. Based on this data, all off-diagonal elements of the target density matrix can be reconstructed. Specifically, for any pair of ${{\alpha }_{ij}}$ and ${{\beta }_{ij}}$, the counts can be divided into four categories based on \SelectiveSummation{}, with their probabilities satisfying:
\begin{equation}
\label{four-classification}
\begin{aligned}
  & {{P}_{1}} = \frac{1}{4} + \frac{1}{2}{{\alpha }_{ij}}, \quad {{P}_{3}} = \frac{1}{4} + \frac{1}{2}{{\beta }_{ij}}, \\ 
  & {{P}_{2}} = \frac{1}{4} - \frac{1}{2}{{\alpha }_{ij}}, \quad {{P}_{4}} = \frac{1}{4} - \frac{1}{2}{{\beta }_{ij}}. \\ 
\end{aligned}
\end{equation}
Let the total number of samples be $N$. Denote the summations of the counts in these four categories as ${{N}_{1}},{{N}_{2}},{{N}_{3}},{{N}_{4}}$. Note that, for simplicity of presentation, the dependence of \( P_k \) and \( N_k \) (\( k = 1, \ldots, 4 \)) on \( i, j \) is omitted here and in subsequent derivations (i.e., they should strictly be \( P_k^{ij} \) and \( N_k^{ij} \)). Define ${{N}_{\alpha }} = {{N}_{1}} + {{N}_{2}}$ and ${{N}_{\beta }} = {{N}_{3}} + {{N}_{4}}$, then the selected ${{\alpha }_{ij}}$ and ${{\beta }_{ij}}$ can be computed using the following equations:
\begin{equation}
\label{expectation-variance}
\begin{aligned}
  & {{{\hat{\alpha }}}_{ij}}=\frac{{{N}_{1}}}{{{N}_{\alpha }}}-\frac{1}{2}, \quad \text{D}({{{\hat{\alpha }}}_{ij}})\approx \frac{1-4\alpha _{ij}^{2}}{2N} ,\\ 
  & {{{\hat{\beta }}}_{ij}}=\frac{{{N}_{3}}}{{{N}_{\beta }}}-\frac{1}{2}, \quad \text{D}({{{\hat{\beta }}}_{ij}})\approx \frac{1-4\beta _{ij}^{2}}{2N} .\\ 
\end{aligned}
\end{equation}
The process can be regarded as a random sampling experiment over four distinct intervals, where the desired results are obtained by statistically analyzing the count of occurrences in each interval. From this perspective, the algorithm can be classified as a Monte Carlo method. In Eq.~\eqref{expectation-variance}, both \(\text{D}({{\hat{\alpha }}_{ij}})\) and \(\text{D}({{\hat{\beta }}_{ij}})\) are less than or equal to \( \frac{1}{2N} \). This procedure can be applied to all \({{\alpha }_{ij}}\) and \({{\beta }_{ij}}\), allowing the algorithm to simultaneously estimate all off-diagonal parameters of the density matrix with the same precision.

Considering the possible cases where \( N_{\alpha}, N_{\beta} = 0 \), there exists a relative error of \( 2^{-N} \) between the strict mathematical expectation \( \text{E}({{\hat{\alpha }}_{ij}}),\text{E}({{\hat{\beta }}_{ij}}) \) and the true values \( {{\alpha }_{ij}},{{\beta }_{ij}} \). However, since the sample size \( N \) is sufficiently large, the relative error \( 2^{-N} \) is negligible, thus it can be considered an unbiased estimate. Under the approximations \( 2^{-N} \approx 0 \) and \( 1/N \approx 1/(N+1) \), the variance of the algorithm can be approximated by the result of Eq.~\eqref{expectation-variance}, which is slightly better than the variance of the method that directly estimates probability using frequency. Therefore, the current method is adopted, and a detailed proof can be found in Appendix~\ref{appendix:A}.

Regarding the reason why all parameters of \( \rho \) are not measured simultaneously, the probability equation system is a linear function of \( \rho \), expressed as $ P=\sum{{{\omega }_{ij}}}{{\rho }_{ij}} $. Among these parameters, diagonal parameters and off-diagonal parameters behave differently. The coefficients of the off-diagonal parameters can be either positive or negative, whereas the coefficients of the diagonal parameters are always positive. This characteristic necessitates non-uniform coefficients for the diagonal parameters to be solvable in the form of equation solving. However, introducing non-uniformity to the coefficients of the diagonal elements inevitably leads to non-uniform coefficients for the off-diagonal parameters as well. This increases the complexity of parameter reconstruction and disrupts the uniformity of errors in the off-diagonal parameters. Moreover, since measuring the diagonal parameters of $\rho$ alone is relatively easy, the current study adopts a strategy of measuring the diagonal and off-diagonal parameters separately.

Since the complete measurement process is divided into two parts: diagonal element measurement and off-diagonal element measurement, the total number of measurements, \( N \), is correspondingly split into \( N_{\text{diag}} \) and \( N_{\text{off-diag}} \). The allocation of these measurements can be considered from two perspectives. First, regarding the error of each matrix element, for any diagonal element \( \rho_{ii} \) and off-diagonal element \( \rho_{ij} \), their measurement variances are given by:
\begin{equation}
\label{eq:two-variances}
\begin{aligned}
  & \text{D}({{{\hat{\rho }}}_{ii}})=\frac{{{P}_{i}}(1-{{P}_{i}})}{{{N}_{\text{diag}}}}\le \frac{1}{4{{N}_{\text{diag}}}} ,\\ 
  & \text{D}({{{\hat{\rho }}}_{ij}})=\text{D}({{{\hat{\alpha }}}_{ij}})+\text{D}({{{\hat{\beta }}}_{ij}})=\frac{1-4\alpha _{ij}^{2}}{2{{N}_{\text{off-diag}}}}+\frac{1-4\beta _{ij}^{2}}{2{{N}_{\text{off-diag}}}}\le \frac{1}{{{N}_{\text{off-diag}}}} .\\ 
\end{aligned}
\end{equation}
If we aim to ensure that all matrix elements of \( \rho \) have the same upper bound on variance, then the ratio should be chosen as 
\( {{N}_{\text{diag}}}:{{N}_{\text{off-diag}}}=1:4 \). Next, we consider the overall error of the density matrix. 

The overall error can be measured using the squared Frobenius norm, \( \left\| \hat{\rho }-\rho  \right\|_{F}^{2} \). When the dimension of \( \rho \) is \( d \), the expectation of \( \left\| \hat{\rho }-\rho  \right\|_{F}^{2} \) satisfies the following:
\begin{equation}
\label{eq:overall-error}
E\left\| \hat{\rho }-\rho  \right\|_{F}^{2}=\sum\limits_{i}{{{({{{\hat{P}}}_{i}}-{{P}_{i}})}^{2}}}+\sum\limits_{i\ne j}{(\Delta \alpha _{ij}^{2}+\Delta \beta _{ij}^{2})}\le \frac{1-\frac{1}{d}}{{{N}_{\text{diag}}}}+\frac{{{d}^{2}}-d}{{{N}_{\text{off-diag}}}}.
\end{equation}
By taking the partial derivative of Eq.~\eqref{eq:overall-error} and setting it to zero, we obtain that when ${{N}_{\text{diag}}}:{{N}_{\text{off-diag}}}=1:d$, $\left\| \hat{\rho }-\rho  \right\|_{F}^{2}$ is minimized, yielding:
\begin{equation}
\label{eq:min-overall-error}
\text{min } E\left\| \hat{\rho }-\rho  \right\|_{F}^{2}=\frac{{{d}^{2}}+d-1-\frac{1}{d}}{N}.
\end{equation}
Thus, for the entire density matrix, if $\text{E}\left\| \hat{\rho }-\rho  \right\|_{F}^{2}\le \varepsilon $, then $N\approx {({{d}^{2}}+d)}/{\varepsilon }\;$, and the sampling complexity is $O({{{d}^{2}}}/{\varepsilon }\;)$.

Furthermore, another more commonly used metric for quantifying the error of density matrices is the trace distance, defined as:
\begin{equation}
\label{eq:trace-distance}
T(\hat{\rho}, \rho) = \frac{1}{2} \mathrm{Tr} \left| \hat{\rho} - \rho \right|.
\end{equation}
For any Hermitian matrix $M$, the following inequality holds:
\begin{equation}
\label{eq:trace-inequality}
\left( \mathrm{Tr} \left| M \right| \right)^2 \leq \mathrm{rank}(M) \cdot \left\| M \right\|_{F}^{2}.
\end{equation}
Substituting Eq.~\eqref{eq:min-overall-error}, we obtain:
\begin{equation}
\label{eq:trace-distance-expectation}
\mathrm{E}\left[ T(\hat{\rho}, \rho) \right] \leq \frac{1}{2}\sqrt{\frac{d^{3} + d^{2} - d - 1}{N}}.
\end{equation}
Therefore, for the trace distance metric, to achieve $\mathrm{E}\left[ T(\hat{\rho}, \rho) \right] \leq \varepsilon$, the sampling complexity is $O(d^3/\varepsilon^2)$, achieving the optimal scaling for single-copy measurements \cite{gebhart_learning_2023,KUENG201788,10353129}.

\subsection{Quantum Circuit Structure} \label{subsec:circuit}

\begin{figure}[H]  
    \centering
    \includegraphics[width=\textwidth]{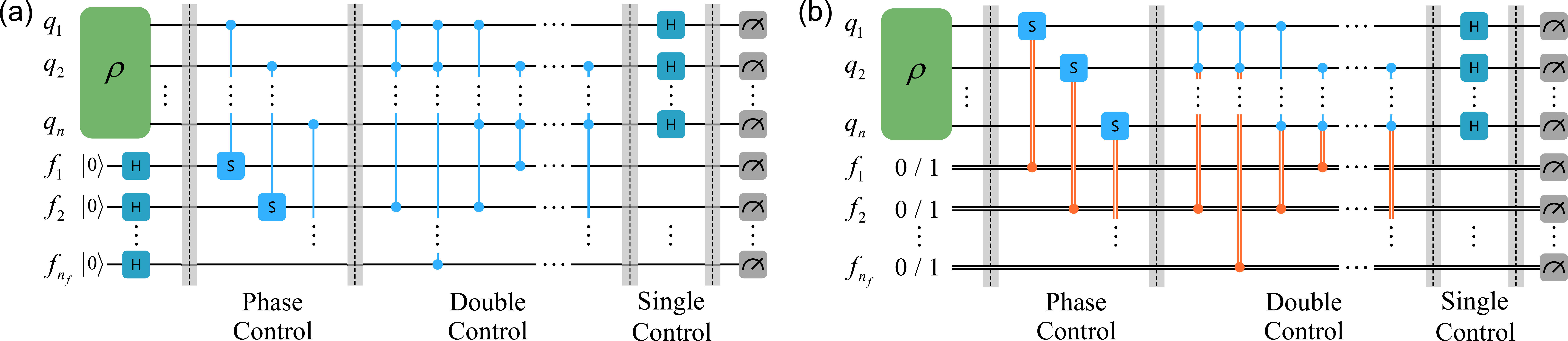}
    \caption{
        (a) shows the circuit diagram of the \OurMethod{} where all auxiliary bits are qubits, while (b) shows the circuit diagram of the \OurMethod{} where all auxiliary bits are classical bits. 
        The circuit includes three types of phase adjustment mechanisms: \phasecontrol{}, \doublecontrol{}, and \singlecontrol{}. 
        The variables $q_1, q_2, \ldots, q_n$ represent the $n$ qubits to be measured, while $f_1, f_2, \ldots, f_{n_f}$ represent the $n_f$ auxiliary (qu)bits, satisfying $n \leq n_f$.
    }
    \label{fig:general-circuit}
\end{figure}

Below, we introduce the quantum circuit structure for \OurMethod{}, which is illustrated in Fig.~\ref{fig:general-circuit}.

For an $n$-qubit system to be measured, the algorithm requires $n_f$ auxiliary (qu)bits, where $n_f \geq n$. As discussed in Section~\ref{subsec:circuit}, the principle of the algorithm is that the density matrix utilizes auxiliary (qu)bits to adjust its phase to achieve the \SpecialProperty{}, which is specifically reflected in the circuit by three types of adjustment mechanisms: \singlecontrol{}, \phasecontrol{}, and \doublecontrol{}. In the scheme where the system to be measured is entangled with auxiliary qubits, as shown in Fig.~\ref{fig:general-circuit}(a), the \phasecontrol{} component consists of $n$ CS gates, each controlled by a measured qubit and acting on an auxiliary qubit. The \doublecontrol{} component is composed of at least $C_n^2$ CCZ gates, where pairs of measured qubits control an auxiliary qubit. The \singlecontrol{} component consists of $n$ Hadamard gates applied to the measured qubits. For CS and CCZ gates, the target and control qubits are interchangeable. In the scheme where the system to be measured is associated with auxiliary classical bits, as shown in Fig.~\ref{fig:general-circuit}(b), the \phasecontrol{} component consists of $n$ S gates controlled by classical bits, and the \doublecontrol{} component is composed of at least $C_n^2$ CZ gates, each controlled by a single classical bit. The \singlecontrol{} component is identical to that in the quantum scheme. In summary, the algorithm requires $O(n^2)$ quantum gates, and the corresponding circuit depth is $O(n)$.

The execution process of the circuit is as follows: First, the auxiliary (qu)bits are initialized. In the quantum scheme, each auxiliary qubit is initialized into an equal-probability superposition state of $\left| 0 \right\rangle$ and $\left| 1 \right\rangle$. In the classical scheme, each  auxiliary bit is input with equal probability of 0 and 1. Then, phase adjustment is performed through the \phasecontrol{}, \doublecontrol{}, and \singlecontrol{} mechanisms. Finally, all target qubits and auxiliary (qu)bits are measured.

Next, we introduce the mechanisms by which various quantum gates adjust the phase, starting with the \phasecontrol{} part. The density matrix $\rho =\sum\limits_{ij}{{{\rho }_{ij}}\left| i \right\rangle \left\langle  j \right|}$ can be viewed as a superposition of components $\left| i \right\rangle \left\langle  j \right|$. The action of quantum gates on the $\left| i \right\rangle \left\langle  j \right|$ components of $\rho$ adjusts the coefficients of the $ij$ terms in the probability equations. The $\left| i \right\rangle \left\langle  j \right|$ components of $\rho$ can be expressed as the tensor product of the components of the qubits to be measured:
\begin{equation}
\label{eq:density-matrix-decomposition}
\left| i \right\rangle \left\langle  j \right| = \left| {{i}_{1}} \right\rangle \left\langle  {{j}_{1}} \right| \otimes \left| {{i}_{2}} \right\rangle \left\langle  {{j}_{2}} \right| \otimes \ldots \otimes \left| {{i}_{n}} \right\rangle \left\langle  {{j}_{n}} \right|.
\end{equation}

Independently considering the components of each qubit will facilitate our subsequent analysis. For a CS gate where the $k$-th measured qubit controls the $m$-th auxiliary qubit, when the auxiliary qubit measurement result is ${{f}_{m}}$, with ${{f}_{m}} \in \{0,1\}$, only the component $\left| {{i}_{k}} \right\rangle \left\langle  {{j}_{k}} \right|\otimes \left| {{f}_{m}} \right\rangle \left\langle  {{f}_{m}} \right|$ contributes to the phase in the $ij$ term of the probability equations. The effect of this CS gate is given by:
\begin{equation}
\label{eq:cs_gate_effect}
\text{CS}(\left| {{i}_{k}} \right\rangle \left\langle  {{j}_{k}} \right|\otimes \left| {{f}_{m}} \right\rangle \left\langle  {{f}_{m}} \right|)={{i}^{{{i}_{k}}\cdot {{f}_{m}}}}{{(-i)}^{{{j}_{k}}\cdot {{f}_{m}}}}\left| {{i}_{k}} \right\rangle \left\langle  {{j}_{k}} \right|\otimes \left| {{f}_{m}} \right\rangle \left\langle  {{f}_{m}} \right|.
\end{equation}
The CS gate introduces a phase of \( {{i}^{{{i}_{k}}\cdot {{f}_{m}}}}{{(-i)}^{{{j}_{k}}\cdot {{f}_{m}}}} \) to the component \( \left| {{i}_{k}} \right\rangle \left\langle  {{j}_{k}} \right|\otimes \left| {{f}_{m}} \right\rangle \left\langle  {{f}_{m}} \right| \), which subsequently appears in the coefficient of the \(ij\) term in the probability equations. The third and fourth columns of Eq.~\eqref{general-equation} represent the effect of \phasecontrol{}. Further analysis reveals that when \( {{i}_{k}},{{j}_{k}} = 0,0 \) or \( 1,1 \), the phase factor equals 1. This scenario can be interpreted as no control action occurring, leaving the outcome unaffected. Conversely, when \( {{i}_{k}},{{j}_{k}} = 0,1 \) or \( 1,0 \), the phase factor is \( {{i}^{{{f}_{m}}}} \) or \( {{(-i)}^{{{f}_{m}}}} \), respectively. This scenario corresponds to an active control effect, imparting a phase of the form \( {{i}^{{{f}_{m}}}} \) to the result.

Next is the \doublecontrol{} mechanism. For a CCZ gate where the ${k}_{1}$-th and ${k}_{2}$-th qubits control the ${m}$-th auxiliary qubit, similar to the previous case, when the auxiliary qubit measurement result is ${{f}_{m}}$, only the component \(\left| {{i}_{{{k}_{1}}}}{{i}_{{{k}_{2}}}} \right\rangle \left\langle  {{j}_{{{k}_{1}}}}{{j}_{{{k}_{2}}}} \right|\otimes \left| {{f}_{m}} \right\rangle \left\langle  {{f}_{m}} \right|\) contributes. The effect of this CCZ gate is given by:
\begin{equation}
\label{eq:ccz_gate_effect}
\text{CCZ}(\left| {{i}_{{{k}_{1}}}}{{i}_{{{k}_{2}}}} \right\rangle \left\langle  {{j}_{{{k}_{1}}}}{{j}_{{{k}_{2}}}} \right|\otimes \left| {{f}_{m}} \right\rangle \left\langle  {{f}_{m}} \right|)={{(-1)}^{{{i}_{{{k}_{1}}}}{{i}_{{{k}_{2}}}}{{f}_{m}}}}{{(-1)}^{{{j}_{{{k}_{1}}}}{{j}_{{{k}_{2}}}}{{f}_{m}}}}\left| {{i}_{{{k}_{1}}}}{{i}_{{{k}_{2}}}} \right\rangle \left\langle  {{j}_{{{k}_{1}}}}{{j}_{{{k}_{2}}}} \right|\otimes \left| {{f}_{m}} \right\rangle \left\langle  {{f}_{m}} \right|.
\end{equation}
CCZ gates induce a phase of ${{(-1)}^{{{i}_{{{k}_{1}}}}{{i}_{{{k}_{2}}}}{{f}_{m}}}}{{(-1)}^{{{j}_{{{k}_{1}}}}{{j}_{{{k}_{2}}}}{{f}_{m}}}}$ on the component $\left| {{i}_{{{k}_{1}}}}{{i}_{{{k}_{2}}}} \right\rangle \left\langle  {{j}_{{{k}_{1}}}}{{j}_{{{k}_{2}}}} \right|\otimes \left| {{f}_{m}} \right\rangle \left\langle  {{f}_{m}} \right|$, which corresponds to the results of the \doublecontrol{} mechanism in Columns 5 and 6 of Eq.~\eqref{general-equation}. It can be observed that the phase becomes ${{(-1)}^{{{f}_{m}}}}$ only when $({{i}_{{{k}_{1}}}}{{i}_{{{k}_{2}}}},{{j}_{{{k}_{1}}}}{{j}_{{{k}_{2}}}})=(11,00),(11,01),(11,10)$ or $(00,11),(01,11),(10,11)$, indicating that the control is activated. In all other cases, the phase remains 1, meaning the control is not activated.

Finally, in the \singlecontrol{} mechanism, for an H gate applied to the $k$-th qubit to be measured, the states $\left| 0 \right\rangle$ and $\left| 1 \right\rangle$ transform into ${\left( \left| 0 \right\rangle +\left| 1 \right\rangle  \right)}/{\sqrt{2}}$ and ${\left( \left| 0 \right\rangle -\left| 1 \right\rangle  \right)}/{\sqrt{2}}$, respectively. These two states have a phase difference of $\pi$. The component of this qubit is $\left| {{i}_{k}} \right\rangle \left\langle  {{j}_{k}} \right|$. When the qubit undergoes the H gate and is subsequently measured with an outcome of ${{q}_{k}}$, 
\begin{equation}
\label{eq:H_gate_effect}
\left\langle  {{q}_{k}} \right|H\left| {{i}_{k}} \right\rangle \left\langle  {{j}_{k}} \right|{{H}^{\dagger }}\left| {{q}_{k}} \right\rangle =\frac{1}{2}{{(-1)}^{{{i}_{k}}\cdot {{q}_{k}}}}{{(-1)}^{{{j}_{k}}\cdot {{q}_{k}}}}.
\end{equation}
This component contributes to the coefficient of the $ij$ term in the probability equations by a factor of ${{(-1)}^{{{i}_{k}}{{q}_{k}}}}{{(-1)}^{{{j}_{k}}{{q}_{k}}}}$, corresponding to the first and second columns in Eq.~\eqref{general-equation}. Further analysis shows that when ${{i}_{k}},{{j}_{k}}=0,0$ or $1,1$, the result is $1$, which can be interpreted as no control being active. When ${{i}_{k}},{{j}_{k}}=0,1$ or $1,0$, the result is ${{(-1)}^{{{q}_{k}}}}$, indicating that control is active.

The effects produced by different quantum gates are independent and can be superimposed. The final result appears in Eq.~\eqref{general-equation} as the product of the individual phase factors.

Next, we introduce the phase adjustment mechanisms in the classical scheme. First, the \singlecontrol{} part of the circuit is identical to that in the quantum scheme. Then, for the \phasecontrol{} part, an $S$ gate controlled by the $m$-th classical auxiliary bit on the $k$-th qubit behaves as follows. When the classical auxiliary bit measurement result is $f_{m} = 0$, the phase remains $1$. When $f_{m} = 1$, the transformation follows: $S\left| i_k \right\rangle \left\langle j_k \right| S^{\dagger} = i^{i_k} (-i)^{j_k} \left| i_k \right\rangle \left\langle j_k \right|$. Considering both cases together, the $S$ gate controlled by this classical bit results in a phase factor of $i^{i_k f_m} (-i)^{j_k f_m}$. Finally, for the \doublecontrol{} part, a CZ gate controlled by the $m$-th classical auxiliary bit on the ${k}_{1}$-th and ${k}_{2}$-th qubits behaves as follows. When the classical auxiliary bit measurement result is $f_m = 0$, the phase remains $1$. When $f_m = 1$, the transformation follows:
\begin{equation}
\label{eq:cz-gate-transformation}
\text{CZ}(\left| {{i}_{{{k}_{1}}}}{{i}_{{{k}_{2}}}} \right\rangle \left\langle  {{j}_{{{k}_{1}}}}{{j}_{{{k}_{2}}}} \right|)={{(-1)}^{{{i}_{{{k}_{1}}}}{{i}_{{{k}_{2}}}}}}{{(-1)}^{{{j}_{{{k}_{1}}}}{{j}_{{{k}_{2}}}}}}\left| {{i}_{{{k}_{1}}}}{{i}_{{{k}_{2}}}} \right\rangle \left\langle  {{j}_{{{k}_{1}}}}{{j}_{{{k}_{2}}}} \right|.
\end{equation}

Combining both cases, the classical-bit-controlled CZ gate results in a phase factor of ${{(-1)}^{{{i}_{{{k}_{1}}}}{{i}_{{{k}_{2}}}}{{f}_{m}}}}{{(-1)}^{{{j}_{{{k}_{1}}}}{{j}_{{{k}_{2}}}}{{f}_{m}}}}$. From the perspective of the final results, the classical scheme exhibits exactly the same phase adjustment mechanism as the quantum scheme. Replacing auxiliary qubits with probabilistic classical auxiliary bits does not introduce any changes in the outcome, allowing both schemes to share the same probability equations. Naturally, auxiliary bits do not necessarily have to be exclusively of one type, as depicted in Fig.~\ref{fig:general-circuit}. They can be mixed, with some being classical bits and others being qubits. The cases illustrated in Fig.~\ref{fig:general-circuit}, where only auxiliary qubits or only classical auxiliary bits are used, represent special cases.

Based on the aforementioned phase adjustment mechanisms, we can appropriately configure the connections of quantum gates to ensure that the probability equations exhibit the \SpecialProperty{} property. In practical applications, the validity of a circuit configuration in enabling the algorithm can be determined through the \GateMatrix{}. The circuit of the algorithm is labeled by the \GateMatrix{}, and data post-processing also relies on the \GateMatrix{}.

\begin{figure}
    \centering
    \includegraphics[width=\textwidth]{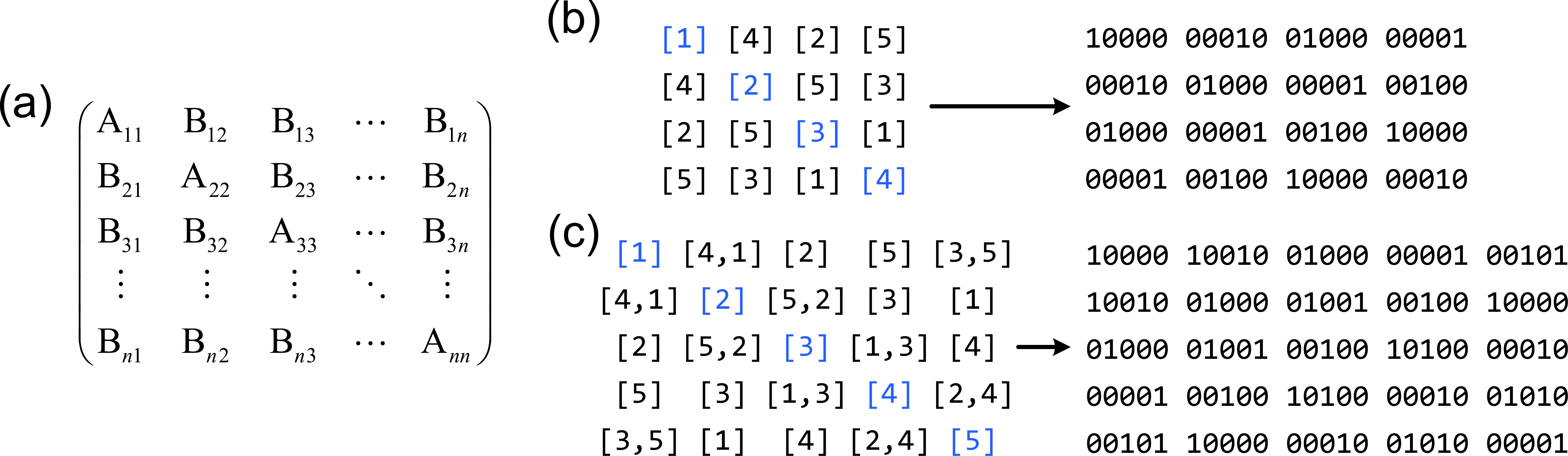}
    \caption{
        (a) illustrates a schematic diagram of the \GateMatrix{}. The diagonal elements ${{\text{A}}_{ii}}$ indicate the connection configuration of CS gates, while the sequences of numbers in the off-diagonal elements ${{\text{B}}_{ij}}$ specify the connection configuration of CCZ gates. 
        (b) presents a \GateMatrix{} for a four-qubit system along with its corresponding binary \GateMatrix{}, which is a special case where each matrix element contains only a single number. 
        (c) shows a \GateMatrix{} for a five-qubit system and its corresponding binary \GateMatrix{}, representing the special case that requires the minimal number of auxiliary bits.
    }
    \label{fig:gate-matrix-schematic}
\end{figure}

The structure of the \GateMatrix{} is illustrated in Fig.~\ref{fig:gate-matrix-schematic}(a). The elements of the \GateMatrix{} are sequences of numbers specifying the connection configuration of CS and CCZ gates in the circuit. The numbering starts from 1, and the numbers in each sequence represent the indices of the auxiliary bits. The sequence ${{\text{A}}_{ii}}$ at the diagonal element of row $i$ and column $i$ indicates the target auxiliary qubits of the CS gates, where the $i$th measured qubit acts as the control qubit. The sequence ${{\text{B}}_{ij}}$ at the off-diagonal element of row $i$ and column $j$ represents the target auxiliary qubits of the CCZ gates, where the $i$th and $j$th measured qubits act as the control qubits. The length of each sequence corresponds to the number of auxiliary qubits serving as target qubits, which also determines the number of controlled quantum gates in the circuit.

Based on the above definitions, several properties of the \GateMatrix{} can be deduced. The dimension of the \GateMatrix{} is equal to the number of measured qubits, and the maximum number appearing in the \GateMatrix{} corresponds to the number of auxiliary qubits. The \GateMatrix{} is symmetric because a CCZ gate with the $i$th and $j$th measured qubits as control qubits is equivalent to a CCZ gate with the $j$th and $i$th measured qubits as control qubits, so the sequences ${{\text{B}}_{ij}}$ and ${{\text{B}}_{ji}}$ must be identical. For the classical scheme, the circuit structure is completely consistent with the quantum scheme, except that the control and target qubits of the quantum gates are interchanged. Therefore, we omit further discussion here.

Although the way numbers are filled in the \GateMatrix{} is not fixed, we define a convention that the number filled in the $i$th diagonal element should be $i$, meaning that the $n$ CS gates in the circuit follow the configuration where the $i$th measured qubit controls the $i$th auxiliary qubit. Alternatively, the numbering should at least be continuous, with the $i$th diagonal element filled with $i+\text{const}$, implying that the CS gate configuration in the circuit has the $i$th measured qubit controlling the $(i+\text{const})$th auxiliary qubit. As will be shown in Section~\ref{subsec:density-reconstruction}, such a convention facilitates data post-processing. Additionally, the indexing of auxiliary bits is entirely arbitrary, as swapping or renaming auxiliary bits is equivalent. Therefore, implementing this convention in practice does not pose any fundamental difficulty.

There are numerous ways to fill the \GateMatrix{} that enable the algorithm to function, i.e., to ensure that the probability equations exhibit the \SpecialProperty{} property. Among these, two specific configurations are particularly notable. One configuration, as illustrated in Fig.~\ref{fig:gate-matrix-schematic}(b), consists of each position in the \GateMatrix{} containing only a single number. Such a configuration is referred to as a single \GateMatrix{}, representing the most structurally minimal scheme. Another configuration, as shown in Fig.~\ref{fig:gate-matrix-schematic}(c), is characterized by the auxiliary bit count satisfying $n_f = n$, which corresponds to the scheme that minimizes the consumption of auxiliary bits. However, there also exist many filling methods that fail to satisfy the algorithm's requirements. To determine whether a given \GateMatrix{} configuration allows the algorithm to function correctly, we can utilize the binary \GateMatrix{}.

The binary \GateMatrix{} can be obtained by converting each sequence in the \GateMatrix{} into a binary number of length $n_f$. Specifically, the numbers in the sequence indicate the positions of ones in the binary number: if a sequence contains a number $k$, then the $k$-th bit (counting from left to right) in the binary number is set to 1. For example, converting the \GateMatrix{} in Fig.~\ref{fig:gate-matrix-schematic}(b) results in the binary \GateMatrix{} shown on the right side of Fig.~\ref{fig:gate-matrix-schematic}(b), and similarly, converting the \GateMatrix{} in Fig.~\ref{fig:gate-matrix-schematic}(c) yields the binary \GateMatrix{} displayed on the right side of Fig.~\ref{fig:gate-matrix-schematic}(c).

The condition for the algorithm to be valid is that the binary \GateMatrix{} must be linearly independent. The linear independence condition for a set of binary numbers can be expressed as follows: for a set of binary numbers $\{{{b}_{1}},{{b}_{2}},\cdots ,{{b}_{n}}\}$, if there does not exist any nontrivial subset such that their bitwise XOR satisfies ${{b}_{{{c}_{1}}}}\oplus {{b}_{{{c}_{2}}}}\oplus \cdots \oplus {{b}_{{{c}_{k}}}}=0$, then this set is linearly independent; otherwise, it is considered linearly dependent. In practical applications, one can efficiently determine whether a set of binary numbers is linearly independent using a method analogous to Gaussian elimination. Furthermore, for a given set of binary vectors $\{{{\nu }^{(1)}},{{\nu }^{(2)}},\cdots ,{{\nu }^{(k)}}\}$, we define their combination as the element-wise XOR of all vectors, that is, the combined result is given by ${{\nu }^{(1)}}\oplus {{\nu }^{(2)}}\oplus \cdots \oplus {{\nu }^{(k)}}$. The linear independence of the binary \GateMatrix{} means that each row of the matrix is linearly independent and that all possible combinations of these rows remain linearly independent. If a binary \GateMatrix{} is linearly independent, then the corresponding circuit's probability equations exhibit the property of \SpecialProperty{}. Since the proof of this claim is relatively intricate, a detailed proof can be found in Appendix~\ref{appendix:B}.

From the above conditions for the establishment of the algorithm, we can derive the following conclusions. In the quantum scheme, any two qubits to be measured must be connected by a CCZ gate; otherwise, there will be zero elements in the binary gate matrix, making it impossible to satisfy the condition of linear independence. Therefore, the algorithm requires at least $C_n^2$ CCZ gates. Similarly, in the classical scheme, any two qubits to be measured must be connected by a CZ gate controlled by a classical bit, and the algorithm requires at least $C_n^2$ such classically controlled CZ gates. Furthermore, the number of auxiliary bits and the number of quantum gates cannot both be minimized simultaneously, meaning that it is impossible to construct a single-gate matrix that satisfies ${n_f} = n$ while maintaining linear independence. This is because, in such a case, to ensure that each row and every pairwise combination of rows remain linearly independent, the gate matrix must be filled in a manner similar to Sudoku, where each row and column contain distinct numbers. However, under this constraint, the combination of all rows in the gate matrix would result in $\{11\ldots 1, 11\ldots 1, \dots, 11\ldots 1\}$, which is linearly dependent.

The problem of finding a circuit configuration that ensures the validity of the algorithm translates into the problem of constructing a \GateMatrix{} that satisfies the condition of linear independence, which we define as the \textit{Gate Matrix Problem}. This problem can be regarded as a special type of Sudoku puzzle. It becomes relatively easy when a large number of auxiliary bits are available but becomes more challenging when the number of auxiliary bits is constrained. The simplest filling method is, of course, to impose no restriction on the number of auxiliary bits and to assign distinct numbers to all elements in the upper triangular part of the \GateMatrix{}. However, this general solution corresponds to the single-gate matrix solution with the maximum number of auxiliary bits. Here, we present two additional solutions. The first is a solution with the minimal number of auxiliary bits, where $n_f = n$, and the second is a single-gate matrix solution with $n_f = 2n-1$ auxiliary bits.

The first solution, which minimizes the number of auxiliary bits, was obtained in the literature on minimal Clifford measurements~\cite{PhysRevApplied.21.064001}. If the classical scheme is interpreted as a process where $\rho$ undergoes unitary evolution via random sampling before measurement, then in this sense, the $n_f = n$ classical scheme circuit of the \OurMethod{} is equivalent to the minimal Clifford measurement circuit set given in~\cite{PhysRevApplied.21.064001}, where the measurement is performed in MUBs. Using this circuit set, it is possible to construct a \GateMatrix{} with $n_f = n$. The detailed method for obtaining this circuit set can be found in~\cite{PhysRevApplied.21.064001}.

The second solution is the single \GateMatrix{} solution with $n_f = 2n-1$. This solution is obtained through cyclic permutations. The first step is to construct the following generator rows:
\begin{equation}
\label{eq:generator-row}
\begin{matrix}
   1 & n+1 & 2 & n+2 & 3 & \cdots  & 2n-2 & n-1 & 2n-1 & n  
\end{matrix}
\end{equation}
By performing $n-1$ cyclic shifts on the generator row in Eq.~\eqref{eq:generator-row}, we construct an $n \times (2n-1)$ matrix by combining the generator row with its cyclic shifts. Extracting the first $n$ columns of this matrix yields an $n \times n$ matrix that serves as a single \GateMatrix{}, ensuring the validity of the algorithm. A detailed proof of this construction can be found in Appendix~\ref{appendix:C}. However, it should be noted that the \GateMatrix{} obtained by this method is not necessarily the structurally simplest one. In fact, there exist some \GateMatrix{} solutions with \( n_f = n+1 \), but such solutions do not exist for all values of \( n \). For example, a \GateMatrix{} solution with \( n_f = n+1 \) can be obtained through cyclic shifts when \( n = 2, 4, 10, 12, 18, 28 \). It is important to emphasize that the number of auxiliary bits does not affect the efficiency of the algorithm. Since in the case of $n_f = n$, the algorithm's efficiency is comparable to the MUBs method, it follows that in any scenario, the \OurMethod{} achieves performance on par with the MUBs method.

\section{Algorithm Procedure} \label{sec:algorithm-process}

\subsection{Density Matrix Reconstruction Algorithm} \label{subsec:density-reconstruction}

The basic idea and procedure of the density matrix reconstruction algorithm are briefly introduced below. For each $ij$, the counts need to be classified into four categories: $+{{\alpha }_{ij}},-{{\alpha }_{ij}},+{{\beta }_{ij}},-{{\beta }_{ij}}$, in order to compute according to Eq.~\eqref{expectation-variance}. Therefore, the density matrix reconstruction algorithm must efficiently perform this classification.

To achieve fast classification, we introduce a binary mask $L$ that marks the phase-$i$ components and another binary mask $LF$ that marks the phase-$(-1)$ components. Given a specific measurement outcome $qf = {{q}_{1}}{{q}_{2}}\cdots {{q}_{n}}{{f}_{1}}{{f}_{2}}\cdots {{f}_{{{n}_{f}}}}$, performing a bitwise AND operation with $L$ results in a binary number. The number of ones in this result represents how many such factors of $i$ are multiplied together in the total phase. Similarly, performing a bitwise AND operation between $qf$ and $LF$ yields another binary number, where the number of ones represents how many factors of $(-1)$ are multiplied together in the total phase. Based on Eq.~\eqref{general-equation}, once the total phase is determined, it becomes straightforward to classify the count ${{N}_{qf}}$ of a given measurement outcome into one of the four categories. By recording the experimentally collected counts as a measurement outcome vector $\mathbf{qf}$ and a corresponding count vector ${{\mathbf{N}}_{\mathbf{qf}}}$, we can efficiently compute the parameters ${{\alpha }_{ij}}$ and ${{\beta }_{ij}}$ using simple bitwise operations and vectorized processing. Iterating over all $ij$ satisfying $i > j$, we obtain all off-diagonal parameters. Finally, by combining these parameters with the diagonal parameters obtained from a separate straightforward measurement and filling them into $\rho$ according to Eq.~\eqref{eq:rho}, the reconstruction process is completed.

Next, we introduce the method for obtaining the $L$ and $LF$ masks. The calculation of the $L$ mask is relatively straightforward. For any given pair $i={{i}_{1}}{{i}_{2}}\cdots {{i}_{n}}$ and $j={{j}_{1}}{{j}_{2}}\cdots {{j}_{n}}$,
\begin{equation}
L = i \oplus j.
\label{eq:mask-L}
\end{equation}

The calculation of the $LF$ mask is relatively complex. It consists of two parts: the $L$ mask and the $F$ mask. Moreover, the $F$ mask is computed from the $F_1$ and $F_2$ masks. Below, we describe the calculation methods for each component step by step.

\begin{figure}
    \centering
    \includegraphics[width=0.85\textwidth]{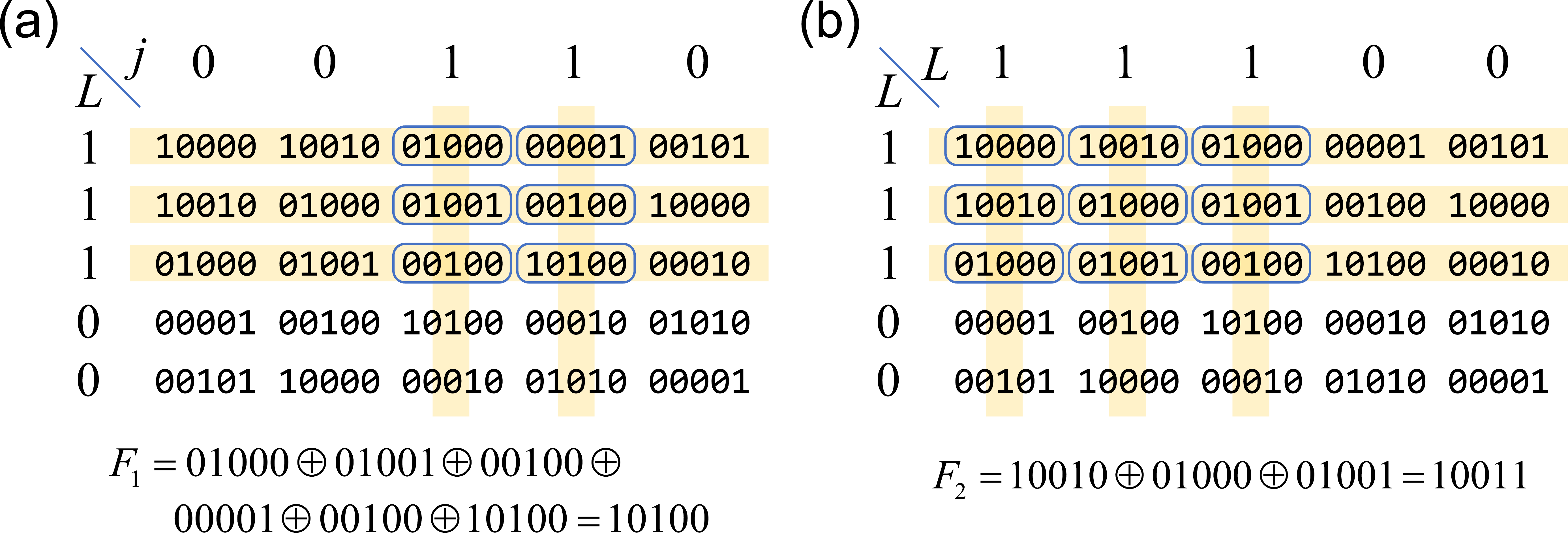}
    \caption{
        The above images illustrate an example of computing $F_1$ and $F_2$ using a binary \GateMatrix{} for a five-qubit system. 
        Panel (a) shows the computation of $F_1$, where the region outlined by the blue box represents the submatrix indexed by $L$ and $j$. 
        The result of the bitwise XOR operation on all binary numbers within this submatrix gives $F_1$. 
        Panel (b) shows the computation of $F_2$, where the blue-boxed region represents the submatrix indexed only by $L$. 
        The result of the bitwise XOR operation on the strictly lower triangular elements of this submatrix gives $F_2$.
    }
    \label{fig:F1_F2_calculation}
\end{figure}

${{F}_{1}}$ and ${{F}_{2}}$ masks are computed using the binary \GateMatrix{}. The $L$ mask serves as the row binary index, while $j$ serves as the column binary index for indexing the binary \GateMatrix{}. The result of applying a bitwise XOR operation to all elements in the indexed submatrix gives ${{F}_{1}}$. Similarly, using $L$ as both the row and column binary index to index the binary \GateMatrix{}, the result of applying a bitwise XOR operation to all strictly lower triangular elements in the indexed submatrix gives ${{F}_{2}}$. Fig.~\ref{fig:F1_F2_calculation} illustrates an example of indexing ${{F}_{1}}$ and ${{F}_{2}}$. In this example, the \GateMatrix{} has $n={{n}_{f}}=5$, and for $i=11010$ and $j=00110$, we obtain $L=11100$. The process of indexing the binary \GateMatrix{} using both $L$ and $j$ is shown in panel (a), where the blue box highlights the indexed submatrix. Applying a bitwise XOR operation to all binary numbers in this submatrix results in ${{F}_{1}}=10100$. The process of indexing the binary \GateMatrix{} using only $L$ is shown in panel (b), where the blue box highlights the indexed submatrix. Applying a bitwise XOR operation to all strictly lower triangular elements in this submatrix results in ${{F}_{2}}=10011$.

After computing ${{F}_{1}}$ and ${{F}_{2}}$, the $F$ mask can be determined using the following equation:
\begin{equation}
\label{eq:F_mask}
F = F_{1} \oplus F_{2}.
\end{equation}
Finally, concatenating $L$ and $F$ together yields the $LF$ mask, i.e.,
\begin{equation}
\label{eq:LF_mask}
LF = (L << {{n}_{f}}) + F.
\end{equation}

The masks marking the phase-$i$ components, $L$, and the phase-$(-1)$ components, $LF$, share a common segment. This arises due to the setting conventions of the \phasecontrol{} part discussed in Section~\ref{subsec:circuit}. According to the discussion in Section~\ref{subsec:circuit} regarding circuit phase adjustment mechanisms, the activation conditions for \singlecontrol{} and \phasecontrol{} are identical. Under the convention where the $k$-th measured qubit in the \phasecontrol{} part is connected to the $(k+\text{const})$-th auxiliary bit, both mechanisms yield similar effects, leading to a shared mask segment. One of the reasons for defining the setting convention of the \phasecontrol{} part is to reduce one mask computation at this step.

Next, the density matrix reconstruction algorithm is presented in detail as pseudocode in Algorithm~\ref{alg:reconstruction}. A reference implementation demonstrating this approach is available in the supplementary materials \cite{long2025efficient}.

\begin{algorithm}


\caption{Density Matrix Reconstruction Algorithm}
\label{alg:reconstruction}

\SetNlSty{}{\footnotesize}{:} 
\SetAlgoNoLine  

\KwIn{Gate Matrix $\mathbf{G}$, Measurement Outcome Vector $\mathbf{qf}$, Count Vector $\mathbf{N}_\mathbf{qf}$, Measured Diagonal Data}
\KwOut{Reconstructed Density Matrix $\rho$}

$\mathbf{G}_b \gets \text{ToBinary}(\mathbf{G})$\;
$n, n_f \gets \text{GetBitInfo}(\mathbf{G})$\;
$\rho \gets \text{zeros}(2^n, 2^n)$\;

\For{$i = 1$ \KwTo $2^n - 1$}{
    \For{$j = 0$ \KwTo $i-1$}{
        $L \gets i \oplus j$\;
        RowIdx $\gets$ \text{ToIndex}$(L)$\;
        ColIdx $\gets$ \text{ToIndex}$(j)$\;

        $F_1 \gets \text{XOR-Submatrix}(\mathbf{G}_b, \text{RowIdx}, \text{ColIdx})$\;
        $F_2 \gets \text{XOR-LowerTriangle}(\mathbf{G}_b, \text{RowIdx})$\;

        $F \gets F_1 \oplus F_2$\;
        $LF \gets (L << n_f) \oplus F$\;

        $\mathbf{N}_{-1} \gets \text{BitCount}(\mathbf{qf} ~\&~ LF) \mod 2$\;
        $\mathbf{N}_i \gets \text{BitCount}(\mathbf{qf} ~\&~ (L << (n_f-n))) \mod 4$\;

        $N_1, N_2, N_3, N_4 \gets \text{Classify}(\mathbf{N_{qf}}, \mathbf{N}_{-1}, \mathbf{N}_i)$\;

        $N_{\alpha} \gets N_1 + N_2 $,\quad$ N_{\beta} \gets N_3 + N_4$\;
        $\alpha_{ij} \gets N_1 / N_{\alpha} - 0.5$,\quad$\beta_{ij} \gets N_3 / N_{\beta} - 0.5$\;
        $\rho[i, j] \gets \alpha_{ij} + i \beta_{ij}$,\quad$\rho[j, i] \gets \alpha_{ij} - i \beta_{ij}$\;
    }
}

Fill $\rho$ diagonal with Measured Diagonal Data\;
\Return $\rho$\;

\end{algorithm}

\subsection{Purity Algorithm} \label{subsec:purity}

The measurement data obtained from the \OurMethod{} contains information about all the off-diagonal elements of the density matrix. However, in some cases, the desired quantities are only specific parameters of the quantum state. In such scenarios, reconstructing $\rho$ and then computing the target parameters may lead to significant computational complexity. Therefore, it is worth exploring methods that directly utilize the measurement data from the \OurMethod{} to calculate various quantum state parameters. This study presents two methods for directly computing purity using the measurement data from the \OurMethod{}. The sampling complexity of both methods scales as $O(2^n)$ with the number of qubits $n$.

Purity is a measure of the degree of mixing of a quantum state and plays an important role in entanglement analysis and noise assessment. The definition of purity is $\text{Tr}(\rho^2)$, with values ranging between 0 and 1. A higher purity indicates a purer quantum state, while a lower purity indicates a more mixed state. The purity of a pure state is equal to 1, while the purity of a mixed state is less than 1. For the density matrix shown in Eq.~\eqref{eq:rho}, we can derive:
\begin{equation}
\text{Tr}(\rho^2) = \sum_{i} P_i^2 + \sum_{i \ne j} \left( \alpha_{ij}^2 + \beta_{ij}^2 \right).
\label{eq:purity}
\end{equation}
It can be seen that the purity consists of two parts: the $P$ part, $\sum_{i} P_i^2$, and the $\alpha \beta$ part, $\sum_{i \ne j} \left( \alpha_{ij}^2 + \beta_{ij}^2 \right)$. Using the data from the \OurMethod{}, we can directly calculate the $\alpha \beta$ part of the purity. The $P$ part of the purity can be calculated based on the results from another simple experiment.

The probability equations of the \OurMethod{} exhibit the property of \SpecialProperty{}. This leads to the fact that for any pair of terms ${{\alpha }_{ij}}$ and ${{\alpha }_{{i}'{j}'}}$, the number of equations where their signs match is equal to the number of equations where their signs differ. Therefore, when computing the sum of all probabilities in the probability equations, $\sum\limits_{qf}{{{({{P}_{qf}}-{1}/{{{2}^{n+{{n}_{f}}}}}\;)}^{2}}}$, the terms $+{{\alpha }_{ij}}{{\alpha }_{{i}'{j}'}}$ and $-{{\alpha }_{ij}}{{\alpha }_{{i}'{j}'}}$ will cancel each other out due to their equal number. Similarly, this result holds for any pair of ${\alpha_{ij}, \beta_{{i}'{j}'}}$ and ${\beta_{ij}, \beta_{{i}'{j}'}}$. Thus, the result of $\sum\limits_{qf}{{{({{P}_{qf}}-{1}/{{{2}^{n+{{n}_{f}}}}}\;)}^{2}}}$ will only contain squared terms $\alpha _{ij}^{2}$ and $\beta _{ij}^{2}$, with no cross terms. This can be clearly seen in Eq.~\eqref{two-qubit-probability-equations}. Therefore, after slight manipulation of the result of $\sum\limits_{qf}{{{({{P}_{qf}}-{1}/{{{2}^{n+{{n}_{f}}}}}\;)}^{2}}}$, we can obtain the following:
\begin{equation}
\sum\limits_{i\ne j}{\alpha _{ij}^{2}+\beta _{ij}^{2}} = 2^{n + n_f} \cdot \sum\limits_{qf} P_{qf}^{2} - 1.
\label{eq:ab_sum}
\end{equation}

For the simple sum of the squares of discrete probabilities, we can estimate it by summing the squares of the counts.
\begin{equation}
\sum\limits_{i} P_i^2 = \frac{\text{E}\left( \sum\nolimits_{i} N_i^2 \right) - N}{N(N-1)},
\label{eq:probability-square-sum}
\end{equation}
where $N$ represents the number of shots in the experiment, and $N_i$ denotes the count number of events associated with the probability $P_i$. The $P$ part of the purity can be calculated using Eq.~\eqref{eq:probability-square-sum}. For the $\alpha\beta$ part, by combining Eq.~\eqref{eq:ab_sum} and Eq.~\eqref{eq:probability-square-sum}, the calculation method for the $\alpha\beta$ part of the purity is derived as follows:
\begin{equation}
\alpha \beta \text{ part} = \sum\limits_{i \neq j} \left( \alpha_{ij}^{2} + \beta_{ij}^{2} \right) = 2^{n + n_f} \cdot \frac{\text{E}\left( \sum\nolimits_{qf} N_{qf}^{2} \right) - N}{N(N - 1)} - 1
\label{eq:alpha-beta-part}.
\end{equation}
Thus, we can directly calculate the purity of the quantum state by summing the squares of the counts, avoiding the complex process of computing the density matrix. Next, we analyze the measurement precision and complexity of this method. First, the variance of the sum of squared counts for any probability is given by
\begin{equation}
\text{D}\left( \sum\nolimits_{i} N_{i}^{2} \right) = 2N(N-1)\sum\limits_{i} P_{i}^{2} - 2N(N-1)(2N-3) \left( \sum\limits_{i} P_{i}^{2} \right)^2 
+ 4N(N-1)(N-2) \sum\limits_{i} P_{i}^{3}
\label{eq:variance-sum-squared-counts}.
\end{equation}

For the $\alpha \beta$ part, from Eq.~\eqref{eq:ab_sum} and the probability constraints of the auxiliary bits, we can obtain

\begin{equation}
\sum\limits_{qf} P_{qf}^{2} = \frac{1 + \alpha \beta \text{ part}}{2^{n + n_f}}, \quad 
\sum\limits_{qf} P_{qf}^{3} < \sum\limits_{qf} P_{qf}^{2} \cdot P_{\max} = \frac{1 + \alpha \beta \text{ part}}{2^{n + n_f}} \cdot \frac{1}{2^{n_f}}
\label{eq:probabilities-squared-cubed}.
\end{equation}

By combining Eq.~\eqref{eq:variance-sum-squared-counts} and~\eqref{eq:probabilities-squared-cubed}, we can obtain that the variance of the $\alpha \beta$ part satisfies
\begin{equation}
\text{D}(\alpha \beta \text{ part}) < 2 \cdot \frac{2^{n + n_f}}{N(N - 1)} (1 + \alpha \beta \text{ part}) - \frac{4N - 6}{N(N - 1)} \left( 1 + \alpha \beta \text{ part} \right)^2 
+ 2^n \cdot \frac{4(N - 2)}{N(N - 1)} (1 + \alpha \beta \text{ part})
\label{eq:variance-alpha-beta-part}.
\end{equation}
It should be noted that the above Eq. \eqref{eq:variance-alpha-beta-part} provides an unattainable upper bound, and the actual variance is much smaller than this upper bound, especially when the cubic terms of $\alpha$ and $\beta$ in $\sum\limits_{qf} P_{qf}^3$ largely cancel out, and when most of the $\alpha$ and $\beta$ parameters in $\rho$ are zero. In fact, the cubic terms of $\alpha$ and $\beta$ in $\sum\limits_{qf} P_{qf}^3$ tend to cancel out, with only a small fraction of quantum states, such as the uniform superposition state, exhibiting large variance. For special quantum states like the GHZ state, where most of the $\alpha$ and $\beta$ parameters are zero, there are no cubic terms of $\alpha$ and $\beta$ in $\sum\limits_{qf} P_{qf}^3$, and the variance is
\begin{equation}
\text{D}{{(\alpha \beta \text{ part})}_{\text{GHZ}}} < \frac{2 \cdot 2^{n + n_f}}{N(N - 1)} (1 + \alpha \beta \text{ part}) - \frac{4N - 6}{N(N - 1)} (1 + \alpha \beta \text{ part})^2 
+ \frac{4(N - 2)}{N(N - 1)} (1 + 3 \cdot \alpha \beta \text{ part}),
\label{eq:variance-ghz}
\end{equation}
which is much smaller than the upper bound in Eq.~\eqref{eq:variance-alpha-beta-part}. A further analysis of the result in Eq.~\eqref{eq:variance-alpha-beta-part} shows that when ${n_f} = n$, it is sufficient to have $N \sim 2^n$ to control the growth of the variance of the $\alpha \beta$ part as $n$ increases. The purity $P$ part is calculated using Eq.~\eqref{eq:probability-square-sum}, and its variance can also be derived from the summation formula in Eq.~\eqref{eq:variance-sum-squared-counts}.
\begin{equation}
\text{D}(P\text{ part}) = \frac{2}{N(N - 1)} \left[ \sum\limits_{i} P_{i}^{2} - (2N - 3) \left( \sum\limits_{i} P_{i}^{2} \right)^2 + 2(N - 2) \sum\limits_{i} P_{i}^{3} \right],
\label{eq:variance-p-part}
\end{equation}
which scales as $O({1}/{N})$ at most. Therefore, considering both parts together, a sample size of $O(2^n)$ is sufficient to control the growth of variance as $n$ increases.
The above method provides an unbiased estimate of the purity using only the sum of squared counts, without reconstructing the density matrix, making it relatively simple in data processing. We refer to this as Method 1. 

However, if higher precision is desired, purity can also be calculated through another more complex post-processing method, which we call Method 2. Based on Eq.~\eqref{four-classification}, we can normalize ${P_1}, {P_2}$ and ${P_3}, {P_4}$ using conditional probabilities, further simplifying the problem. The events corresponding to ${P_1}, {P_2}$ are the measurements of $\alpha$, and the events corresponding to ${P_3}, {P_4}$ are the measurements of $\beta$. For the $\alpha$ part, under the condition of measuring $\alpha$, we can define:
\begin{equation}
\begin{aligned}
    P_1' &= \frac{P_1}{P_1 + P_2} = \frac{1}{2} + \alpha_{ij}, \\
    P_2' &= \frac{P_2}{P_1 + P_2} = \frac{1}{2} - \alpha_{ij}.
\end{aligned}
\label{eq:normalized-parts}
\end{equation}
Therefore, for the calculation of the purity, we can obtain the $\alpha_{ij}^2$ terms as:
\begin{equation}
2\alpha_{ij}^2 = \frac{1}{2} - P_1' P_2'.
\label{eq:alpha-ij-squared}
\end{equation}

Define $N_{\alpha} = N_1 + N_2$ as the number of occurrences of the measurement $\alpha$, and then we can estimate the value of $2 P_1' P_2'$ using the following equation:
\begin{equation}
2 P_1' P_2' = \frac{2 \, \text{E}(N_1 N_2)}{N_{\alpha} (N_{\alpha} - 1)}
\label{eq:estimated-purity}.
\end{equation}
Additionally, we can compute the variance of $2 N_1 N_2$ as:
\begin{equation}
D(2 N_1 N_2) = \frac{1}{2} N_{\alpha} (N_{\alpha} - 1) + 2 N_{\alpha} (N_{\alpha} - 1) (N_{\alpha} - 2) \cdot 2 \alpha_{ij}^2 - 2 N_{\alpha} (N_{\alpha} - 1) (2 N_{\alpha} - 3) \cdot 4 \alpha_{ij}^4
\label{eq:variance-two-N1-N2}.
\end{equation}

The $\beta$ part also satisfies Eq.~\eqref{eq:variance-two-N1-N2}, where we simply replace $\alpha$ with $\beta$, and 1 and 2 with 3 and 4, respectively. For the complete purity $\alpha \beta \text{ part}$, similar to the density matrix reconstruction algorithm, we can iterate over all $i > j$ for the $ij$ pairs and calculate the corresponding $\alpha_{ij}^2$ and $\beta_{ij}^2$ in the same way. We can then obtain the $\alpha \beta \text{part}$, and the specific calculation method is as follows:
\begin{equation}
2 \alpha_{ij}^2 + 2 \beta_{ij}^2 = 1 - \frac{2 \, \text{E}(N_1 N_2)}{N_{\alpha} (N_{\alpha} - 1)} - \frac{2 \, \text{E}(N_3 N_4)}{N_{\beta} (N_{\beta} - 1)}
\label{eq:alpha-beta-squared},
\end{equation}
\begin{equation}
\alpha \beta \text{ part} = \sum\limits_{i \neq j} \alpha_{ij}^2 + \beta_{ij}^2 = \sum\limits_{i > j} \left( 2 \alpha_{ij}^2 + 2 \beta_{ij}^2 \right)
\label{eq:alpha-beta-part2}.
\end{equation}

To analyze the measurement precision, based on Eq.~\eqref{eq:variance-two-N1-N2}, assuming that the $\alpha_{ij}^2$ and $\beta_{ij}^2$ are independent, and substituting ${N_{\alpha}} \approx {N_{\beta}} \approx \frac{N}{2}$ and neglecting the fourth-order terms, we can obtain the measurement variance for Method 2 as:
\begin{equation}
\text{D}(\alpha \beta \text{ part}) \approx 2 \frac{4^n - 2^n}{N^2} + \frac{4}{N} \alpha \beta \text{ part} < 2 \frac{4^n}{N^2} + \frac{4}{N}
\label{eq:variance-alpha-beta-part-method2}.
\end{equation}

From the variance upper bound of Method 2 in Eq.~\eqref{eq:variance-alpha-beta-part-method2}, we can see that the sampling complexity of Method 2 is the same as that of Method 1, i.e., $O(2^n)$, but the measurement precision of Method 2 is much higher than that of Method 1. Method 2 can be considered to reach the Heisenberg limit, where the standard deviation follows a $1/N$ relationship as the number of samples $N$ increases. When the sample size $N \ll d^2$, the term $\frac{2 \cdot 4^n}{N^2}$ in the variance upper bound dominates, and the measured standard deviation will approximately follow the $1/N$ relation. However, the variance upper bound of Method 2 does not strictly hold. In the theoretical analysis, we assumed that the $\alpha_{ij}^2$ and $\beta_{ij}^2$ are independent, but this condition is not actually satisfied. Nevertheless, in the subsequent simulation experiments, we can observe that the actual measurement variance is very close to the variance upper bound, and this upper bound still holds significant reference value.

\section{Numberical Simulation} \label{sec:simulation-verification}

In the previous chapters, we have discussed in detail the basic principles and circuit structure of the \OurMethod\ for measuring the quantum state density matrix, and analyzed the performance and computational resource requirements of the algorithm. At the same time, we have proposed two methods for measuring quantum state purity and analyzed the variance performance of the measurement results for both methods. This chapter will focus on presenting and analyzing the simulation results based on the theoretical derivations and solution designs mentioned above. Through a series of simulation experiments, we will validate the accuracy of the theoretical analysis and evaluate the effectiveness of the algorithms for density matrix reconstruction and purity measurement. Specifically, we will validate the expectation and variance of the \OurMethod, compare the reconstruction accuracy of \OurMethod\ and SQST, numerically verify the complexity scaling of trace distance, and compare the accuracy and performance of the two purity measurement methods.

\subsection{Simulation of Density Matrix Reconstruction} \label{subsec:density-matrix-reconstruction-simulation}

Based on the discussion in Section~\ref{subsec:principle} regarding the algorithm's principles, we know that the density matrix of an $n$-qubit system is uniquely determined by ${{2}^{n}}$ diagonal parameters ${P_{i}}$ and ${{4}^{n}}-{{2}^{n}}$ off-diagonal parameters ${{\alpha }_{ij}}$, ${{\beta }_{ij}}$. The \OurMethod\ can simultaneously measure all off-diagonal parameters of the density matrix with the same precision limit. It can be considered that the algorithm provides unbiased estimates for each ${{\alpha }_{ij}}$, ${{\beta }_{ij}}$, and the relationship between the measurement variance and the number of measurements follows Eq.~\eqref{expectation-variance}.

\begin{figure}
    \centering
    \includegraphics[width=\textwidth]{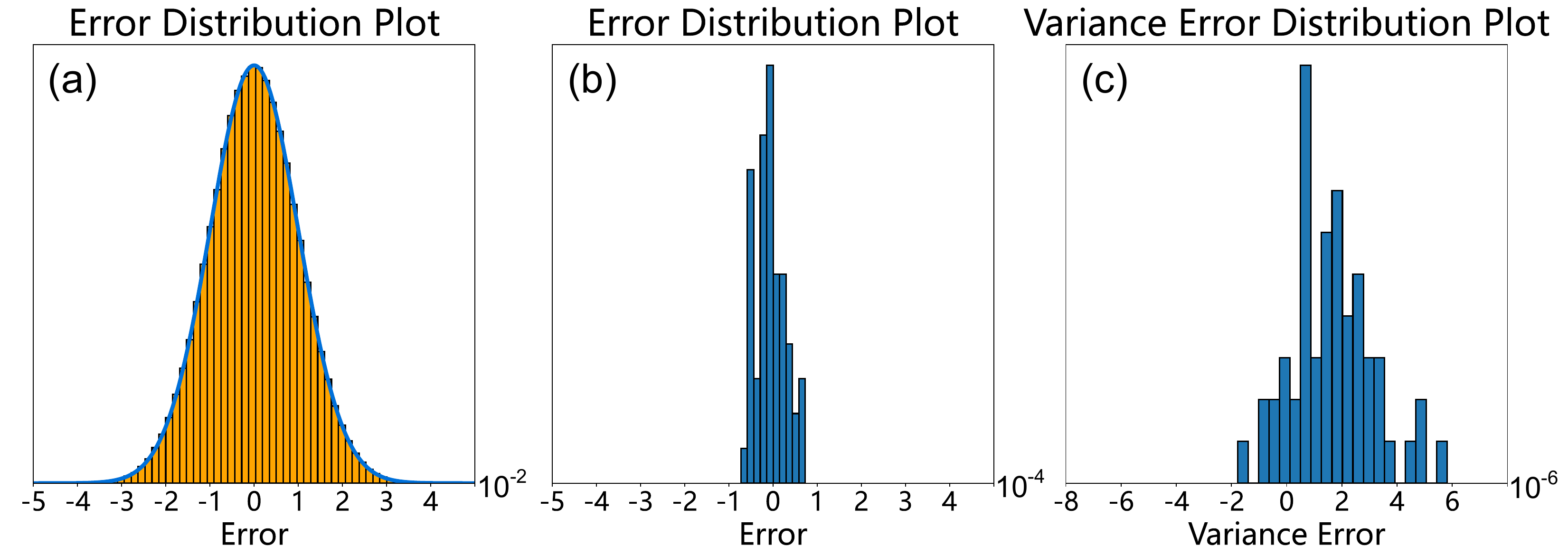}
    \caption{
        (a) The orange part in (a) shows the frequency histogram of the errors in the $\alpha$, $\beta$ parameters for 10-qubit with 5000 samples. The blue curve above the histogram represents the theoretical normal distribution curve of the errors. 
        (b) The frequency histogram of the error between the estimated expectation values and the true values for the $\alpha$, $\beta$ parameters of a 3-qubit system. These errors are small and concentrated around 0, hence it can be considered an unbiased estimate. 
        (c) The frequency histogram of the error between the estimated variance and the theoretical variance for the $\alpha$, $\beta$ parameters of a 3-qubit system. Although there is slight deviation, the overall error is small, so Eq.~\eqref{expectation-variance} can be considered a good estimate.
    }
    \label{fig:simulation-figure-1}
\end{figure}

To verify the effectiveness of the \OurMethod{} and the expectation and variance of the measurements of each ${\alpha_{ij}}$ and ${\beta_{ij}}$, we conducted two sets of simulation experiments. First, we chose a random quantum state of 10 qubits and measured it with a sampling number of $N = 5000$. The density matrix of this quantum state has ${{4}^{10}} - {{2}^{10}} = 1047552$ non-diagonal parameter elements. According to the theoretical analysis, the upper bound for the common variance of these parameters is $1/(2N) = 0.0001$, meaning the measurement standard deviation $\sigma \leq 0.01$. By comparing the measurement results with the true values, we obtain the measurement errors of each parameter: $\Delta \alpha_{ij} = {\hat{\alpha}}_{ij} - \alpha_{ij}$ and $\Delta \beta_{ij} = {\hat{\beta}}_{ij} - \beta_{ij}$. The frequency histograms of the measurement errors for each parameter were plotted and compared with the normal distribution curve with $\mu = 0$ and $\sigma = 0.01$, as shown in Fig.~\ref{fig:simulation-figure-1}(a). It can be observed that the two match very well, indicating that the algorithm successfully measured all 1047552 non-diagonal elements of the density matrix with the same precision upper bound.

To further verify the measurement expectation and variance in Eq.~\eqref{expectation-variance}, we designed another set of simulation experiments. We selected a random quantum state of 3 qubits and measured it with a sampling number of $N = 500$. The entire process was repeated 1000000 times, yielding 1000000 independent measurement results. From these, we estimated the measurement expectation and variance for each parameter. Comparing the results with the theoretical values, Fig.~\ref{fig:simulation-figure-1}(b) shows the frequency histogram of the errors between the measured expectation and the theoretical expectation for each parameter, while Fig.~\ref{fig:simulation-figure-1}(c) shows the frequency histogram of the errors between the measured variance and the theoretical estimated variance for each parameter. From both figures, it can be seen that the results of the simulation experiments are consistent with the theoretical predictions. The measured expectations and variances of all parameters are very close to the theoretical values, verifying the correctness of the theoretical analysis.

\begin{figure}[H]  
    \centering
    \includegraphics[width=0.49\textwidth]{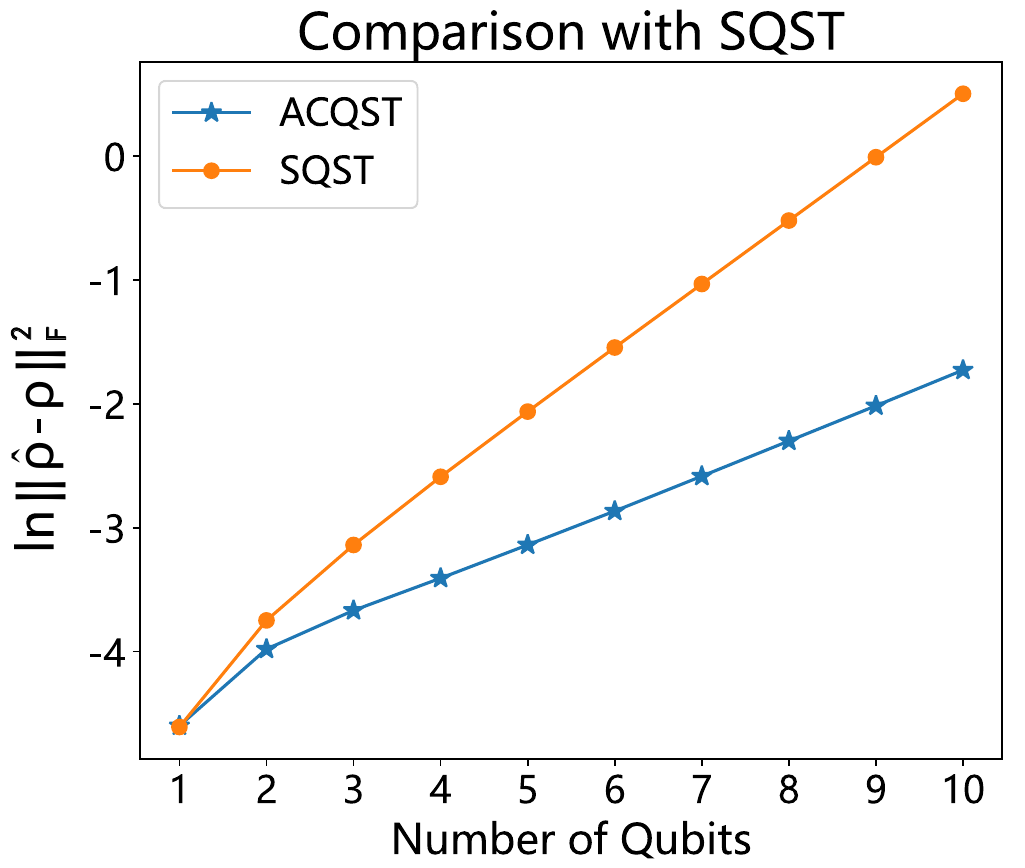}
    \caption{
        The above figure compares the results of \OurMethod{} and SQST. The blue circles represent the results of the \OurMethod{}, while the orange circles represent the results of SQST. The vertical axis shows the $\ln$ of the Frobenius norm squared, and the horizontal axis represents the number of qubits.
    }
    \label{fig:comparison_DMMA_ST}
\end{figure}

In order to demonstrate the efficiency of \OurMethod{} in quantum state tomography, this section also compares the performance differences between \OurMethod{} and SQST through simulation experiments. We selected a set of random quantum states with qubit numbers ranging from 1 to 10, measured them using \OurMethod{} and SQST with the same number of samples, and then compared the Frobenius norm squared of the measurement results $\left\| \hat{\rho }-\rho  \right\|_{\text{F}}^{2}$, plotting the logarithm.

The SQST requires measurements under ${3^n}$ different Pauli basis combinations, and we chose to measure 100 times for each combination, so the total number of samples is $100 \cdot 3^n$. The \OurMethod{} requires measurements of diagonal and off-diagonal elements. Based on the discussion in Section~\ref{subsec:principle}, we set the sampling ratio between the two measurements as $1 : 2^n$. The entire measurement process is repeated multiple times, and the average value of $\left\| \hat{\rho }-\rho  \right\|_{\text{F}}^{2}$ is taken as the final result, which is shown in Fig.~\ref{fig:comparison_DMMA_ST}. It can be seen that the error of \OurMethod{} is significantly lower than that of SQST, and as the number of qubits increases, this advantage further amplifies, with an optimization of about ${(\frac{4}{5})}^n$. This result demonstrates the efficiency of the \OurMethod{} in quantum state tomography, especially when the system has a larger number of qubits.

\begin{figure}[H]  
    \centering
    \includegraphics[width=0.5\textwidth]{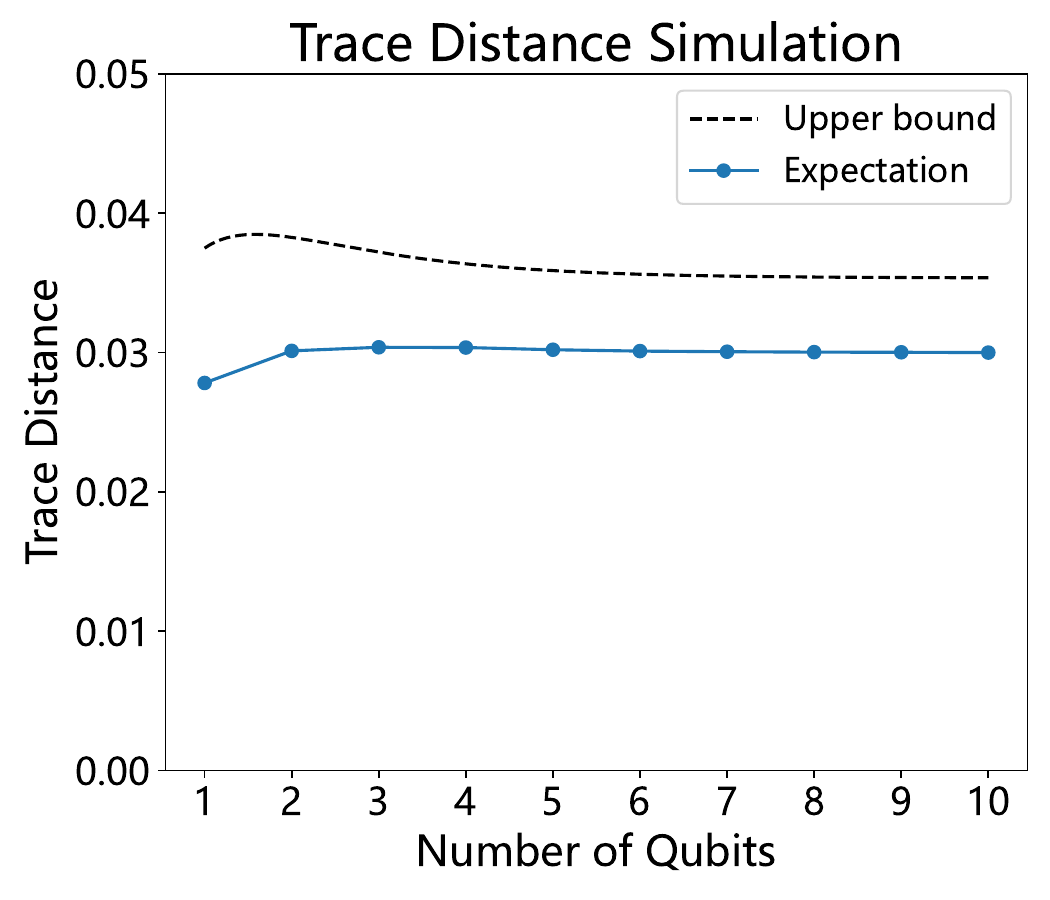}
    \caption{
        The figure above shows the trace distance simulation results for \OurMethod{}. The blue dots represent the expected trace distance of \OurMethod{}, while the black dashed line represents the theoretical upper bound of the expected trace distance. The vertical axis indicates the trace distance between the reconstructed density matrix and the true density matrix, and the horizontal axis represents the number of qubits.
    }
    \label{fig:trace distance}
\end{figure}

To verify the discussion on the upper bound of the trace distance and its complexity in the theoretical analysis, and to further evaluate the effectiveness of the proposed method in quantum state reconstruction, this section also designs and implements numerical simulation experiments based on the trace distance metric. First, we randomly generate quantum states for different numbers of qubits ranging from 1 to 10, set the ratio of the number of samples for diagonal and off-diagonal measurements to be \(1:d\), and measure them with \(N = 200 \cdot d^3\) samples, where \(d\) is the dimension of the density matrix. The entire measurement process is repeated multiple times, and the average value is taken as the final result of the expected trace distance. The results and the theoretical upper bound are plotted as shown in Fig.~\ref{fig:trace distance}. It can be seen that the theoretical upper bound of the expected trace distance is correct, and the sampling number of \(O(d^3)\) can indeed control the growth of the trace distance. The simulation results are consistent with the theoretical predictions, verifying the correctness of the theoretical analysis and the effectiveness of \OurMethod.

\subsection{Simulation of Purity Measurement} \label{subsec:purity-measurement-simulation}

In the discussion of Section~\ref{subsec:purity}, this study presents two methods based on the \OurMethod{} circuit to measure the purity of quantum states, and provides detailed error analysis for both methods. For both pure and mixed states, the measurement complexity of both methods is $O({{2}^{n}})$, and Method 2's measurement accuracy surpasses the standard quantum limit, achieving the Heisenberg limit. Based on the discussion in Section~\ref{subsec:purity}, we know that the purity of a quantum state can be divided into two parts: the $P$ part and the $\alpha \beta$ part. The measurement error is primarily contributed by the $\alpha \beta$ part. Therefore, in the subsequent simulation experiments, we will mainly focus on analyzing the $\alpha \beta$ part of the purity and set the $P$ part to its ideal value to more clearly demonstrate the variance performance of each methods.

\begin{figure}[H]  
    \centering
    \includegraphics[width=\textwidth]{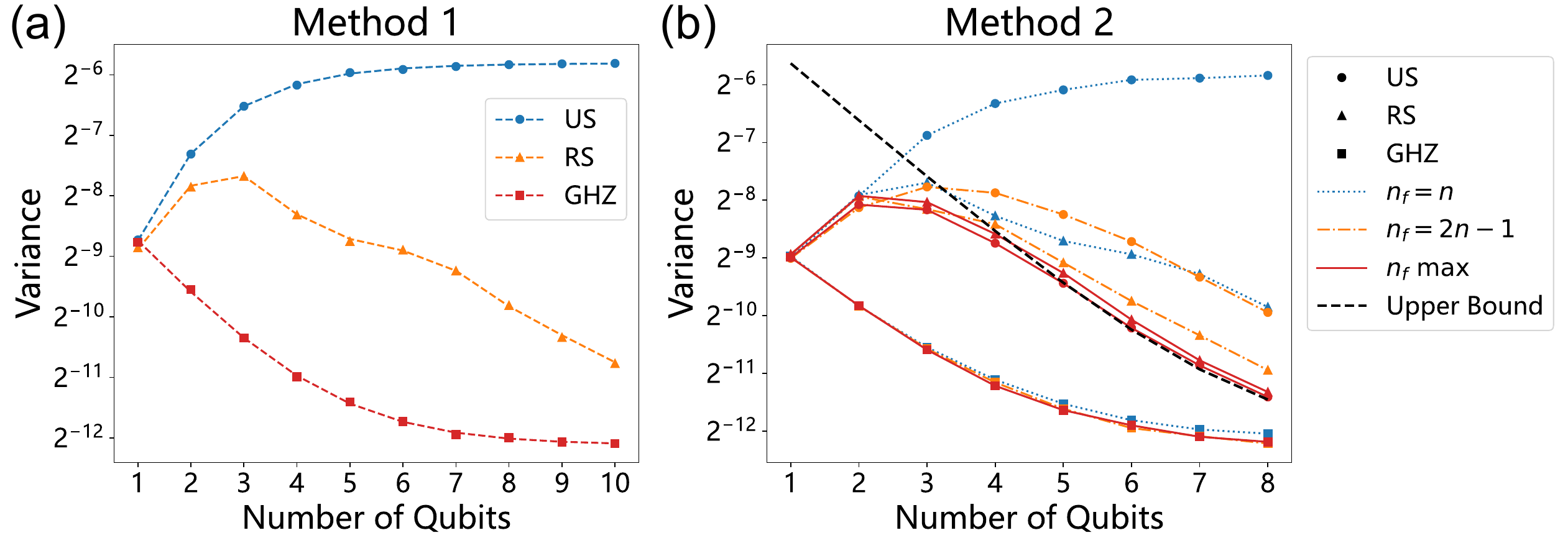}
    \caption{
        (a) The variance of the purity measurement using Method 1 for different numbers of qubits. The blue dots represent the uniform superposition state (US), the orange triangles represent the random state (RS), and the red squares represent the GHZ state. The dashed line represents the theoretical variance. 
        (b) The variance of the purity measurement using Method 2 for different numbers of qubits. The circles represent the uniform superposition state (US), the triangles represent the random state (RS), and the squares represent the GHZ state. The blue, orange, and red colors correspond to auxiliary bit numbers of $n$, $2n-1$, and the maximum value, respectively. The black dashed line represents the theoretical variance upper bound. When the number of auxiliary bits $n_f$ is at its maximum, the measurement variance is close to the theoretical variance upper bound.
    }
    \label{fig:purity_variance_comparison}
\end{figure}

To validate the effectiveness of the two methods and the discussion of measurement variance in the theoretical analysis, two sets of numerical simulations were performed. In the first set of simulations, we selected quantum states with qubit numbers ranging from 1 to 10, including the uniform superposition state, random state, and GHZ state. Dephasing noise was added to the first qubit to adjust the purity of the quantum state. The sample size was set to $N = 100 \cdot 2^n$, and Method 1 was used to measure the purity of the above quantum states. The entire measurement process was repeated 10,000 times to obtain 10,000 measurement results. Based on these measurement results and the true value of the purity, the variance was estimated and compared with the theoretical variance. The results are shown in Fig.~\ref{fig:purity_variance_comparison}(a).

It can be seen that, with a sample size of $100 \cdot 2^n$, the measurement variance changes with the increase of qubits but eventually stabilizes. This indicates that the sampling complexity is indeed $O(2^n)$. Among them, the variance of the random state gradually approaches the curve of the GHZ state as the number of qubits increases. This is because the third-order terms of $\alpha$ and $\beta$ in the random state $\sum P_{qf}^3$ tend to cancel each other out, causing the variance to decrease and eventually approach the GHZ state where all the third-order terms of $\alpha$ and $\beta$ are zero. From the three theoretical curves in Fig.~\ref{fig:purity_variance_comparison}(a), it can be observed that the simulation results are consistent with the theoretical predictions, validating the correctness of the theoretical derivations in previous sections.

In the second set of simulations, we selected the same series of quantum states as in the first set of simulations and used Method 2 to measure their purity under the conditions of ${{n}_{f}} = n$, ${{n}_{f}} = 2n - 1$, and the maximum value of ${{n}_{f}}$. The measurement was repeated 10,000 times, and the variance was estimated based on the measurement results and the true purity values. The measured variance was plotted and compared with the purity variance upper bound given by Eq.~\eqref{eq:variance-alpha-beta-part-method2}, as shown in Fig.~\ref{fig:purity_variance_comparison}(b).

It can be seen that the results do not fully match the theory, as the theoretical analysis assumes the independence of all $\alpha_{ij}^2$ and $\beta_{ij}^2$ terms, but in reality, this independence is not satisfied. However, we can still observe that the result for ${{n}_{f}} = 2n - 1$ performs better than for ${{n}_{f}} = n$, and the result for the maximum ${{n}_{f}}$ performs better than for ${{n}_{f}} = 2n - 1$. This indicates that as ${{n}_{f}}$ increases, the independence between all $\alpha_{ij}^2$ and $\beta_{ij}^2$ terms improves and aligns more closely with the theory. When ${{n}_{f}}$ is at its maximum, although the measured variance for some quantum states exceeds the theoretical variance upper bound, it is still very close, and the theoretical analysis remains valuable. By combining the results in Fig.~\ref{fig:purity_variance_comparison}(a) and Fig.~\ref{fig:purity_variance_comparison}(b), it can be observed that Method 2 shows significantly better performance in measurement accuracy compared to Method 1, which is consistent with the theoretical predictions.

\begin{figure}
    \centering
    \includegraphics[width=0.5\textwidth]{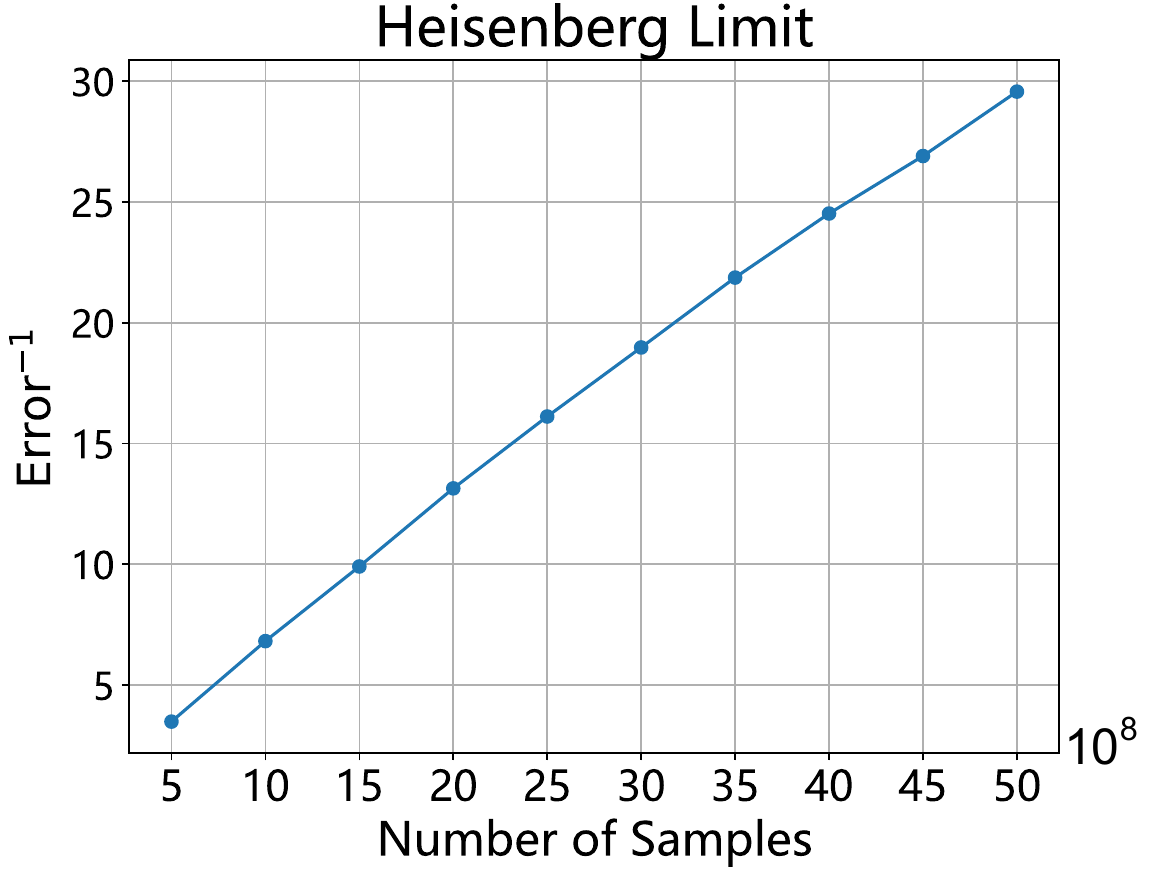}
    \caption{
        The above figure presents the measurement results of the 8-qubit uniform superposition state under sample sizes of $5\cdot {{2}^{8}},10\cdot {{2}^{8}},\cdots ,50\cdot {{2}^{8}}$. The measurement process was repeated 10,000 times to estimate the measurement error $\Delta p$ of Method 2 for purity. The vertical axis represents the reciprocal of the error, $1/\Delta p$, while the horizontal axis denotes the number of samples.
    }
    \label{fig:purity_measurement_error}
\end{figure}

According to the theoretical derivation in the purity measurement section, we know that the error in Method 2's purity measurement approximately follows the Heisenberg limit. To further verify this conclusion, we designed and conducted the following numerical experiments. We selected an 8-qubit uniform superposition state and added dephasing noise to the first qubit to adjust the purity of the quantum state. Method 2 was then used to measure the purity of the quantum state under a range of different sample sizes. The measurement error $\Delta p$ was obtained by repeated the experiment sufficiently, and the results were plotted as $1/\Delta p$.

Ideally, the error satisfies $\Delta p \propto 1/N$, and thus $1/\Delta p \propto N$. Therefore, if Method 2's purity measurement accuracy reaches the Heisenberg limit, $1/\Delta p$ will be linearly proportional to the sample size $N$. As shown in Fig.~\ref{fig:purity_measurement_error}, within the sample size range of $5\cdot {{2}^{8}},10\cdot {{2}^{8}},\cdots ,50\cdot {{2}^{8}}$, the reciprocal of the error $1/\Delta p$ is linearly related to the sample size $N$. This demonstrates that, within a certain range, Method 2's measurement sensitivity for purity reaches the Heisenberg limit. The results from the numerical experiments are consistent with the theoretical predictions, further confirming the validity of the theoretical derivations presented in previous sections.


\begin{figure}[H]  
    \centering
    \includegraphics[width=\textwidth]{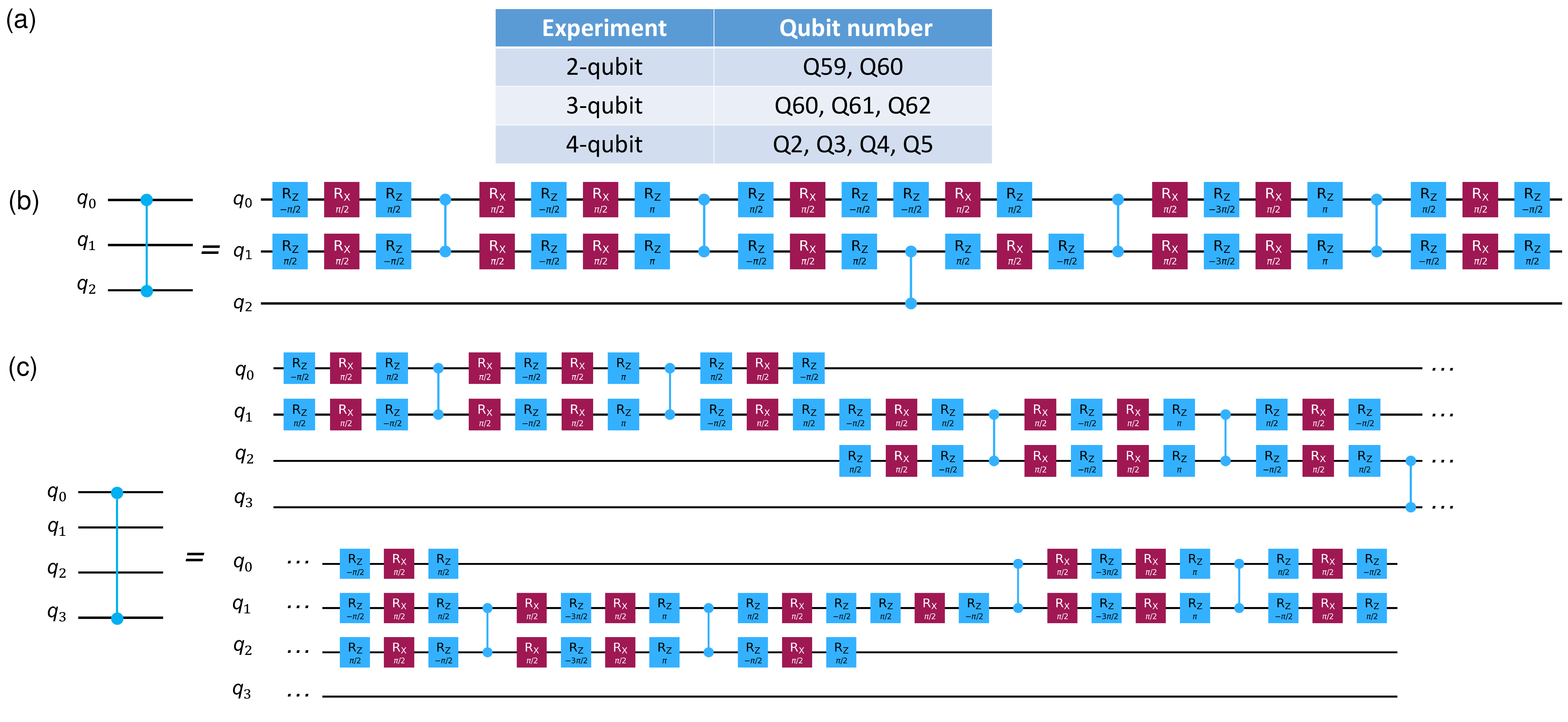}
    \caption{
        (a) The selected qubits in the experiment on quantum cloud hardware \textit{Quafu, Baihua}.  (b) The deposition of CZ gates for qubits separated by a distance of 2.  (c) The deposition of CZ gates for qubits separated by a distance of 3.
    }
    \label{fig:5-1}
\end{figure}

\section{Experimental Results} \label{sec:experimental-results}

We experimentally demonstrate our algorithm on the quantum cloud platform \textit{Quafu Superconducting Quantum Computing}~\cite{QuafuSQC}. The experiments are performed on one-dimensional chains of $N=2$, 3, and 4 qubits, with the specific qubit index listed in Fig.~\ref{fig:5-1}(a). To implement CZ gates between non-adjacent qubits, we decompose them into sequences of adjacent CZ gates and single-qubit gates. This decomposition requires 5 and 9 CZ gates for qubits separated by a distance of 2 and 3, respectively, as depicted in Fig.~\ref{fig:5-1}(b) and Fig.~\ref{fig:5-1}(c). 

\begin{figure}[H]  
    \centering
    \includegraphics[width=\textwidth]{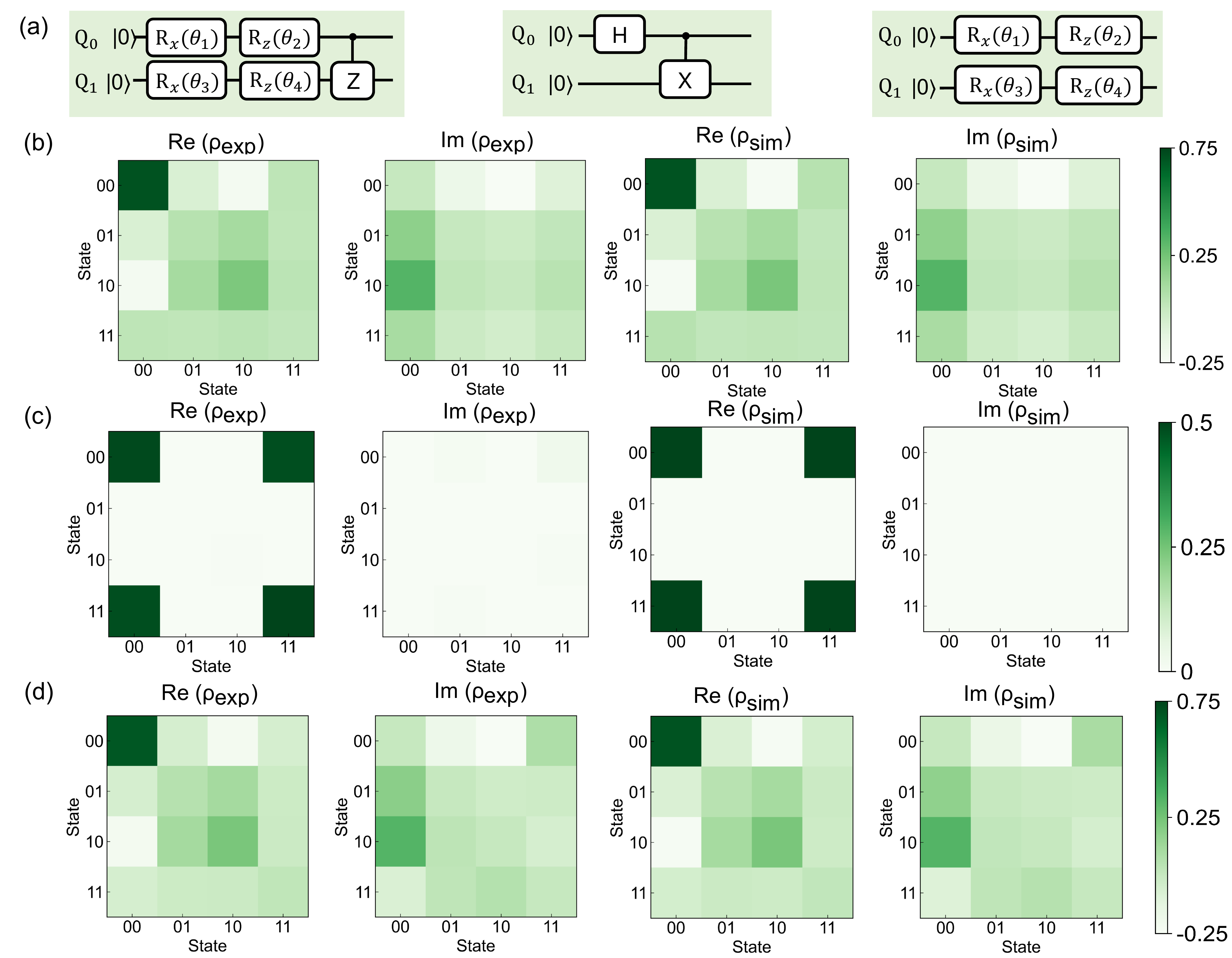}
    \caption{
        The results for 2-qubit experiment. (a) The initialization circuits for preparing the qubits in the arbitrary entangled state (left), Bell state (middle), and arbitrary tensor state (right). The rotation angles $\vec{\theta }$ in single qubit gates are chosen as $\vec{\theta }=\left[ -\frac{\pi }{3},\frac{\pi }{5},\frac{\pi }{6},\frac{\pi }{7} \right]$. (b-d) The real part (Re) and imaginary part (Im) of density matrix obtained from experiment ($\rho_{\text{exp}}$), with qubits initialized to arbitrary entangled state, Bell state, arbitrary tensor state in order. The corresponding density matrix obtained from simulation ($\rho_{\text{sim}}$) are plotted as a reference. The fidelities are 99.84\%, 99.10\% and 99.32\%, respectively. 
    }
    \label{fig12}
\end{figure}

\begin{figure}[H]  
    \centering
    \includegraphics[width=\textwidth]{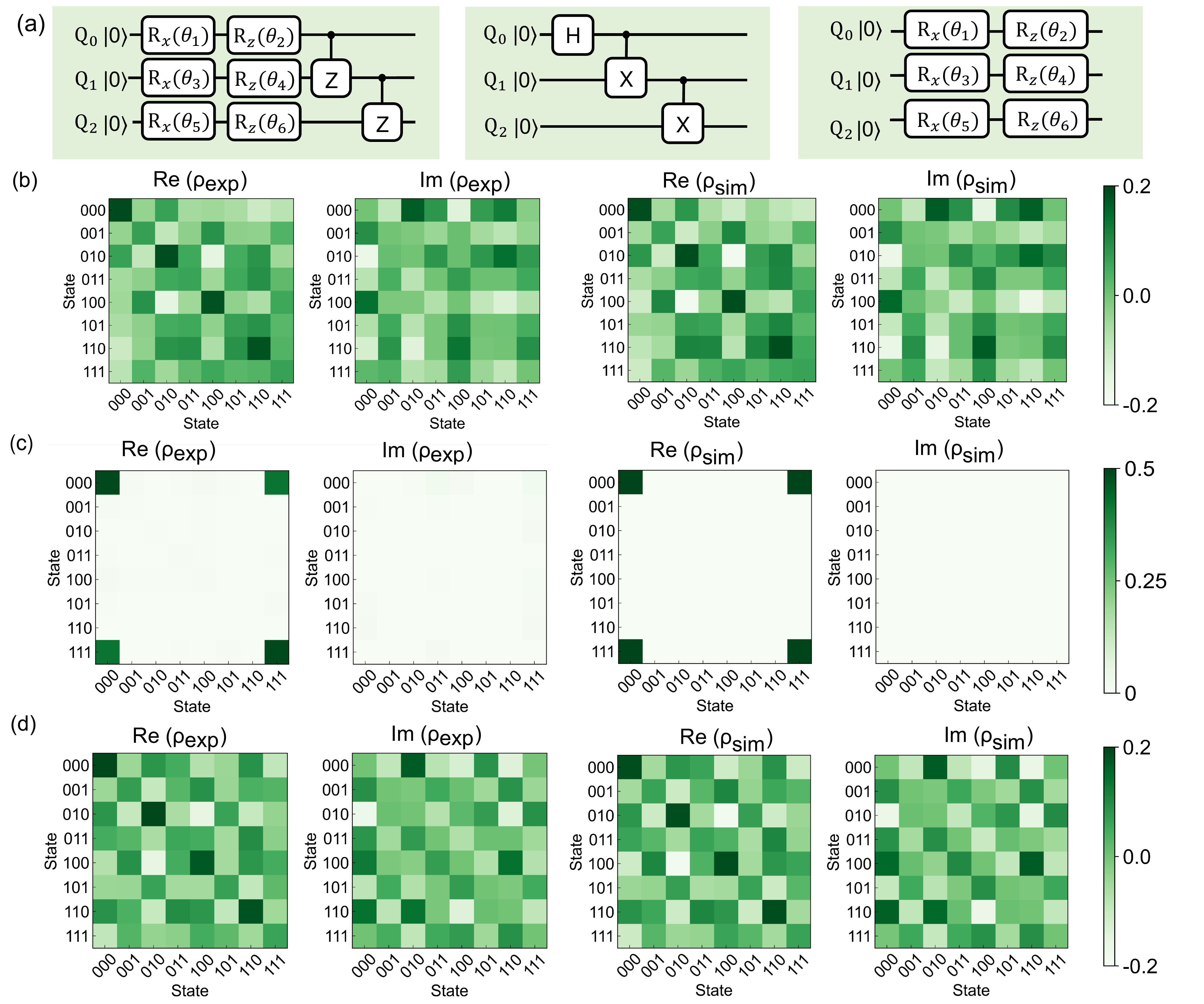}
    \caption{
         The results for 3-qubit experiment. (a) The initialization circuits for preparing the qubits in the arbitrary entangled state (left), GHZ state (middle), and arbitrary tensor state (right). The rotation angles $\vec{\theta }$ in single qubit gates are chosen as $\vec{\theta }=\left[ -\frac{\pi }{2},\frac{\pi }{5},\frac{\pi }{2},\frac{\pi }{7},-\frac{\pi }{3},\frac{\pi }{6} \right]$. (b-d) The real part (Re) and imaginary part (Im) of density matrix obtained from experiment ($\rho_{\text{exp}}$), with qubits initialized to arbitrary entangled state, GHZ state, arbitrary tensor state in order. The corresponding density matrix obtained from simulation ($\rho_{\text{sim}}$) are plotted as a reference. The fidelities are 93.15 \%, 95.38\%, and 94.86\%, respectively.
    }
    \label{fig13}
\end{figure}

\begin{figure}[H]  
    \centering
    \includegraphics[width=\textwidth]{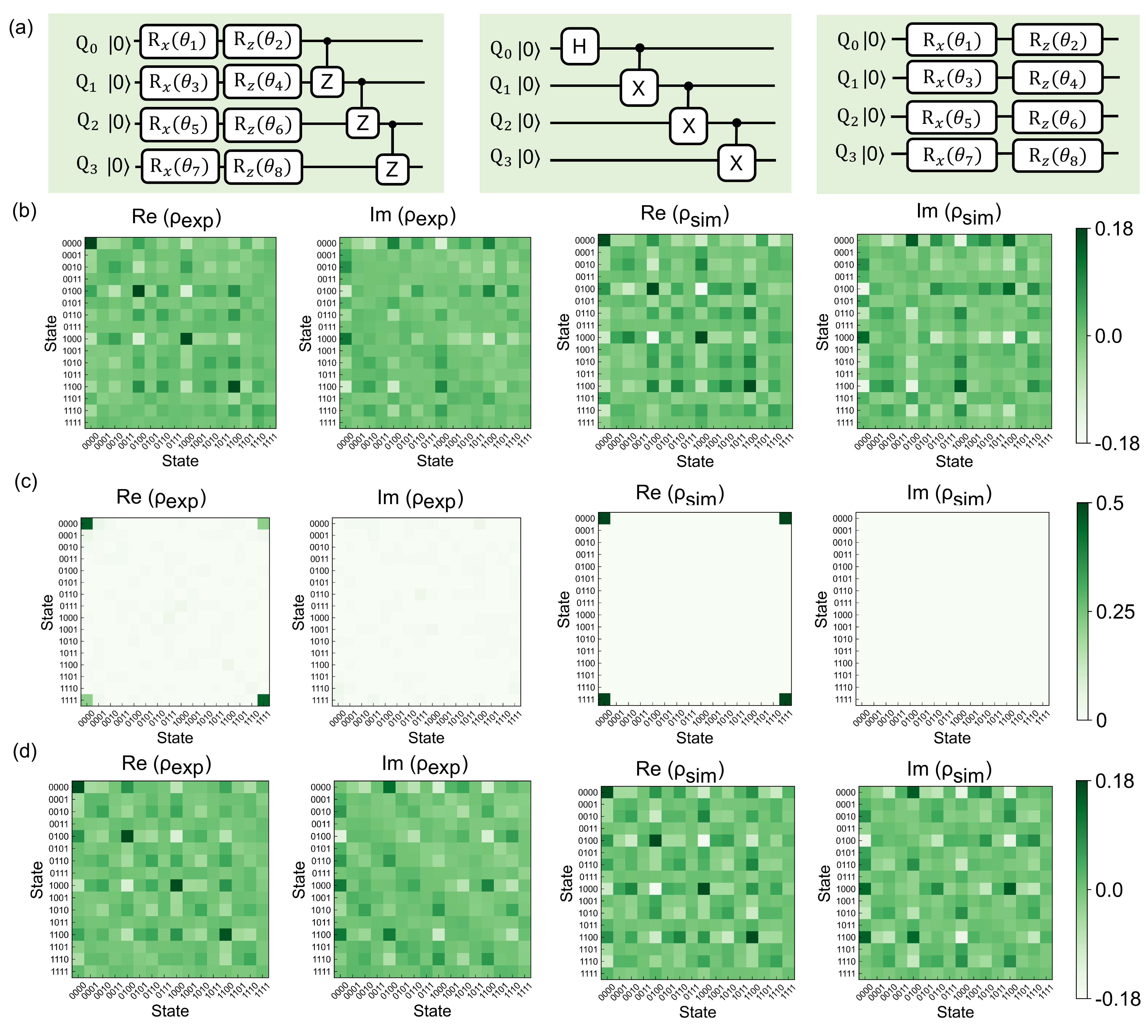}
    \caption{
        The results for 4-qubit experiment. (a) The initialization circuits for preparing the qubits in the arbitrary entangled state (left), GHZ state (middle), and arbitrary tensor state (right). The rotation angles $\vec{\theta }$ in single qubit gates are chosen as $\vec{\theta }=\left[ -\frac{\pi }{2},\frac{\pi }{5},\frac{\pi }{2},\frac{\pi }{7},-\frac{\pi }{3},\frac{\pi }{6},-\frac{\pi }{5},\frac{\pi }{3} \right]$. (b-d) The real part (Re) and imaginary part (Im) of density matrix obtained from experiment ($\rho_{\text{exp}}$), with qubits initialized to arbitrary entangled state, GHZ state, arbitrary tensor state in order. The corresponding density matrix obtained from simulation ($\rho_{\text{sim}}$) are plotted as a reference. The fidelities are 81.92 \%, 81.78\%, and 84.77\%, respectively. 
    }
    \label{fig14}
\end{figure}

To demonstrate generality, in each experiment, the qubits are initialized to three states: the arbitrary entangled state, the GHZ (Bell) state, and the arbitrary tensor state. The corresponding initialization circuits are shown in Fig.~\ref{fig12}, Fig.~\ref{fig13}, and Fig.~\ref{fig14}(a), respectively. Each circuit is repeated with 10240 shots. The density matrices obtained from the experiment are shown in Fig.~\ref{fig12}, Fig.~\ref{fig13}, and Fig.~\ref{fig14}(b-d), respectively. In 2-qubit experiment, the fidelities of state tomography are over 99\%. The errors mainly originate from the CZ gates, with single gate error rates around ~2\%. As the number of qubits increases, more CZ gates are required in the state initialization and readout. The average fidelities of the 3 and 4 quantum state tomography are 94.46\% and 82.82\%, respectively.

\begin{table}[H] 
\caption{Fidelity between the measured density matrices and ideal density matrices\label{tab:fidelity}}
\begin{ruledtabular}
\begin{tabular}{l c c c}
 & \textbf{Arbitrary entangle} & \textbf{GHZ} & \textbf{Arbitrary tensor} \\
\hline
2-qubit & 99.84\% & 99.10\% & 99.32\% \\
3-qubit & 93.15\% & 95.38\% & 94.86\% \\
4-qubit & 81.92\% & 81.78\% & 84.77\% \\
\end{tabular}
\end{ruledtabular}
\end{table}

The experimental results bring out the potential challenges of the Density Matrix Measurement Algorithm in practical quantum devices. In this algorithm, CZ gates between every possible pair of qubits are required. Their realization becomes more complex as the number of qubits increases. To overcome this problem, gates with lower error rates are required. Besides, qubits with higher connectivity are beneficial for the experimental application of the algorithm. 

\section{Conclusion} \label{sec:conclusion}

This study proposes a quantum state tomography method based on auxiliary systems, referred to as the \OurMethod{}. This method extends the measurement dimensions by using entanglement between the system to be measured and a quantum auxiliary system, or by using correlation with a probabilistic classical auxiliary system, to more efficiently acquire the information of the quantum state to be measured. Through three phase control mechanisms in the circuit, the density matrix can adjust its own phase using auxiliary (qu)bits, allowing the probability equations for the measurement outcomes to exhibit the \SpecialProperty{}. This allows for the estimation of all non-diagonal parameters of the density matrix by simply performing selective summation of the counts. The application of the algorithm centers around the \GateMatrix{}, which can label different circuit setups and determine whether a particular setup can make the algorithm work. The \GateMatrix{} also plays a crucial role in data post-processing. In this paper, we also provide general solutions for three types of \GateMatrix{} and some special solutions for single \GateMatrix{}.

The \OurMethod{} requires only two measurement settings to reconstruct the quantum state, greatly simplifying the experimental procedure. The information of the density matrix can be divided into diagonal and non-diagonal parameters. All diagonal parameters of $\rho$ can be measured with a single trivial measurement, and the \OurMethod{} algorithm can measure all non-diagonal parameters of the density matrix with a variance precision of ${1}/{2N}$ in a single measurement setting. In terms of resource consumption, the \OurMethod{} requires $O(n^2)$ quantum gates and has a total sampling complexity of $O(d^2)$. Compared to traditional method, this algorithm provides an efficient and practical solution for many-body quantum state tomography, with fewer sampling requirements and acceptable resource consumption.

In addition, this study presents two methods based on the \OurMethod{} circuit to measure the purity of quantum states. The sampling complexity of both methods for any quantum state is $O({{2}^{n}})$. However, Method 1 only requires squaring and summing the counts measured by the circuit to obtain the $\alpha \beta$ part of the purity, without the need to reconstruct the density matrix, making the data processing simpler. While Method 2 has a data processing complexity comparable to that of reconstructing the density matrix, its measurement accuracy surpasses the standard quantum limit and reaches the Heisenberg limit, showing superior error performance.

Despite significant progress, the required number of samples and the complexity of data post-processing still grow exponentially, and quantum state tomography faces challenges when dealing with large-scale quantum systems. This study introduces a special Sudoku problem—\GateMatrixproblem{}. By filling in the \GateMatrix{}, corresponding algorithm circuits can be designed. Although we have provided three general solutions, the \GateMatrixproblem{} as a mathematical problem remains unsolved. The general solution for ${n_f} = n$, though it minimizes the number of auxiliary bits, does not minimize the number of quantum gates. The general solution for ${n_f} = 2n-1$, while minimizing the number of quantum gates, does not achieve the optimal number of auxiliary bits. These issues still remain open for further exploration. Additionally, methods for directly using the measurement data from the \OurMethod{} algorithm to calculate other quantum state parameters, aside from purity, are also worth exploring. Although the density matrix reconstructed by the current method satisfies Hermiticity, it may exhibit negative eigenvalues. To address this issue, corresponding maximum likelihood estimation methods are still worth further investigation.

Future research will need to explore more efficient state tomography methods and further optimize existing technologies to meet the needs of more complex quantum systems. However, this study demonstrates the tremendous potential of the auxiliary-system-based tomography method in improving measurement efficiency and reducing experimental complexity, and still provides a practical solution for the field of quantum state tomography.

\begin{acknowledgments}
We thank Zhaohui Wei for helpful discussions. This work is supported by 
the Beijing Institute of Technology Research Fund Program under Grant No. 2024CX01015, the National Natural Science Foundation of China (Grant Nos. 12441502, 92365206, 12404560) and the Innovation Program for Quantum Science and Technology (No. 2021ZD0301802).

\end{acknowledgments}

\section*{DATA AVAILABILITY} 
The source code supporting this study is available in Ref.~\cite{long2025efficient}.

\bibliography{ref}

\appendix  

\section{} \label{appendix:A}  
This study employs the current density matrix reconstruction algorithm (see Eq.~\eqref{expectation-variance}) rather than the direct frequency-estimation method because the current approach obtains smaller variance without almost any cost (neither additional computational expense nor performance degradation). We now analyze both methods' performance. First, for the direct frequency-estimation method, as discussed in Section~\ref{subsec:principle}, we consider the following system of equations:
\begin{equation}
\label{appendix four-classification}
\begin{aligned}
  & {{P}_{1}} = \frac{1}{4} + \frac{1}{2}{{\alpha }_{ij}}, \quad {{P}_{3}} = \frac{1}{4} + \frac{1}{2}{{\beta }_{ij}} ,\\ 
  & {{P}_{2}} = \frac{1}{4} - \frac{1}{2}{{\alpha }_{ij}}, \quad {{P}_{4}} = \frac{1}{4} - \frac{1}{2}{{\beta }_{ij}} .\\ 
\end{aligned}
\end{equation}
Denote the occurrence counts of events ${{P}_{1}},{{P}_{2}},{{P}_{3}},{{P}_{4}}$ as ${{N}_{1}},{{N}_{2}},{{N}_{3}},{{N}_{4}}$ respectively. Using the event frequencies as probability estimates, we obtain:
\begin{equation}
\label{frequency-estimation}
{{\hat{\alpha }}_{ij}}=\frac{{{N}_{1}}-{{N}_{2}}}{N}, \quad {{\hat{\beta }}_{ij}}=\frac{{{N}_{3}}-{{N}_{4}}}{N}.
\end{equation}
where $N$ is the total sample size. We compute the expectation and variance of this method:
\begin{equation}
\label{expectation-variance-derivation}
\begin{aligned}
  & \text{E}({{\hat{\alpha }}_{ij}}) = \frac{\text{E}({{N}_{1}}) - \text{E}({{N}_{2}})}{N} = \frac{N{{P}_{1}} - N{{P}_{2}}}{N} = {{\alpha }_{ij}} ,\\ 
  & \text{D}({{\hat{\alpha }}_{ij}}) = \frac{\text{D}({{N}_{1}} - {{N}_{2}})}{{{N}^{2}}} = \frac{\text{E}(N_{1}^{2} + N_{2}^{2} - 2{{N}_{1}}{{N}_{2}}) - {{\left[ \text{E}({{N}_{1}} - {{N}_{2}}) \right]}^{2}}}{{{N}^{2}}} \\ 
  & \qquad = \frac{\left[ N(N-1)P_{1}^{2} + N{{P}_{1}} \right] + \left[ N(N-1)P_{2}^{2} + N{{P}_{2}} \right] - 2N(N-1){{P}_{1}}{{P}_{2}} - {{(N{{P}_{1}} - N{{P}_{2}})}^{2}}}{{{N}^{2}}} \\ 
  & \qquad = \frac{N({{P}_{1}} + {{P}_{2}}) - N{{({{P}_{1}} - {{P}_{2}})}^{2}}}{{{N}^{2}}} \\ 
  & \qquad = \frac{1 - 2\alpha_{ij}^{2}}{2N} .\\ 
\end{aligned}
\end{equation}

Similarly, $\text{E}({{\hat{\beta }}_{ij}})={{\beta }_{ij}}$ and $\text{D}({{\hat{\beta }}_{ij}})=\frac{1-2\beta _{ij}^{2}}{2N}$, which shows that the direct frequency-estimation method is an unbiased estimator. For the current method, the events corresponding to ${{P}_{1}},{{P}_{2}}$ can be viewed as independent measurements of $\alpha$, while those corresponding to ${{P}_{3}},{{P}_{4}}$ can be considered as independent measurements of $\beta$. Under the condition of measuring $\alpha$, we obtain the following conditional probabilities:
\begin{equation}
\label{conditional-probabilities}
\begin{aligned}
  & {{P}_{1}}^{\prime} = \frac{1}{2} + {{\alpha }_{ij}} ,\\ 
  & {{P}_{2}}^{\prime} = \frac{1}{2} - {{\alpha }_{ij}} .\\ 
\end{aligned}
\end{equation}
Consequently we derive:
\begin{equation}
\label{new-result}
{{\hat{\alpha }}_{ij}} = \frac{{{N}_{1}}}{{{N}_{\alpha }}} - \frac{1}{2},
\end{equation}
where ${{N}_{\alpha }} = {{N}_{1}} + {{N}_{2}}$. However, considering the case when ${{N}_{\alpha }} = 0$, the actual expression for ${{\hat{\alpha }}_{ij}}$ should be:
\begin{equation}
\label{strict-result}
{{\hat{\alpha }}_{ij}} = \left\{ 
\begin{aligned} 
&\frac{{{N}_{1}}}{{{N}_{\alpha }}} - \frac{1}{2} & , & {{N}_{\alpha }} \ne 0 \\
&0 & , & {{N}_{\alpha }} = 0 
\end{aligned} 
\right.
\end{equation}

At this point, we can calculate that under the condition of known ${{N}_{\alpha }}$, the expectation of ${{\hat{\alpha }}_{ij}}$ is:
\begin{equation}
\label{conditional-expectation}
\text{E}\left({{\hat{\alpha }}_{ij}} \mid {{N}_{\alpha }}\right) = \left\{ 
\begin{aligned} 
&\alpha_{ij} & , & {{N}_{\alpha }} \neq 0 \\
&0 & , & {{N}_{\alpha }} = 0 
\end{aligned} 
\right.
\end{equation}
It is straightforward to see that for a sample size of $N$, $P({{N}_{\alpha }}=0) = {1}/{2^{N}}$. Therefore, we can calculate the expectation of the current method ${{\hat{\alpha }}_{ij}}$ as:
\begin{equation}
\label{strict-expectation}
\begin{aligned}
  & \text{E}({{{\hat{\alpha }}}_{ij}})=\sum\limits_{{{N}_{\alpha }}}{\text{E}({{{\hat{\alpha }}}_{ij}}\left| {{N}_{\alpha }} \right.)\cdot P({{N}_{\alpha }})}=0+\sum\limits_{{{N}_{\alpha }}\ne 0}{{{\alpha }_{ij}}\cdot P({{N}_{\alpha }})} \\ 
  & \qquad ={{\alpha }_{ij}}\cdot \sum\limits_{{{N}_{\alpha }}\ne 0}{P({{N}_{\alpha }})}={{\alpha }_{ij}}\left[ 1-P({{N}_{\alpha }}=0) \right] \\ 
  & \qquad =\left( 1-\frac{1}{{{2}^{N}}} \right){{\alpha }_{ij}} .\\ 
\end{aligned}
\end{equation}
It can be seen that the current method is not strictly unbiased, having a relative error of ${1}/{2^{N}}$. However, since $N$ is generally large, ${1}/{2^{N}}$ is extremely small, so it can be considered as effectively unbiased. We now analyze the variance of the current method. When ${{N}_{\alpha }} \ne 0$:
\begin{equation}
\label{conditional-variance}
\begin{aligned}
  & \text{D}({{\hat{\alpha }}_{ij}}\left| {{N}_{\alpha }} \right.) = \text{D}\left( \left. \frac{{{N}_{1}}}{{{N}_{\alpha }}} - \frac{1}{2} \right| {{N}_{\alpha }} \right) = \frac{\text{D}(\left. {{N}_{1}} \right| {{N}_{\alpha }})}{N_{\alpha }^{2}} \\
  & \qquad = \frac{{{N}_{\alpha }}{{P}_{1}}^{\prime }{{P}_{2}}^{\prime }}{N_{\alpha }^{2}} = \frac{1}{{{N}_{\alpha }}}\left(\frac{1}{4} - \alpha_{ij}^{2}\right) .\\ 
\end{aligned}
\end{equation}
Neglecting the probability ${1}/{2^{N}}$ case where ${{N}_{\alpha }}=0$, we have:
\begin{equation}
\label{variance-estimation}
\text{D}({{\hat{\alpha }}_{ij}}) \approx \sum\limits_{{{N}_{\alpha }}\ne 0}{\text{D}({{{\hat{\alpha }}}_{ij}}\left| {{N}_{\alpha }} \right.)\cdot P({{N}_{\alpha }})}=(\frac{1}{4}-\alpha _{ij}^{2})\sum\limits_{{{N}_{\alpha }}\ne 0}{\frac{1}{{{N}_{\alpha }}}\cdot \frac{N!}{{{N}_{\alpha }}!{{N}_{\beta }}!}{{\left( \frac{1}{2} \right)}^{N}}}.
\end{equation}
Considering that when $N$ is sufficiently large, $N \approx N+1$ and ${{N}_{\alpha }} \approx {{N}_{\alpha }}+1$, we have:
\begin{equation}
\label{variance-approximation}
\text{D}({{\hat{\alpha }}_{ij}}) \approx \left(\frac{1}{4}-\alpha _{ij}^{2}\right)\frac{2}{(N+1)}\sum\limits_{{{N}_{\alpha }}\ne 0}{\frac{(N+1)!}{({{N}_{\alpha }}+1)!{{N}_{\beta }}!}{{\left( \frac{1}{2} \right)}^{N+1}}} \approx \frac{1-4\alpha _{ij}^{2}}{2N}.
\end{equation}

For ${{\hat{\beta }}_{ij}}$, by the same reasoning we have 
$\text{E}({{\hat{\beta }}_{ij}}) = \left( 1 - \frac{1}{{{2}^{N}}} \right){{\beta }_{ij}}$,
$\text{D}({{\hat{\beta }}_{ij}}) \approx \left( \frac{1 - 4\beta_{ij}^{2}}{2N} \right)$.
From Eq.~\eqref{expectation-variance-derivation} and Eq.~\eqref{variance-approximation}, 
it can be seen that the current method exhibits superior variance performance compared to the direct frequency-estimation approach.

\section{} \label{appendix:B}  
Based on the discussion of three phase control mechanisms in the quantum circuit in Section~\ref{subsec:circuit}, the system of probability equations takes the form:
\begin{widetext}
\begin{equation}
\label{general-equation1}
\resizebox{\textwidth}{!}{$
\renewcommand{\arraystretch}{0.7} 
\setlength{\arraycolsep}{1pt} 
\begin{array}{cccc}
   P({{q}_{1}}{{q}_{2}}\ldots {{q}_{n}}{{f}_{1}}{{f}_{2}}\ldots {{f}_{{{n}_{f}}}})=\frac{1}{{{2}^{n+{{n}_{f}}}}}+\frac{1}{{{2}^{n+{{n}_{f}}}}}2\operatorname{Re}\sum\limits_{i>j}{({{\alpha }_{ij}}+i{{\beta }_{ij}})} & {{(-1)}^{{{i}_{1}}{{q}_{1}}}}{{(-1)}^{{{j}_{1}}{{q}_{1}}}} & {{i}^{{{i}_{1}}{{f}_{1}}}}{{(-i)}^{{{j}_{1}}{{f}_{1}}}} & {{(-1)}^{{{i}_{1}}{{i}_{2}}{{x}_{1}}}}{{(-1)}^{{{j}_{1}}{{j}_{2}}{{x}_{1}}}}  \\
   {} & {{(-1)}^{{{i}_{2}}{{q}_{2}}}}{{(-1)}^{{{j}_{2}}{{q}_{2}}}} & {{i}^{{{i}_{2}}{{f}_{2}}}}{{(-i)}^{{{j}_{2}}{{f}_{2}}}} & {{(-1)}^{{{i}_{1}}{{i}_{3}}{{x}_{2}}}}{{(-1)}^{{{j}_{1}}{{j}_{3}}{{x}_{2}}}}  \\
   {} & \vdots  & \vdots  & \vdots   \\
   {} & {{(-1)}^{{{i}_{n}}{{q}_{n}}}}{{(-1)}^{{{j}_{n}}{{q}_{n}}}} & {{i}^{{{i}_{n}}{{f}_{n}}}}{{(-i)}^{{{j}_{n}}{{f}_{n}}}} & {}  \\
\end{array}.
$}
\end{equation}
\end{widetext}
where ${{q}_{1}}{{q}_{2}}\ldots {{q}_{n}}{{f}_{1}}{{f}_{2}}\ldots {{f}_{{{n}_{f}}}}$ are the readout values of $n$ measured qubits and ${{n}_{f}}$ auxiliary bits. By enumerating ${{q}_{1}}{{q}_{2}}\ldots {{q}_{n}}{{f}_{1}}{{f}_{2}}\ldots {{f}_{{{n}_{f}}}}$ from $00\ldots 000\ldots 0$ to $11\ldots 111\ldots 1$, we obtain the complete set of probability equations. To study the \SpecialProperty{} properties of the probability equations, we use $i,j$ as markers to partition the equations column-wise and independently analyze the ${{\alpha }_{ij}}$ and ${{\beta }_{ij}}$ terms. The phase in Eq.~\eqref{general-equation1} is denoted as $L$, where $L$ is a function of ${{q}_{1}}{{q}_{2}}\ldots {{q}_{n}}{{f}_{1}}{{f}_{2}}\ldots {{f}_{{{n}_{f}}}}$. Each $i,j$ pair corresponds to one column in the probability equations and its associated $L$. Specifically, this can be written as ${{L}_{ij}}({{q}_{1}}{{q}_{2}}\ldots {{q}_{n}}{{f}_{1}}{{f}_{2}}\ldots {{f}_{{{n}_{f}}}})$. Proper configuration of quantum gates enables global \SpecialProperty{} of the probability equations, achieved by tuning $L$ column-wise.

\begin{figure}[H]  
    \centering
    \includegraphics[width=\textwidth]{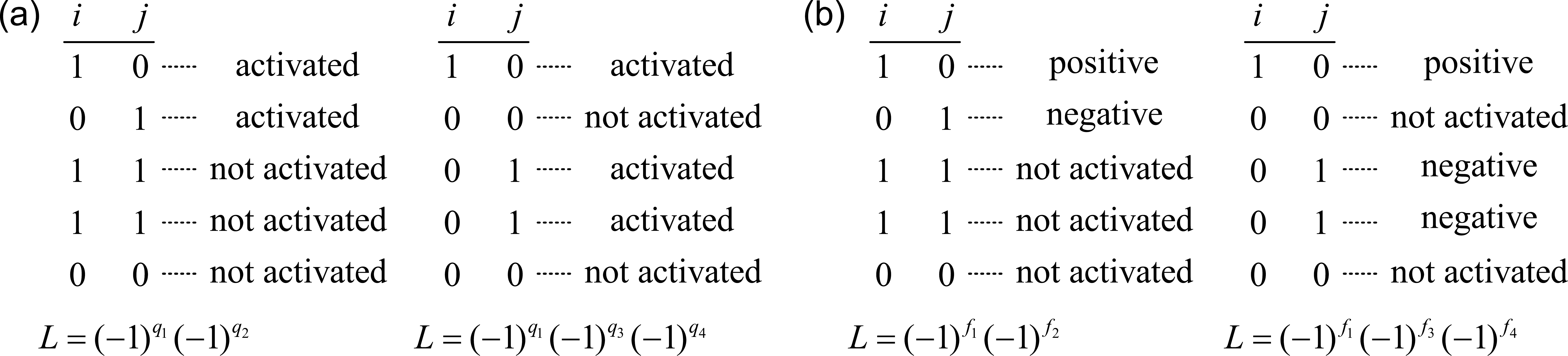}
    \caption{
        The figure shows two examples: \(i = 10110, j = 01110\) and \(i = 10000, j = 00110\). Panel (a) illustrates the process and result of the \singlecontrol{}, while panel (b) illustrates the process and result of the \phasecontrol{}.
    }
    \label{fig:appendix_1}
\end{figure}

There exist numerous approaches to achieve this result, with the method presented in this paper representing the simplest solution currently obtained in our study. To further illustrate this, we employ a pictorial explanation. For ease of discussion, we vertically write \(i\) and \(j\) in binary form, with \(i\) preceding \(j\). According to the discussion in Section~\ref{subsec:circuit}, \singlecontrol{} is governed by a pair of \(i_k,j_k\) that makes \(L\) possess the term \((-1)^a\), where \(a\) is one of \(q_1,q_2,\ldots,q_n\). However, following our defined specification, it should be \(q_k\). For the \(L\) of each \(i,j\) column, when \(i_k=0,j_k=0\) or \(i_k=1,j_k=1\), we have \((-1)^{i_k q_k}(-1)^{j_k q_k}=1\), indicating that the control is not activated. In all other cases, \((-1)^{i_k q_k}(-1)^{j_k q_k}=(-1)^{q_k}\), indicating that the control is activated. \phasecontrol{} is governed by a pair of \(i_k,j_k\) that makes \(L\) possess the term \(i^a\), where \(a\) is one of \(f_1,f_2,\ldots,f_{n_f}\). Following our defined specification, it should be \(f_k\). For the \(L\) of each \(i,j\) column, when \(i_k=0,j_k=0\) or \(i_k=1,j_k=1\), we have \(i^{i_k f_k}(-i)^{j_k f_k}=1\), indicating that the control is not activated. When \(i_k=1,j_k=0\), we have \(i^{i_k f_k}(-i)^{j_k f_k}=i^{f_k}\), indicating positive control activation; when \(i_k=0,j_k=1\), we have \(i^{i_k f_k}(-i)^{j_k f_k}=(-i)^{f_k}\), indicating negative control activation. We further illustrate the processes of \singlecontrol{} and \phasecontrol{} through two examples, as shown in Fig.~\ref{fig:appendix_1}.

\begin{figure}[H]  
    \centering
    \includegraphics[width=0.32\textwidth]{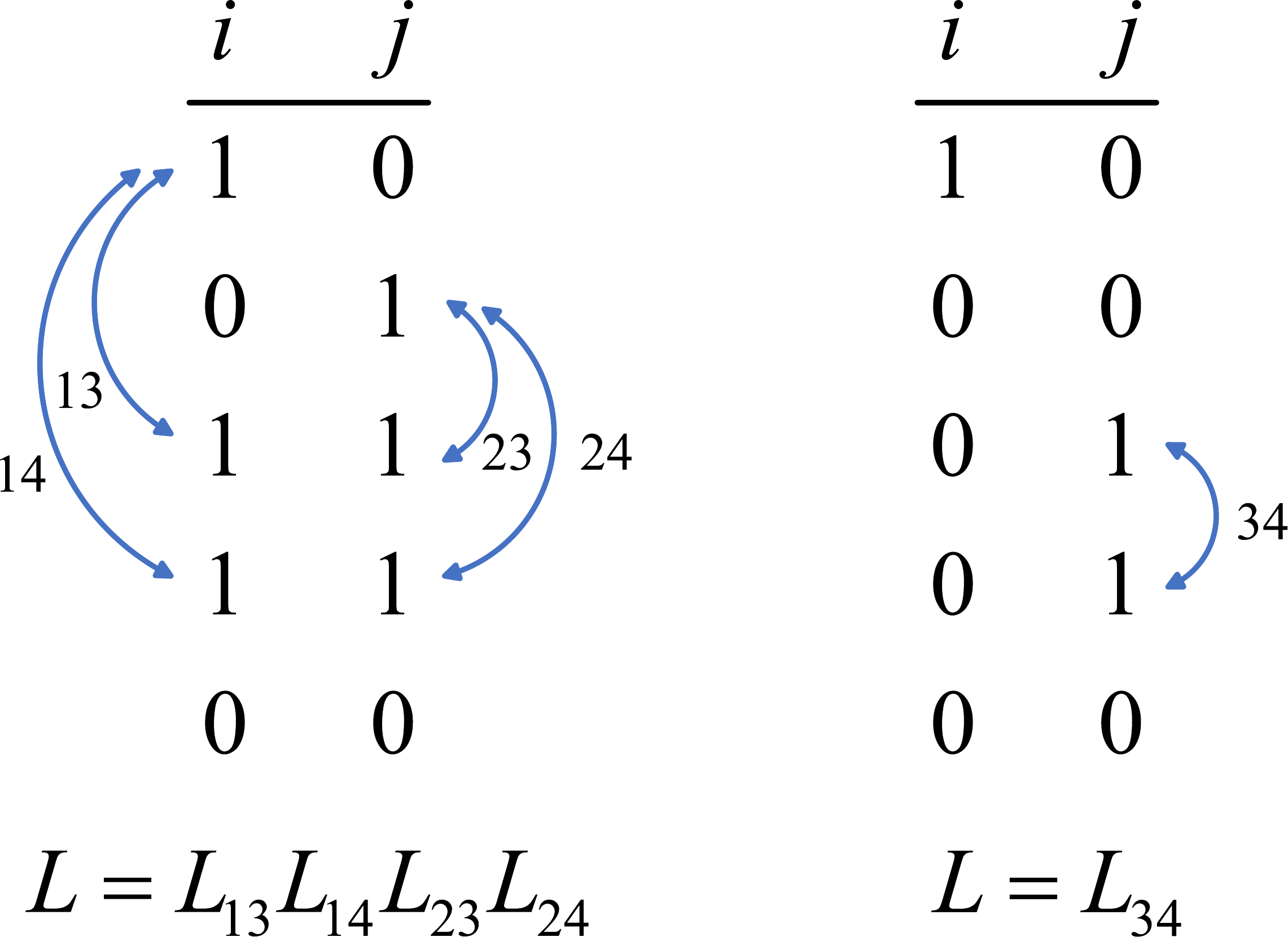}
    \caption{
        The figure illustrates the process and results under \doublecontrol\ for two examples: \(i = 10110, j = 01110\) and \(i = 10000, j = 00110\). Only the controls corresponding to the bit pairs connected by arrows are activated.
    }
    \label{fig:appendix_2}
\end{figure}

The \doublecontrol{} is governed by two pairs of \(i_{k_1}, j_{k_1}\) and \(i_{k_2}, j_{k_2}\), which make \(L\) possess the process of \((-1)^a (-1)^b (-1)^c \ldots\), where \(a, b, c, \ldots\) are a set of \(f_1, f_2, \ldots, f_{n_f}\). For the \(L\) of each \(i,j\) column, only when
\begin{equation}
\label{control-conditions}
\left( \begin{matrix}
   i_{k_1} & j_{k_1}  \\
   i_{k_2} & j_{k_2}  \\
\end{matrix} \right)
=
\left( \begin{matrix}
   1 & 0  \\
   1 & 1  \\
\end{matrix} \right),\ 
\left( \begin{matrix}
   1 & 1  \\
   1 & 0  \\
\end{matrix} \right),\ 
\left( \begin{matrix}
   1 & 0  \\
   1 & 0  \\
\end{matrix} \right),\ 
\left( \begin{matrix}
   0 & 1  \\
   1 & 1  \\
\end{matrix} \right),\ 
\left( \begin{matrix}
   1 & 1  \\
   0 & 1  \\
\end{matrix} \right),\ 
\left( \begin{matrix}
   0 & 1  \\
   0 & 1  \\
\end{matrix} \right),
\end{equation}
the control is activated, while in all other cases the control is not activated. According to the discussion in Section~\ref{subsec:circuit}, the number and specific auxiliary bits used for the \doublecontrol{} can be arbitrarily chosen. Therefore, for the \doublecontrol{}, the primary focus is on whether the control is activated for \(k_1 k_2\), and the effect of the gate is denoted as \(L_{k_1 k_2}\). Eventually, the \doublecontrol{} is given by \(L = \prod L_{k_1 k_2}\). In the two examples mentioned above, as shown in Fig.~\ref{fig:appendix_2}.

Finally, the complete \(ij\) column corresponding \(L\) is the product of the \(L\)s from each of the three controls.
\begin{equation}
\label{control-combination}
L = {{L}_{\text{Single}}} \cdot {{L}_{\text{Phase}}} \cdot {{L}_{\text{Double}}}.
\end{equation}

\subsection{Properties of the Current \textit{L}}
\label{subsec:L_properties}
According to the current settings of \singlecontrol{}, \phasecontrol{}, and \doublecontrol{}, the \(L\) corresponding to each \(ij\) has some properties. Since all \(ij\) satisfy \(i \neq j\), the control of \singlecontrol{} and \phasecontrol{} will definitely be activated, and the results generated by \singlecontrol{} and \phasecontrol{} will not cancel each other out, nor will they be canceled out by the result of \doublecontrol{}. Therefore, the current \(L\) has \textbf{Property I}: The \(L\) of all \(ij\) contains negative signs and contains \(i\). That is, \(L\) must include at least one \((-1)^a\) and \(i^b\), where \(a\) is one of \(q_1, q_2, \ldots, q_n\), and \(b\) is one of \(f_1, f_2, \ldots, f_{n_f}\). At the same time, since the control conditions for \singlecontrol{} and \phasecontrol{} are the same, the current \(L\) has \textbf{Property II}: For any two \(L_1\) and \(L_2\) with the same \singlecontrol{} part, their \phasecontrol{} parts are aligned. That is, the auxiliary bits of the \(i\) part in \(L_1\) and \(L_2\) are the same. For example, if the \(i\) part of \(L_1\) is \(i^{f_1} (-i)^{f_2} i^{f_5}\), and the \(i\) part of \(L_2\) is \(i^{f_1} i^{f_2} (-i)^{f_5}\), they are different, but both are on the same auxiliary bits \(f_1, f_2, f_5\).

\subsection{Mathematical Treatment of \textit{L}}
\label{sec:math_treatment_L} 
Before further introducing the conditions that \(L\) should satisfy, it is necessary to first explain some mathematical treatments of \(L\). Let's start with the decomposition of \(L\). \(L\) can be arbitrarily divided into multiple \(L_k\), each of which has the same form as \(L\) (that is, a product of power functions with bases \(1, -1, i, -i\) and exponents being 01 variables), and \(L = \prod L_k\). There are many ways to decompose \(L\), one of which is bit-wise decomposition. The variables in the exponents of \(L\) correspond to a qubit or classical bit, with the variable's value being the readout of that (qu)bit. Conversely, each bit corresponds to a power function in \(L\). Although terms like \(1^a\) can be omitted, they still exist. We can assign different bits to different \(L_k\). For example,
\begin{equation}
\label{example1}
\begin{aligned}
L &= (-1)^a (-1)^b i^c (-i)^d (-1)^e ,\\
L_1 &= (-1)^a (-1)^b (-1)^e, \quad L_2 = i^c (-i)^d.
\end{aligned}
\end{equation}

In Eq.~\eqref{example1}, \(L\) is decomposed into the non-\(i\) part \(L_1\) and the \(i\) part \(L_2\) through bit-wise decomposition. Bit-wise decomposition can divide \(L\) into several independent parts, which facilitates theoretical analysis.

Another special decomposition method is the negative-signs decomposition. A single term \((-i)^a\) can be split into the product of two terms \((-1)^a\) and \(i^a\). Therefore, we can decompose all the \(-1\) terms into a single \(L_k\). For example,
\begin{equation}
\label{example2}
\begin{aligned}
L &= (-1)^{a} (-1)^{b} i^{c} (-i)^{d} (-1)^{e} ,\\
L_1 &= (-1)^{a} (-1)^{b} (-1)^{d} (-1)^{e}, \quad L_2 = i^{c} \cdot i^{d}.
\end{aligned}
\end{equation}

In Eq.~\eqref{example2}, \(L\) is decomposed into the pure negative part \(L_1\) and the pure \(i\) part \(L_2\) through \signdecomposition{}. It can be seen that \(L_1\) and \(L_2\) are now pure and simple, which facilitates the processing by a computer.

Next, we discuss the combination of \(L\). The process of multiplying some subparts \(L_k\) together to obtain a new \(L'_k\) is called combination. By combining all \(L_k\), we can obtain \(L\), that is, \(L = \prod L_k\). Each quantum gate operation corresponds to an \(L_k\). Combining some operations can be helpful for designing quantum gates in algorithms. Conversely, decomposing \(L\) according to operations can also be helpful for understanding the algorithms. Decomposition and combination are the basic mathematical methods for handling \(L\).

\subsection{\SpecialProperty\ Conditions} \label{sec:self_retaining_cancellation}
Only \(L\) that satisfies specific conditions can achieve \SpecialProperty\ for parameters during \SelectiveSummation. The following explains these conditions. First, the basic property that should be satisfied is that \(\alpha\) and \(\beta\) in the same column should \SpecialProperty. The result of \SelectiveSummation\ for \(\alpha\) should not contain \(\beta\) terms, and the result of \SelectiveSummation\ for \(\beta\) should not contain \(\alpha\) terms. At the same time, the number of equations for \SelectiveSummation\ of \(\alpha\) should be equal to that for \(\beta\), otherwise the measurement sensitivity for \(\alpha\) and \(\beta\) will be different. Finally, an implicit constraint that must be satisfied is that the sum of all terms in the same column should be zero; otherwise, it will violate the normalization of the probability equations. Next, we detail the \SpecialProperty\ conditions.

\textbf{Condition I:} The \SpecialProperty\ condition for \(\alpha\) and \(\beta\) in the same column is that \(L\) contains negative signs and \(i\).

After factoring out the constant coefficients, the \(\alpha, \beta\) terms in any \(ij\) column of the probability equations can be written as \(\operatorname{Re}\, \, (\alpha + i\beta) \cdot L\), where \(L\) can take four possible values: \(1, -1, i, -i\). Let the number of rows (equations) for these values be \(n_1, n_{-1}, n_i, n_{-i}\), respectively. When \(L\) equals \(1\) or \(-1\), the result of \(\operatorname{Re}\, \, (\alpha + i\beta) \cdot L\) will contain only \(\alpha\) and not \(\beta\); when \(L\) equals \(i\) or \(-i\), the result of \(\operatorname{Re}\, \, (\alpha + i\beta) \cdot L\) will contain only \(\beta\) and not \(\alpha\). This ensures that the result of \SelectiveSummation\ for \(\alpha\) will not contain \(\beta\) terms, and the result of \SelectiveSummation\ for \(\beta\) will not contain \(\alpha\) terms.

Since for \(L = i^{a_1} i^{a_2} \ldots i^{a_n}\), the total number of rows is \(2^n\), that is, \(a_1 a_2 \ldots a_n\) ranges from \(00\ldots0\) to \(11\ldots1\). It can be calculated that
\begin{equation}
\label{C}
\begin{aligned}
n_1 + n_{-1} &= C_n^0 + C_n^2 + C_n^4 + \cdots \\
n_i + n_{-i} &= C_n^1 + C_n^3 + C_n^5 + \cdots
\end{aligned}
\end{equation}
Because
\begin{equation}
\label{CC}
(1-1)^n = \left(C_n^0 + C_n^2 + C_n^4 + \cdots\right) - \left(C_n^1 + C_n^3 + C_n^5 + \cdots\right) = 0.
\end{equation}
Thus, \(n_1 + n_{-1} = n_i + n_{-i}\). Moreover, multiplying \(L\) by any \((-1)^{b_1} (-1)^{b_2} \ldots (-1)^{b_k}\) does not change the ratio between \(n_1 + n_{-1}\) and \(n_i + n_{-i}\). Therefore, as long as \(L\) contains \(i\), it satisfies \(n_1 + n_{-1} = n_i + n_{-i}\). At this point, the number of equations for \SelectiveSummation\ of \(\alpha\) is equal to that for \(\beta\), ensuring that the algorithm has the same measurement sensitivity for \(\alpha\) and \(\beta\).

On this basis, if \(L\) also contains negative signs, then decomposing \(L\) into the non-\(i\) part \(L_1\) and the \(i\) part \(L_2\) yields that the number of rows \(N_{-1}\) where \(L_1\) takes the value \(-1\) is equal to the number of rows \(N_1\) where it takes the value \(1\). For \(L_2\), the values are \(1, -1, i, -i\), with their respective row counts denoted as \(n'_1, n'_{-1}, n'_i, n'_{-i}\). Then, for the overall \(L\),
\begin{equation}
\label{nnnn}
\begin{aligned}
n_1 &= N_1 \cdot n'_1 + N_{-1} \cdot n'_{-1} ,\\
n_{-1} &= N_1 \cdot n'_{-1} + N_{-1} \cdot n'_1 ,\\
n_i &= N_1 \cdot n'_i + N_{-1} \cdot n'_{-i} ,\\
n_{-i} &= N_1 \cdot n'_{-i} + N_{-1} \cdot n'_i.
\end{aligned}
\end{equation}

Combining the conditions \(N_1 = N_{-1}\) and \(n'_1 + n'_{-1} = n'_i + n'_{-i}\), we can obtain \(n_1 = n_{-1} = n_i = n_{-i}\). Therefore, when summing all terms in the same column, the number of \(+\alpha\) terms is equal to the number of \(-\alpha\) terms, and the number of \(+\beta\) terms is equal to the number of \(-\beta\) terms. The result of the summation is zero, satisfying the normalization constraint of the probability equations.

Thus, the \SpecialProperty\ condition for the same \(ij\) column is that \(L\) contains negative signs and \(i\). This is a fundamental condition, and subsequent discussions on the \SpecialProperty\ condition between any two columns are based on this. According to the discussion in Section~\ref{subsec:L_properties} on the properties of the current \SpecialProperty\ column, it can be noted that, based on the current circuit setup, this condition has already been satisfied.

\textbf{Condition II:} One of the \SpecialProperty\ conditions between any two columns is:
\[
\operatorname{Re}\,\, (\alpha_1 + i\beta_1) \cdot L_1 \longleftrightarrow \operatorname{Re}\,\, (\alpha_2 + i\beta_2) \cdot L_2
\]
If the common non-\(i\) part of \(L_1\) and \(L_2\) contains negative signs and is not identical, then \SpecialProperty\ is guaranteed between the two columns.

\(L_1\) and \(L_2\) can be decomposed into their respective common non-\(i\) parts through bit-wise decomposition. The bits corresponding to this part in \(L_1\) and \(L_2\) have power functions with bases of \(1\) or \(-1\), not \(i\) or \(-i\). For example,
\begin{equation}
\label{example3}
\begin{aligned}
L_1 &= i^{b_1} (-i)^{b_2} (-1)^{b_3} \cdot (-1)^{a_1} (-1)^{a_2} ,\\
L_2 &= (-1)^{b_1} (-i)^{b_2} i^{b_3} \cdot (-1)^{a_2}.
\end{aligned}
\end{equation}
For \(L_1\), the common non-\(i\) part is \((-1)^{a_1} (-1)^{a_2}\), and for \(L_2\), it is \(1^{a_1} (-1)^{a_2}\). Both contain negative signs and are not identical. Therefore, the columns \(L_1\) and \(L_2\) in the example exhibit \SpecialProperty. After \SelectiveSummation\ of the \(L_1\) column, the result for the \(L_2\) column is zero, and vice versa.

Next, we prove this condition. Let the common non-\(i\) parts of \(L_1\) and \(L_2\) be \(L'_1\) and \(L'_2\), respectively, and the corresponding set of bits for these parts be \(A\), while the remaining parts correspond to another set of bits \(B\). Since \(L'_1\) and \(L'_2\) both contain negative signs and are not identical, when iterating over all possible configurations of \(A\), in the rows where \(L'_1 = 1\), \(L'_2\) has half of the rows being \(1\) and the other half being \(-1\); similarly, in the rows where \(L'_1 = -1\), \(L'_2\) has half of the rows being \(1\) and the other half being \(-1\). For any configuration of \(B\), keeping \(B\) fixed, the results of \(\operatorname{Re}\, (\alpha_1 + i\beta_1) \cdot L_1\) and \(\operatorname{Re}\, (\alpha_2 + i\beta_2) \cdot L_2\) will be \(\pm \alpha\) or \(\pm \beta\). Iterating over \(A\), regardless of whether we are performing \SelectiveSummation\ on the \(+\alpha\) or \(-\alpha\) terms, or the \(+\beta\) or \(-\beta\) terms in the \(L_1\) column, the summation result for the \(L_2\) column will always consist of two parts, one positive and one negative, which cancel each other out, resulting in a sum of zero.

Therefore, for the complete iteration over \(AB\), performing \SelectiveSummation\ on the \(L_1\) column will always yield a summation result of zero for the \(L_2\) column. Viewing \(L_2\) as \(L_1\), the same logic applies: performing \SelectiveSummation\ on the \(L_2\) column will always yield a summation result of zero for the \(L_1\) column. Thus, the two columns corresponding to \(L_1\) and \(L_2\) exhibit \SpecialProperty.

\textbf{Condition III:} The second \SpecialProperty\ condition between any two columns is:
\[
\operatorname{Re}\,\, (\alpha_1 + i\beta_1) \cdot L_1 \longleftrightarrow \operatorname{Re}\,\, (\alpha_2 + i\beta_2) \cdot L_2
\]
When \textbf{Property II} is satisfied, the common \(i\)-bit parts of \(L_1\) and \(L_2\) are different and not complex conjugates of each other.

\textbf{Condition II} and \textbf{Condition III} are parallel conditions. \(L_1\) and \(L_2\) will exhibit \SpecialProperty\ as long as they satisfy either one of these conditions. Therefore, in the following proof, we assume that \(L_1\) and \(L_2\) do not satisfy \textbf{Condition II}. At this point, due to \textbf{Property II}, the \(i\) parts of \(L_1\) and \(L_2\) are aligned, and the non-\(i\) parts are identical. We first prove that when \(L_1\) and \(L_2\) are complex conjugates of each other, the columns \(L_1\) and \(L_2\) do not exhibit \SpecialProperty.

\begin{figure}[H]  
    \centering
    \includegraphics[width=0.36\textwidth]{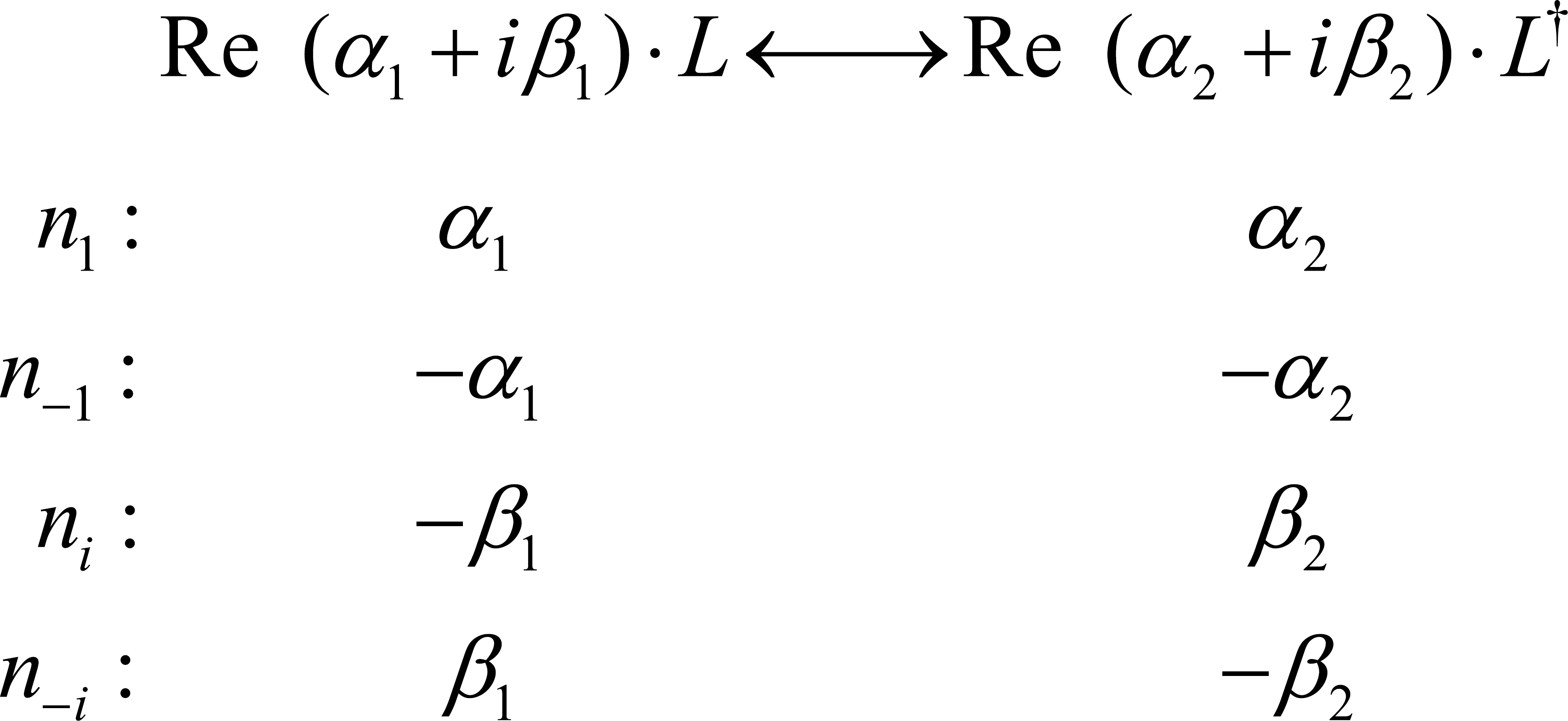}
    \caption{
        The figure shows the results of exhaustive enumeration by category, traversing all possible cases of the expression, which can be divided into four categories in total.
    }
    \label{fig:appendix_3}
\end{figure}

Let \(L_1 = L\) and \(L_2 = L^\dagger\), where \(L\) takes values \(1, -1, i, -i\) with respective row counts \(n_1, n_{-1}, n_i, n_{-i}\). By exhaustively enumerating all cases by category (summing identical rows and combining them into one category), we can obtain the results shown in Fig.~\ref{fig:appendix_3}.

As shown in Fig.~\ref{fig:appendix_3}, there are four categories of situations. This is because the values of \(L\) and \(L^\dagger\) correspond one-to-one. In each row, when \(L = 1\), \(L^\dagger = 1\); when \(L = -1\), \(L^\dagger = -1\); when \(L = i\), \(L^\dagger = -i\); and when \(L = -i\), \(L^\dagger = i\). From the results of the exhaustive enumeration by category, it is easy to see that for \SelectiveSummation\ of \(\alpha\) in the \(L_1\) column, the summation result in the \(L_2\) column is \(n_1 \alpha_2\); for \SelectiveSummation\ of \(-\alpha\) in the \(L_1\) column, the summation result in the \(L_2\) column is \(-n_{-1} \alpha_2\); for \SelectiveSummation\ of \(\beta\) in the \(L_1\) column, the summation result in the \(L_2\) column is \(n_{-i} \beta_2\); and for \SelectiveSummation\ of \(-\beta\) in the \(L_1\) column, the summation result in the \(L_2\) column is \(n_1 \alpha_2\). Thus, the \(L_1\) column and the \(L_2\) column do not exhibit \SpecialProperty.

\begin{figure}[H]  
    \centering
    \includegraphics[width=0.46\textwidth]{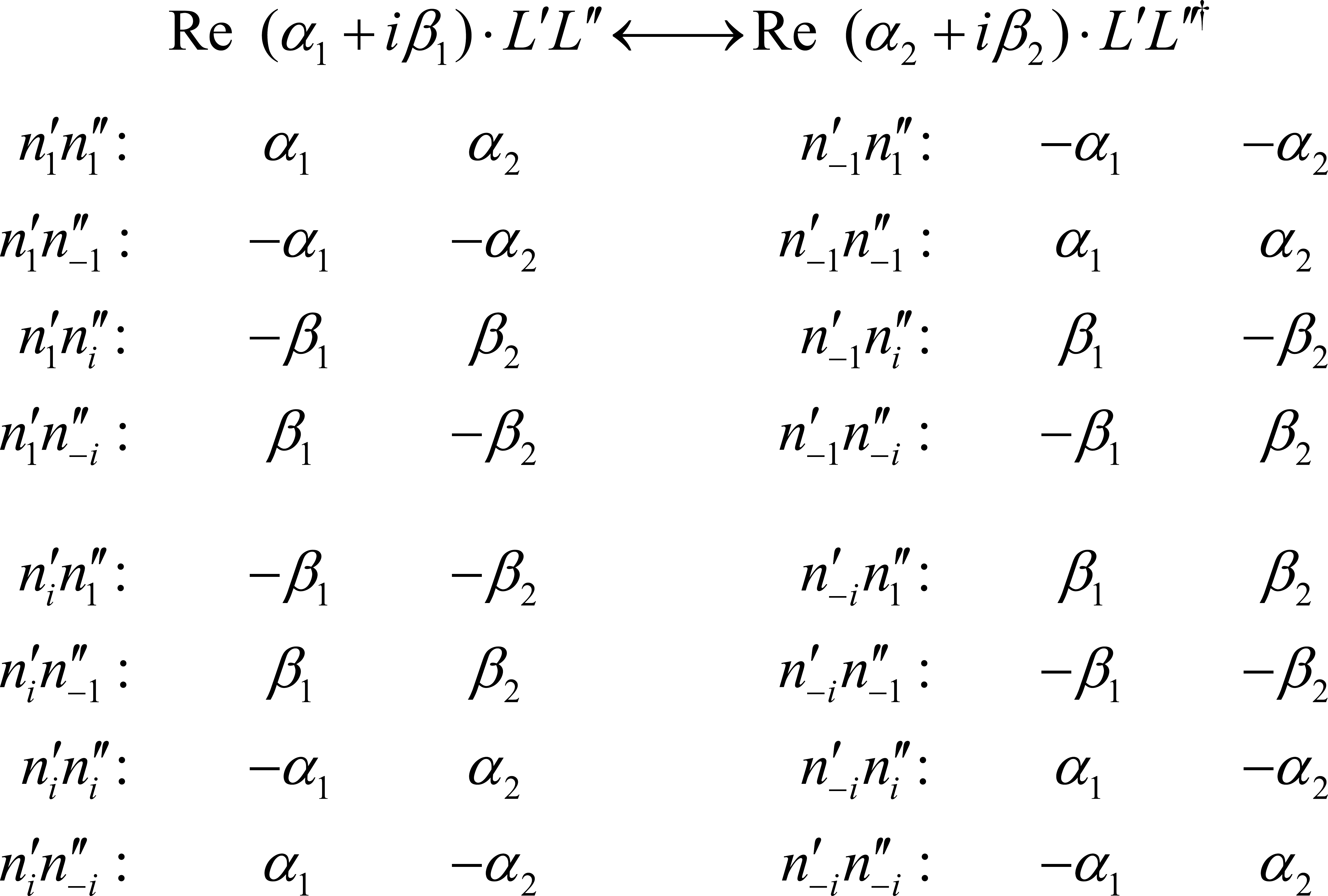}
    \caption{
        The figure shows the results of exhaustive enumeration by category, traversing all possible cases of the expression, which can be divided into 16 categories in total.
    }
    \label{fig:appendix_4}
\end{figure}

Next, we prove that when the common \(i\)-bit parts of \(L_1\) and \(L_2\) are different and not complex conjugates, the \(L_1\) column and the \(L_2\) column exhibit \SpecialProperty. At this point, \(L_1\) and \(L_2\) can be decomposed into identical parts and conjugate parts through bit-wise decomposition. Let \(L_1 = L' L''\) and \(L_2 = L' L''^\dagger\), where \(L'\) contains negative signs and \(i\), with row counts satisfying \(n'_1 = n'_{-1} = n'_i = n'_{-i}\), and \(L''\) contains \(i\), with row counts satisfying \(n''_1 + n''_{-1} = n''_i + n''_{-i}\). After exhaustively enumerating all cases by category, we can obtain the results shown in Fig.~\ref{fig:appendix_4}.

From the results of the exhaustive enumeration by category in Fig.~\ref{fig:appendix_4} and the row count relationships of \({L}'\) and \({L}''\), it is easy to see that performing \SelectiveSummation\ on \(\alpha\), \(-\alpha\), \(\beta\), and \(-\beta\) in the \(L_1\) column, the summation results in the \(L_2\) column are
\begin{equation}
\label{sum-to-zero}
\begin{aligned}
\alpha: \quad & n'_1 n''_1 \alpha_2 + n'_{-1} n''_{-1} \alpha_2 + n'_i n''_{-i} (-\alpha_2) + n'_{-i} n''_i (-\alpha_2) = 0 ,\\
-\alpha: \quad & n'_1 n''_{-1} (-\alpha_2) + n'_{-1} n''_1 (-\alpha_2) + n'_i n''_i \alpha_2 + n'_{-i} n''_{-i} \alpha_2 = 0 ,\\
\beta: \quad & n'_1 n''_{-i} (-\beta_2) + n'_{-1} n''_i (-\beta_2) + n'_i n''_{-1} \beta_2 + n'_{-i} n''_1 \beta_2 = 0 ,\\
-\beta: \quad & n'_1 n''_i \beta_2 + n'_{-1} n''_{-i} \beta_2 + n'_i n''_1 (-\beta_2) + n'_{-i} n''_{-1} (-\beta_2) = 0.
\end{aligned}
\end{equation}
By regarding \(L_2\) as \(L_1\), it follows that performing \SelectiveSummation\ on the \(L_2\) column will definitely result in a summation of zero for the \(L_1\) column. Thus, the columns \(L_1\) and \(L_2\) exhibit \SpecialProperty.

The above three conditions constitute the \SpecialProperty\ conditions. Among them, the first condition is a fundamental requirement that each \(L\) corresponding to \(ij\) must satisfy, which is already guaranteed by the current circuit setup. The second and third conditions are parallel conditions. If \(L_1\) and \(L_2\) satisfy either of these conditions, the two columns will exhibit \SpecialProperty. The process of designing the algorithmic circuit involves properly setting the quantum gates so that any two \(L\)s corresponding to all \(ij\) satisfy either the second or the third condition.

\subsection{Classification of Columns} \label{sec:classification_of_columns} 
\begin{figure}[H]  
    \centering
    \includegraphics[width=0.7\textwidth]{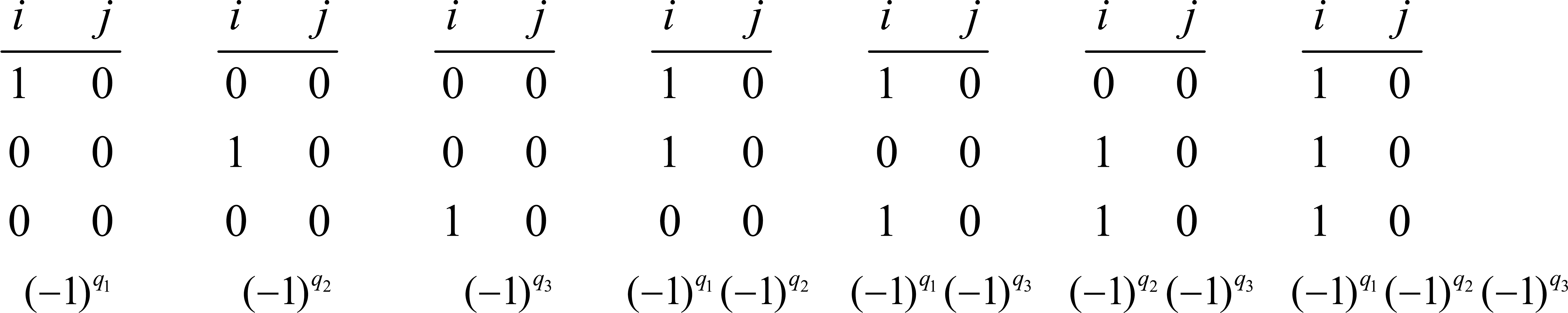}
    \caption{
    The figure illustrates an example with three qubits. There are a total of 7 classes and their corresponding 7 types of \singlecontrol\ results.
    }
    \label{fig:appendix_5}
\end{figure}

Next, we discuss how to ensure that any two \(L\)s corresponding to \(ij\) satisfy the \SpecialProperty\ condition. To simplify this problem, we first classify each \(ij\) column. Based on the setup of the \singlecontrol\ part in the current circuit and \SpecialProperty\ condition two, we can classify each \(ij\) column according to the result of the \singlecontrol. Since two columns with different \singlecontrol\ results definitely satisfy \SpecialProperty\ condition two, we group columns with the same \singlecontrol\ result into one class. Specifically, for \(n\)-bit binary numbers \(i\) and \(j\), let \(j = 0\), and the range of \(i\) is from \(00\ldots01\) to \(11\ldots11\). We can use such \(ij\) as the label for classification and define them as the first in their respective classes. In this way, we classify \(\frac{(4^n - 2^n)}{2}\) columns into \(2^n - 1\) classes. For example, when \(n = 3\), we can classify all 28 columns into the 7 classes shown in Fig.~\ref{fig:appendix_5}, each with different \singlecontrol\ results.

\begin{figure}[H]  
    \centering
    \includegraphics[width=0.7\textwidth]{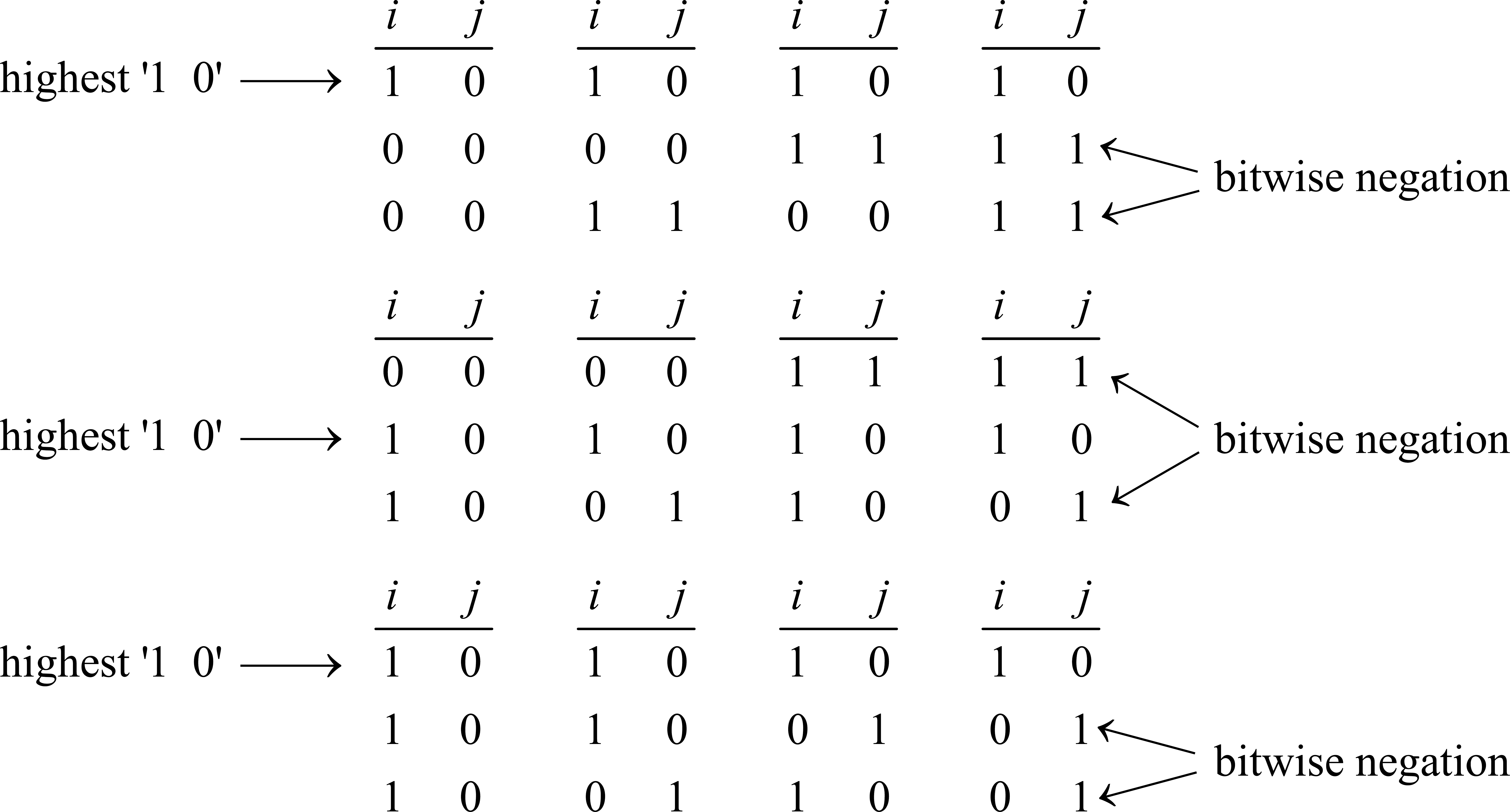}
    \caption{
    The figure shows three examples where all \(ij\) pairs in a class are obtained from the first \(ij\).
    }
    \label{fig:appendix_6}
\end{figure}

For the first \(ij\) in any class, by applying bitwise negation to each bit other than the highest '1 0', all \(ij\) pairs in that class can be obtained. Each class contains \(2^{n-1}\) \(ij\) pairs. For example, the three examples shown in Fig.~\ref{fig:appendix_6} list all \(ij\) pairs in their respective classes using this method.

Classifying each \(ij\) column based on the result of the \singlecontrol\ has two main advantages. First, because any two columns from different classes are guaranteed to exhibit \SpecialProperty, this classification simplifies our task from ensuring \SpecialProperty\ between any two \(ij\) columns to ensuring it only within the same class. The second advantage is that the bits corresponding to the \singlecontrol\ in \(L\) are the target measurement qubits. This classification method has already utilized all the information from the \singlecontrol. Since both the \phasecontrol\ and \doublecontrol\ operate on auxiliary bits, we can focus solely on the auxiliary bits part of \(L\) after bit-wise decomposition in subsequent analyses, which further simplifies our problem.

\subsection{Judgment Based on Pure Negative Parts} \label{sec:judgment_pure_negative}

The current problem is how to set the quantum gates so that any two columns within each class exhibit \SpecialProperty. What needs to be considered is the auxiliary bit part \(L_f\) obtained after bit-wise decomposition of the complete phase \(L_{ij}\). To further simplify the problem, we decompose \(L_f\) into the pure negative part \(F\) and the pure \(i\) part \(L\), both of which correspond to \(n_f\) auxiliary bits (note that the naming of \(L\) here might cause confusion; \(L\) here refers only to the pure \(i\) part of the auxiliary bit part, which is the precursor of the mask \(L\) mentioned in Section~\ref{subsec:density-reconstruction}. Therefore, we start to keep it consistent from here on, and all subsequent \(L\)s will be the same as here, no longer representing the complete \(L_{ij}\)). Columns within the same class have the same \singlecontrol\ result, and according to Property Two, their \phasecontrol\ parts are aligned, so their pure \(i\) parts \(L\) are the same. In judging whether any two columns within the same class exhibit \SpecialProperty, we should focus on the \(F\) of each column. At this point, \SpecialProperty\ conditions two and three can be simplified to the negative judgment condition.

\textbf{Negative Judgment Condition:} The \SpecialProperty\ condition between any two columns within the same class is:
\[
\operatorname{Re}\,\, (\alpha_1 + i\beta_1) \cdot F_1 L \longleftrightarrow \operatorname{Re}\,\, (\alpha_2 + i\beta_2) \cdot F_2 L
\]
where \(F_1\) and \(F_2\) are different, and the differing bits are not exactly the same as the bits of \(L\).

It is easy to prove that for any \(L = i^{a_1} i^{a_2} \ldots i^{a_k}\), defining \(F_L = (-1)^{a_1} (-1)^{a_2} \ldots (-1)^{a_n}\), we have \(F_L \cdot L = L^\dagger\). When the differing bits of \(F_1\) and \(F_2\) are exactly the same as the bits of \(L\), \(F_1\) and \(F_2\) differ by \(F_L\), i.e., \(F_1 = F_2 \cdot F_L\). In this case, \(F_1 L = F_2 F_L L = (F_2 L)^\dagger\), which does not satisfy condition three. Except for this situation, as long as \(F_1\) and \(F_2\) are different, the relationship between \(F_1 L\) and \(F_2 L\) will definitely satisfy at least one of conditions two and three. Therefore, such a simplification can be made.

After classifying the columns in this way, we can more conveniently use the negative judgment condition to determine whether the columns in the same class exhibit \SpecialProperty.

\subsection{Linear Independence Condition} \label{sec:linear_independence_condition}
Next, we discuss how to achieve \SpecialProperty\ among columns of the same classification using the pure negative part \(F\). The pure negative part \(F\) has two sources: one is from \doublecontrol, and the other is from the negative control in \phasecontrol. Before further discussion, we need to clarify the concepts of \heterobit\ and \homobit. When \(i\) and \(j\) are written as \(n\)-bit binary numbers, for a pair \((i_k, j_k)\), if \(i_k \neq j_k\), then the \(k\)-th bit is called a \heterobit; if \(i_k = j_k\), then the \(k\)-th bit is called a \homobit. According to the discussion at the beginning of this chapter, \heterobit s are crucial for the activation of control. \heterobit s directly determine the occurrence of \phasecontrol\ and whether it is positive or negative control. Additionally, \doublecontrol\ also revolves around \heterobit s. \doublecontrol\ is controlled by two bits, but when \doublecontrol\ occurs, at least one of the bits must be a \heterobit. From the definitions of classifications and \heterobit s, it can be seen that a classification corresponds to a set of \heterobit s, and a set of \heterobit s also corresponds to a single classification.

Next, we first consider an example with only a single \heterobit\ to illustrate how to achieve \SpecialProperty\ among columns of the same classification. Initially, we focus on making the \(F\) of each column different, and we will address issues related to \(F_L\) later.

\begin{figure}[H]  
    \centering
    \includegraphics[width=0.7\textwidth]{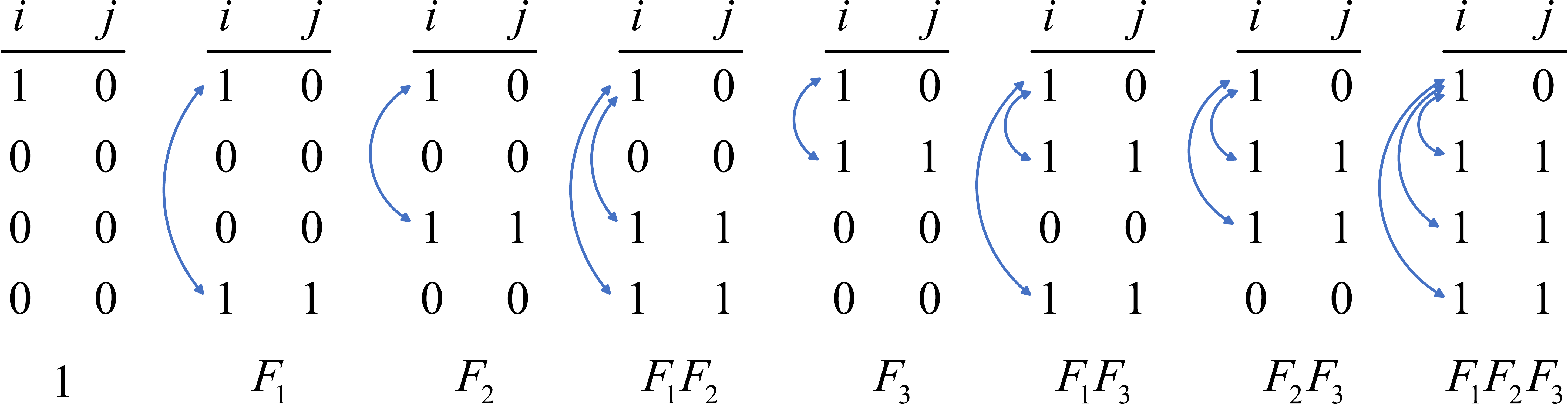}
    \caption{
    The figure shows an example of the class for four qubits where the first \(ij\) pair is \(i = 1000, j = 0000\).
    }
    \label{fig:appendix_7}
\end{figure}

In the example of Fig.~\ref{fig:appendix_7}, the \(F\) of each column is brought by \doublecontrol, where \(F_1 = L_{14}\), \(F_2 = L_{13}\), \(F_3 = L_{12}\), and the \(F\) of other columns are combinations of \(F_1\), \(F_2\), and \(F_3\). We can define a set of \(F\)s to be linearly independent: for a set of \(F_1, F_2, \ldots, F_n\), if any \(F_k\) cannot be represented by a combination of the other \(F\)s, then this set of \(F\)s is linearly independent. We can also equivalently define that, for a set of \(F_1, F_2, \ldots, F_n\), if there does not exist any \(F_{a_1} F_{a_2} \ldots F_{a_k} = 1\), then this set of \(F\)s is linearly independent; otherwise, it is linearly dependent. In the example above, if \(F_1\), \(F_2\), and \(F_3\) are linearly independent, then the \(F\)s of all columns in the class will be distinct from each other; otherwise, it would violate the definition of linear independence because there exists a combination of \(F_1\), \(F_2\), and \(F_3\) equal to 1.

In the general discussion, we let the CS gate controlled by the \(k\)-th bit act on the \(s_k\)-th auxiliary bit, requiring only that each \(s_k\) bit be distinct from each other, without requiring \(s_k\) to be \(f_k\). For \doublecontrol, we still denote the result of a set of gates controlled by the \(k_1\)-th and \(k_2\)-th bits as \(L_{k_1 k_2}\).

\begin{figure}[H]  
    \centering
    \includegraphics[width=0.7\textwidth]{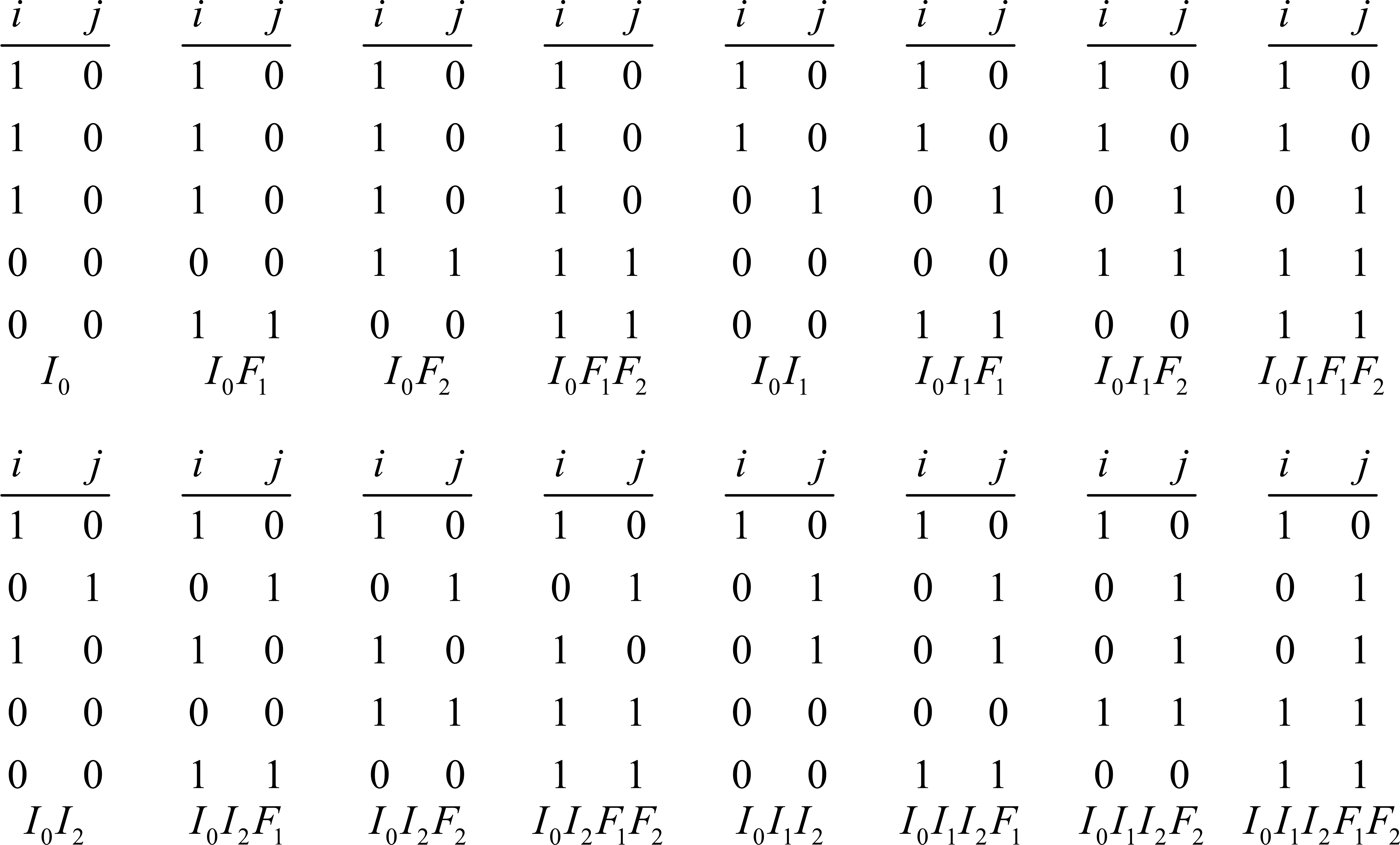}
    \caption{
    The figure shows an example of the class for five qubits where the first \(ij\) pair is \(i = 11100, j = 00000\). The \(F\) for each column are represented by \(I_0\), the various \(F_G\), and the various \(I_M\).
    }
    \label{fig:appendix_8}
\end{figure}

Now we discuss the general case. Each class has a set of determined \heterobit s \(x_1, x_2, \ldots, x_k\). We define \(I_0\) as the result of \doublecontrol\ between all pairs of \heterobit s, that is, \(I_0 = \prod_{a \neq b} L_{ab}\), where \(a, b \in \{x_1, x_2, \ldots, x_k\}\). We define \(I_M\) as the result of combining the negative part of the negative control at the \(M\)-th \heterobit\ from the bottom (counting from \(x_k\) to \(x_1\)) with the \doublecontrol\ between this \heterobit\ and all other \heterobit s, that is, \(I_M = S_m \cdot \prod_{m \neq a} L_{ma}\), where \(m\) is the \(M\)-th \heterobit\ from the bottom, \(a \in \{x_1, x_2, \ldots, x_k\}\), and \(S_m\) is the negative part of the negative control \((-1)^{s_m}\). We define \(F_G\) as the result of \doublecontrol\ between the \(G\)-th \homobit\ from the bottom and all \heterobit s, that is, \(F_G = \prod_{g \neq a} L_{ga}\), where \(g\) is the \(G\)-th \homobit\ from the bottom, and \(a \in \{x_1, x_2, \ldots, x_k\}\). An example is shown in Fig.~\ref{fig:appendix_8}.

The condition for \SpecialProperty\ among columns in the same class is that \(F_1, F_2, \ldots, F_{n-k}, I_1, I_2, \ldots, I_{k-1}, F_L\) are linearly independent. This condition is called the linear independence condition. Note that when $k = n$, the condition contains no $F_G$ terms, and when $k = 1$, it contains neither $I_M$ terms nor the $I_0$ term.

\subsection{Binary Processing} \label{sec:binary_processing}
After decomposing \(L_f\) into the pure negative part \(F\) and the pure \(i\) part \(L\), we can use a binary approach to make it easier for computers to process. Suppose the pure negative part \(F\) corresponds to \(m\) bits \(a_1, a_2, \ldots, a_m\). Then we can represent \(F\) with an \(m\)-bit binary number \(F^{(2)}\), where the part of \(F\) with \((-1)^{a_k}\) is set to 1, and the part of \(F\) with \(1^{a_k}\) is set to 0 to obtain \(F^{(2)}\). For example, in Eq.~\eqref{F-example}, \(F\) corresponds to 6 bits.
\begin{equation}
\label{F-example}
\begin{aligned}
 & F={{(-1)}^{{{a}_{1}}}}{{(-1)}^{{{a}_{2}}}}{{(-1)}^{{{a}_{5}}}}={{(-1)}^{{{a}_{1}}}}{{(-1)}^{{{a}_{2}}}}{{1}^{{{a}_{3}}}}{{1}^{{{a}_{4}}}}{{(-1)}^{{{a}_{5}}}}{{1}^{{{a}_{6}}}} ,\\ 
 & {{F}^{(2)}}=110010 .\\
\end{aligned}
\end{equation}

Similarly, for the pure \(i\) part \(L\), suppose it corresponds to \(m\) bits \(a_1, a_2, \ldots, a_m\). Then we can also represent \(L\) with an \(m\)-bit binary number \(L^{(2)}\), where the part of \(L\) with \(i^{a_k}\) is set to 1, and the part of \(L\) with \(1^{a_k}\) is set to 0 to obtain \(L^{(2)}\).

Representing \(F\) and \(L\) as binary numbers facilitates processing by computers. First, consider the combination calculation for the pure negative part. For \(F = F_1 \cdot F_2\), at any given bit position, there are four possible calculation processes and corresponding results:
\begin{equation}
\label{four-possibilities}
\begin{array}{*{35}{l}}
   {{(-1)}^{a}}\cdot {{(-1)}^{a}}=1 & \,\,\,\,\,1\oplus 1=0  \\
   {{(-1)}^{a}}\cdot 1={{(-1)}^{a}} & \,\,\,\,\,1\oplus 0=1  \\
   1\cdot {{(-1)}^{a}}={{(-1)}^{a}} & \,\,\,\,\,0\oplus 1=1  \\
   1\cdot 1=1 & \,\,\,\,\,0\oplus 0=0  \\
\end{array}
\end{equation}

From Eq.~\eqref{four-possibilities}, it can be seen that these four possibilities correspond to the XOR operation of binary numbers. According to the previous definition of the binary representation of the pure negative part, we have \(F^{(2)} = F_1^{(2)} \oplus F_2^{(2)}\). Therefore, after representing the pure negative part as a binary number, we can equivalently handle the combination of a series of \(F_k\) using a series of XOR operations on \(F_k^{(2)}\).

The binary representation also facilitates the judgment of the linear independence condition. We can define a set of binary numbers to be linearly independent: for a set of binary numbers \(B_1, B_2, \ldots, B_n\), if there does not exist any \(B_{a_1} \oplus B_{a_2} \oplus \ldots \oplus B_{a_k} = 0\), then this set of binary numbers is linearly independent; otherwise, it is linearly dependent. By treating each binary number as a row and each bit within the binary numbers as a column, and replacing the operations of multiplying by coefficients and adding/subtracting with XOR, we can use Gaussian elimination to efficiently determine whether a set of binary numbers is linearly independent. If the rank equals the number of rows, the set is linearly independent; if the rank is less than the number of rows, the set is linearly dependent (since a rank less than the number of rows implies the existence of a zero row, meaning there exists \(B_{a_1} \oplus B_{a_2} \oplus \ldots \oplus B_{a_k} = 0\)). In this way, we can equivalently determine whether the linear independence condition is satisfied by judging the relationship between a series of binary numbers.

The binary representation also facilitates the calculation of the specific results of \(F\) and \(L\) for given readings. For a specific pair \(\alpha_{ij}\) and \(\beta_{ij}\), we can divide the \(2^{n + n_f}\) possible readings into 4 events through \SelectiveSummation, and estimate the values of \(\alpha_{ij}\) and \(\beta_{ij}\) based on the counts of these four events in the measurements. However, to determine which of the four events the count \(N_{q_1 q_2 \ldots q_n f_1 f_2 \ldots f_{n_f}}\) for a specific reading \(q_1 q_2 \ldots q_n f_1 f_2 \ldots f_{n_f}\) should be assigned to, we need to know the specific value of \(\operatorname{Re}\,\,(\alpha_{ij} + i \beta_{ij}) \cdot L_{ij}\), which can be \(+\alpha\), \(-\alpha\), \(+\beta\), or \(-\beta\). This, in turn, requires us to know the specific values of \(F\) and \(L\) for the given reading. Given a specific reading, the value of \(F\) is the product of a series of \(1\)s and \(-1\)s. At any bit position \(a\), there are four possible calculation processes and results.
\begin{equation}
\label{four-operations}
\begin{matrix}
   {{1}^{0}}=1 & \,\,\,\,\,0\And 0=0  \\
   {{1}^{1}}=1 & \,\,\,\,\,0\And 1=0  \\
   {{(-1)}^{0}}=1 & \,\,\,\,\,1\And 0=0  \\
   {{(-1)}^{1}}=-1 & \,\,\,\,\,1\And 1=1  \\
\end{matrix}
\end{equation}

In Eq.~\eqref{four-operations}, these four possibilities correspond to the AND operation of binary numbers. The calculation result at bit position \(a\) is \(-1\) only if \(F\) has \((-1)^a\) at position \(a\) and the reading of \(a\) is 1; otherwise, the result is 1. According to the previous definition of the binary representation, the number of "1"s in \(F^{(2)} \And a_1 a_2 \ldots a_m\) indicates how many \(-1\)s are multiplied together in \(F\). Therefore, we can quickly determine whether the value of \(F\) is 1 or \(-1\) based on the number of "1"s in \(F^{(2)} \And a_1 a_2 \ldots a_m\). Similarly, the number of "1"s in \(L^{(2)} \And a_1 a_2 \ldots a_m\) indicates how many \(i\)s are multiplied together in \(L\), allowing us to quickly determine whether the value of \(L\) is \(1\), \(-1\), \(i\), or \(-i\).

\subsection{\GateMatrix{} and Circuit Design} \label{sec:gate_matrix_circuit_design}
The linear independence condition can be represented with the aid of a matrix.
\begin{equation}
\label{original-matrix}
\begin{matrix}
   S_1 & L_{12} & L_{13} & L_{14} & L_{15} & \cdots & L_{1n} \\
   L_{12} & S_2 & L_{23} & L_{24} & L_{25} & \cdots & L_{2n} \\
   L_{13} & L_{23} & S_3 & L_{34} & L_{35} & \cdots & L_{3n} \\
   L_{14} & L_{24} & L_{34} & S_4 & L_{45} & \cdots & L_{4n} \\
   L_{15} & L_{25} & L_{35} & L_{45} & S_5 & \cdots & L_{5n} \\
   \vdots & \vdots & \vdots & \vdots & \vdots & \ddots & \vdots \\
   L_{1n} & L_{2n} & L_{3n} & L_{4n} & L_{5n} & \cdots & S_n \\
\end{matrix}\,\,\begin{matrix}
   | \\
   | \\
   | \\
   | \\
   | \\
   | \\
   | \\
\end{matrix}\,\,\begin{matrix}
   S_1 \\
   S_2 \\
   S_3 \\
   S_4 \\
   S_5 \\
   \vdots \\
   S_n \\
\end{matrix}
\end{equation}

For any given class, let the \heterobit s be \(x_1, x_2, \ldots, x_k\). We combine the rows \(x_1, x_2, \ldots, x_k\) of the matrix in Eq.~\eqref{original-matrix} column-wise, that is, we combine the elements in the same column of the matrix in Eq.~\eqref{original-matrix}. The combination results of columns \(x_2, x_3, \ldots, x_k\) correspond to the \(I_M\) defined in the linear independence condition, the combination result of the last column \(S\) corresponds to the \(F_L\) defined in the linear independence condition, and the combination results of the other columns, except for columns \(x_2, x_3, \ldots, x_k\) and the last column, correspond to the \(F_G\) defined in the linear independence condition. Therefore, the linear independence condition is satisfied as long as the combination results, excluding the \(x_1\) column, are linearly independent. Regarding the unused \(x_1\) column, due to symmetry, the combination results of columns \(x_1, x_2, \ldots, x_k\), \(F_{x_1}, F_{x_2}, \ldots, F_{x_n}\), must satisfy Eq.~\eqref{F-constraint}.
\begin{equation}
\label{F-constraint}
{{F}_{{{x}_{1}}}} \cdot {{F}_{{{x}_{2}}}} \cdot \ldots \cdot {{F}_{{{x}_{n}}}} = {{F}_{L}}.
\end{equation}
Therefore, the combination result of column \(x_1\), \(F_{x_1}\), can be used to replace the combination result of the last column \(S\), \(F_L\). Hence, the last column of the matrix can be removed.
\begin{equation}
\label{gate-matrix}
\begin{matrix}
   {{S}_{1}} & {{L}_{12}} & {{L}_{13}} & {{L}_{14}} & {{L}_{15}} & \cdots  & {{L}_{1n}}  \\
   {{L}_{12}} & {{S}_{2}} & {{L}_{23}} & {{L}_{24}} & {{L}_{25}} & \cdots  & {{L}_{2n}}  \\
   {{L}_{13}} & {{L}_{23}} & {{S}_{3}} & {{L}_{34}} & {{L}_{35}} & \cdots  & {{L}_{3n}}  \\
   {{L}_{14}} & {{L}_{24}} & {{L}_{34}} & {{S}_{4}} & {{L}_{45}} & \cdots  & {{L}_{4n}}  \\
   {{L}_{15}} & {{L}_{25}} & {{L}_{35}} & {{L}_{45}} & {{S}_{5}} & \cdots  & {{L}_{5n}}  \\
   \vdots  & \vdots  & \vdots  & \vdots  & \vdots  & \ddots  & \vdots   \\
   {{L}_{1n}} & {{L}_{2n}} & {{L}_{3n}} & {{L}_{4n}} & {{L}_{5n}} & \cdots  & {{S}_{n}}  \\
\end{matrix}
\end{equation}

The condition for the entire algorithm to be valid is that, for any set of rows \(x_1, x_2, \ldots, x_k\) in the matrix of Eq.~\eqref{gate-matrix}, the combination results after column-wise combination must be linearly independent. This condition is referred to as the entire matrix being linearly independent. The pure negative part can not only be represented by binary numbers but also by "gates." According to the discussion in the circuit section~\ref{subsec:circuit}, the bit on the power of \(S_k\) corresponds to the auxiliary bit connected by the CS gate controlled by the \(k\)-th target qubit, and the bits on the power of \(L_{k_1 k_2}\) correspond to the auxiliary bits connected by several CCZ gates controlled by the \(k_1\)-th and \(k_2\)-th target qubits. Therefore, each pure negative part in the matrix can be represented by the serial numbers of the auxiliary bits connected by CS and CCZ gates. Such a matrix is called a \GateMatrix. The \GateMatrix\ not only directly reflects the setup of quantum gates but also makes it easier to convert the serial numbers of auxiliary bits into binary numbers, which can then be more easily used by a computer to determine whether the entire \GateMatrix\ is linearly independent.

Thus, the process of designing the circuit has shifted from the original method of setting gates and calculating a set of \(F\)s for each class, and then judging whether the algorithm is valid based on their linear independence, to the current method of filling in the numbers (serial numbers of auxiliary bits) in the \GateMatrix, converting the \GateMatrix\ into a binary \GateMatrix, and then judging whether the entire \GateMatrix\ is linearly independent to determine the validity of the algorithm.

\subsection{Rapid Classification Based on the \GateMatrix{}} \label{sec:rapid_classification}
According to the discussion in Section~\ref{sec:binary_processing}, for a specific pair \(\alpha_{ij}\) and \(\beta_{ij}\), to divide the counts from \(N\) experiments into four outcomes \(N_1, N_2, N_3, N_4\), one can quickly achieve this by performing bitwise AND operations between the binary representations of \(F\) and \(L\) and the readings \(q_1 q_2 \ldots q_n f_1 f_2 \ldots f_{n_f}\) from the \(N\) measurements. To this end, it is also necessary to have a fast method to obtain the binary representations \(F^{(2)}\) and \(L^{(2)}\) for the \(ij\) column.

Firstly, for the pure \(i\) part \(L^{(2)}\), after writing \(i\) and \(j\) in binary form and determining the pure \(i\) part \(L\) based on whether each bit is a \heterobit\ (whether the \phasecontrol\ is active, regardless of positive or negative control) and the connection of controlled phase gates, \(L\) is then converted to its binary form \(L^{(2)}\). However, this method is relatively inefficient. If the setting specification for the \phasecontrol\ part is adopted, that is, the \(k\)-th qubit controls the \(k\)-th auxiliary bit for each CS gate, then \(L^{(2)}\) can be quickly obtained by shifting \(i \oplus j\). This is because, in this case, the order of \(L^{(2)}\) will definitely match the \heterobit s of \(i, j\), so simply shifting the result of \(i \oplus j\) to the correct position of the auxiliary bits yields \(L^{(2)}\). In the introduction to the algorithm in the main text, \(L^{(2)}\) is obtained by \((i \oplus j) << (n - n_f)\). The setting specification for \phasecontrol\ can be generalized to the \(k\)-th qubit controlling the \((k + c)\)-th auxiliary bit, where \(c\) is a constant, and \(L^{(2)}\) can still be obtained by XOR and shifting, albeit with a different number of shifts. Specifically, when the \phasecontrol\ is located on the last \(n\) auxiliary bits, no shifting is required, and \(L^{(2)} = i \oplus j\).

Next, we consider the pure negative part. \(F\) is not the pure negative part of the entire \(L_{ij}\), but rather the pure negative part of the auxiliary bit part \(L_f\). The target qubit part \(L_q\) is purely negative and is also easy to calculate. It is the result of the \singlecontrol\ part of the algorithm. As long as the \(k\)-th bit of \(i, j\) is a \heterobit, the control will be activated, introducing \((-1)^{q_k}\) into \(L_q\). Therefore, \(L_q^{(2)}\) can be directly obtained by the XOR of \(i\) and \(j\), that is, \(L_q^{(2)} = i \oplus j\). Regarding \(F^{(2)}\), as discussed in Section~\ref{sec:linear_independence_condition}, \(F^{(2)}\) is a combination of \(I_0\) and \(F_1, F_2, \ldots, F_{n-k}, I_1, I_2, \ldots, I_{k-1}\) for its class. Relative to the first \(ij\) in the class, for each bit that is negated in the current \(ij\), \(F^{(2)}\) is the result of combining \(I_0\) with the corresponding \(I_M\) and \(F_G\) for those negated bits. Therefore, \(F^{(2)}\) can be calculated with the aid of the binary \GateMatrix.
\begin{equation}
\label{binary-gate-matrix}
\begin{matrix}
   S_{1}^{(2)} & L_{12}^{(2)} & L_{13}^{(2)} & L_{14}^{(2)} & L_{15}^{(2)} & \cdots  & L_{1n}^{(2)}  \\
   L_{12}^{(2)} & S_{2}^{(2)} & L_{23}^{(2)} & L_{24}^{(2)} & L_{25}^{(2)} & \cdots  & L_{2n}^{(2)}  \\
   L_{13}^{(2)} & L_{23}^{(2)} & S_{3}^{(2)} & L_{34}^{(2)} & L_{35}^{(2)} & \cdots  & L_{3n}^{(2)}  \\
   L_{14}^{(2)} & L_{24}^{(2)} & L_{34}^{(2)} & S_{4}^{(2)} & L_{45}^{(2)} & \cdots  & L_{4n}^{(2)}  \\
   L_{15}^{(2)} & L_{25}^{(2)} & L_{35}^{(2)} & L_{45}^{(2)} & S_{5}^{(2)} & \cdots  & L_{5n}^{(2)}  \\
   \vdots  & \vdots  & \vdots  & \vdots  & \vdots  & \ddots  & \vdots   \\
   L_{1n}^{(2)} & L_{2n}^{(2)} & L_{3n}^{(2)} & L_{4n}^{(2)} & L_{5n}^{(2)} & \cdots  & S_{n}^{(2)}  \\
\end{matrix}
\end{equation}

The contents of the matrix in Eq.~\eqref{binary-gate-matrix} are all binary numbers. According to the discussion in Section~\ref{sec:gate_matrix_circuit_design}, let the \heterobit s of the current \(ij\) be \(x_1, x_2, \ldots, x_k\). Then, by performing column-wise XOR operations on the rows \(x_1, x_2, \ldots, x_k\) of the binary \GateMatrix, the combination results, excluding the \(x_1\) column, correspond to \(F_1, F_2, \ldots, F_{n-k}, I_1, I_2, \ldots, I_{k-1}\). \(x_1, x_2, \ldots, x_k\) serve as row indices, and since they are the \heterobit s of \(i, j\), they can be easily obtained from \(i \oplus j\). Additionally, a set of column indices is required. Since, relative to the first \(ij\), the \(k\)-th bit of the current \(ij\) is negated, \(F^{(2)}\) will include the combination result of the \(k\)-th column. Because the \(j\) of the first \(ij\) is 0, the negated bits can be directly determined from the current \(ij\)'s \(j\), thereby obtaining the column indices. If the \(k\)-th bit of \(j\) is 1, then \(k\) is added to the column indices.

In summary, by performing XOR operations on all elements indexed by the row and column indices in the binary \GateMatrix, we obtain the part of \(F^{(2)}\) excluding \(I_0\), which we denote as \(F'^{(2)}\). For \(I_0\), using the row indices as new column indices, the elements indexed by the row and new column indices in the binary \GateMatrix\ form a sub-square matrix. According to the definition of \(I_0\), \(I_0^{(2)}\) is the result of XORing all the upper triangular elements of the sub-square matrix, excluding the diagonal elements. Finally, \(F^{(2)} = I_0 \oplus F'^{(2)}\). Concatenating \(L_q^{(2)}\) and \(F^{(2)}\) yields the \(LF\) mask in the main text, that is, \(LF = (L_q^{(2)} << n_f) \And F^{(2)}\).

In this way, for the task of calculating a specific pair \(\alpha_{ij}\) and \(\beta_{ij}\), we quickly obtain the \(F^{(2)}\) and \(L^{(2)}\) for the \(ij\) column through some binary calculations. We then iterate over the counts obtained from the experiment and classify each count \(N_{q_1 q_2 \ldots q_n f_1 f_2 \ldots f_{n_f}}\) based on the calculation result of \(\operatorname{Re}\,\,(\alpha_{ij} + i \beta_{ij}) \cdot L_{ij}\) under \(q_1 q_2 \ldots q_n f_1 f_2 \ldots f_{n_f}\): if the result is \(+\alpha\), it is assigned to \(N_1\); if \(-\alpha\), to \(N_2\); if \(+\beta\), to \(N_3\); and if \(-\beta\), to \(N_4\). After iterating through all the experimental counts, we obtain \(N_1, N_2, N_3, N_4\), and the entire process can be efficiently implemented through vectorized operations. With \(N_1, N_2, N_3, N_4\) obtained, we can then calculate the specific \(\alpha_{ij}\) and \(\beta_{ij}\). By iterating through all \(ij\) pairs in this manner, we achieve the calculation of all off-diagonal elements of the density matrix based on the experimental results.

\section{} \label{appendix:C}  
For any \GateMatrix\ solution with \(n_f = 2n - 1\) given in Section~\ref{subsec:circuit}, we can rename the auxiliary bits to transform the \GateMatrix\ into the following form:
\begin{equation}
\label{C-matrix1}
\begin{matrix}
1 & 2 & 3 & 4 & \cdots & n \\
2 & 3 & 4 & 5 & \cdots & n+1 \\
3 & 4 & 5 & 6 & \cdots & n+2 \\
4 & 5 & 6 & 7 & \cdots & n+3 \\
\vdots & \vdots & \vdots & \vdots & \ddots & \vdots \\
n & n+1 & n+2 & n+3 & \cdots & 2n-1 \\
\end{matrix}
\end{equation}
Therefore, after transforming the \GateMatrix\ into a binary \GateMatrix, any row combination will yield \(n\) binary numbers of the following form:
\begin{equation}
\label{C-matrix2}
\begin{matrix}
   {{B}_{1}}: & {{b}_{1}} & {{b}_{2}} & {{b}_{3}} & \cdots  & {{b}_{n}} & 0 & 0 & \cdots  & 0  \\
   {{B}_{2}}: & 0 & {{b}_{1}} & {{b}_{2}} & {{b}_{3}} & \cdots  & {{b}_{n}} & 0 & \cdots  & 0  \\
   {{B}_{3}}: & 0 & 0 & {{b}_{1}} & {{b}_{2}} & {{b}_{3}} & \cdots  & {{b}_{n}} & \cdots  & 0  \\
   {} & {} & {} & {} & \vdots  & {} & {} & {} & {} & {}  \\
   {{B}_{n}}: & 0 & 0 & \cdots  & 0 & {{b}_{1}} & {{b}_{2}} & {{b}_{3}} & \cdots  & {{b}_{n}}  \\
\end{matrix}
\end{equation}
Among them, \(b_1, b_2, b_3, \ldots, b_n\) are \(n\) binary numbers, the specific values of which depend on the row combination, but at least one of them is 1. It can be proven that \(B_1, B_2, B_3, \ldots, B_n\) are linearly independent. Since the highest "1" bit in \(B_1\) is "0" in \(B_2, B_3, \ldots, B_n\), \(B_1\) cannot be obtained by combining \(B_2, B_3, \ldots, B_n\). Similarly, if \(B_1, B_2, \ldots, B_k\) cannot be obtained by combining \(B_{k+1}, B_{k+2}, \ldots, B_n\), then \(B_{k+1}\) cannot be obtained by combining \(B_{k+2}, B_{k+3}, \ldots, B_n\). Conversely, if \(B_{k+1}\) can be obtained by combining the arrays other than \(B_{k+1}\), the combination will not include \(B_1, B_2, \ldots, B_k\), otherwise it would contradict the initial assumption. Therefore, if \(B_1, B_2, \ldots, B_k\) cannot be obtained by combining \(B_{k+1}, B_{k+2}, \ldots, B_n\), then \(B_{k+1}\) cannot be obtained by combining the other arrays.

At this point, by mathematical induction, it is known that any one of \(B_1, B_2, B_3, \ldots, B_n\) cannot be obtained by combining the other numbers, and \(B_1, B_2, B_3, \ldots, B_n\) are linearly independent. Therefore, the \GateMatrix\ solutions with \(n_f = 2n - 1\) are all linearly independent \GateMatrix s that can make the algorithm valid.



\end{document}